\documentclass[aps,prb,twocolumn,groupedaddress,preprintnumbers,amsmath,amssymb,floatfix,superscriptaddress,longbibliography]{revtex4-1}

\usepackage{morefloats}
\usepackage[dvips]{graphicx}
\usepackage{subcaption}
\usepackage{caption}
\usepackage{color}
\usepackage{epsfig,graphicx,amsfonts,amsbsy}
\usepackage{amsmath,amsfonts,amsthm,amssymb}
\usepackage{appendix}
\usepackage{bbm}
\usepackage{makeidx}
\usepackage{url}
\usepackage{verbatim}
\usepackage{dsfont}
\usepackage{orcidlink}
\usepackage[usenames,dvipsnames,svgnames]{xcolor}
\usepackage{hyperref}
\hypersetup{colorlinks=true, linkcolor=Blue, citecolor=Blue,urlcolor=Blue}

\usepackage{tabularx}

\makeatletter

\usepackage[version=3]{mhchem} 
\usepackage{feynmf}
\usepackage[retainorgcmds]{IEEEtrantools}
\usepackage{bbold}
\usepackage{xcolor}
\usepackage{nicefrac, xfrac}

\begin{document}

\title{Superconductivity in multi-Weyl semimetals: Conditions for the
coexistence of topological and conventional phases}
\author{Alonso Tapia}
\affiliation{Facultad de F\'isica, Pontificia Universidad Cat\'olica de Chile, Vicu\~{n}a Mackenna 4860, Santiago, Chile}
\author{Enrique Mu\~noz~\orcidlink{0000-0003-4457-0817}}
\email{ejmunozt@uc.cl}
\affiliation{Facultad de F\'isica, Pontificia Universidad Cat\'olica de Chile, Vicu\~{n}a Mackenna 4860, Santiago, Chile}
\affiliation{Center for Nanotechnology and Advanced Materials CIEN-UC, Avenida Vicuña Mackenna 4860, Santiago, Chile}

\date{\today}
\begin{abstract}
    In this work, we explore the possible emergence of superconducting phases in a multi-Weyl semimetal. In particular, we show that the presence of a pair of Weyl nodes with chirality $|\nu| \ge 1$ leads to an effective description of the intra-nodal pairings in terms of monopole harmonics, in contrast to inter-nodal pairings that preserve the angular dependence of conventional spherical harmonics. Therefore, we explore the conditions for the competition and/or coexistence between both types of superconducting phases, and we identify the presence of the so-called "topological repulsion" mechanism, which was previously reported in the context of simple Weyl semimetals. We identified the critical temperatures corresponding to the monopole and conventional superconducting phases, and calculated the specific heat as a function of temperature, thus showing that this thermodynamical parameter may provide an experimental probe to determine the chirality index $\nu$ in the material.
\end{abstract}
\maketitle
\noindent \section{Introduction}

Since their theoretical prediction, Weyl semimetals have been studied with great interest \cite{Murakami_2007, Burkov_Balents_2011, Wan_et_al_2011} due to their unusual properties. The presence of pairs of gapless nodal points connecting the valence and conduction bands leads to different quasiparticle excitations~\cite{Wehling_014,Chiu_016,Bradlyn_016,Armitage_018}. In single Weyl semimetals (WSMs), the nodal points exhibit nearly linear dispersion where quasiparticles are pseudorelativistic Weyl fermions~\cite{Burkov_018,Armitage_018}, as experimentally verified~\cite{Xu_015,Grassano_018,Zhang_016,Arnold_016,Shekhar_015} in transition metal monopnictides (TaAs\cite{Lv_et_al_2015(TaAs), Xu_et_al_2015(TaAs), Zhang_et_al_2017(TaAs)}, TaP\cite{Xu_et_al_2015(TaP), Xu_et_al_2016(TaP)}, NbAs\cite{Xu_et_al_2015(NbAs)} and NbP), and  in other materials such as ZrTe$_{5}$ \cite{Li_et_al_2016} and Cd$_{3}$As$_{2}$~\cite{Moll_et_al_2016}. Each Weyl node is a monopolar source of Berry curvature, and hence they are protected from being gapped since their monopole charge $\mathcal{C} = \pm 1$ (chirality) is a topological invariant~\cite{Burkov_018,Armitage_018}, thus involving “spin-momentum locking” for Weyl fermions. Among the remarkable properties of these materials are the presence of Fermi arcs, the chiral anomaly, and the chiral magnetic effect~\cite{Burkov_018,Armitage_018,Turner_et_al_2013}. 

Recently, another type of WSM was proposed, known as multi-Weyl \cite{Fang_et_al_2012, Huang_et_al_2016, Ahn_et_al_2016}. These materials have multiple Weyl nodes (double or triple) with an anisotropic dispersion relationship that is only linear in one direction, and quadratic or cubic in the other two directions. Therefore, the different chiral charges and the semi-linear dispersion of the electronic bands imply modified spin-momentum locking properties, as well as the possible existence of novel strongly correlated phases in the presence of interactions~\cite{Lai_PRB_2015,Jian_PRB_2015,Huang_PRB_2015,Pyatkovskiy_PRB_2016}. Examples of candidate materials are $Sr Si_2$~\cite{Huang_et_al_2016,Singh_SCIREP_2018}, $HgCr_2Se_4$~\cite{Guan_PRL_2015,Jian_PRB_2015}, $RhAs_3$~\cite{Zhu_PRB_2018}, $Rb(MoTe)_3$~\cite{Liu_PhysRevX.7.021019} and $Tl(MoTe)_3$~\cite{Liu_PhysRevX.7.021019}.

The topological features of the band structure in WSMs suggest the possible existence of diverse pairing states~\cite{Murakami_03,Meng_012,Cho-2012,Li_018,Sun_020}. In particular, the monopole superconductor (MSC), recently proposed as a pairing state between the two Fermi surfaces (FSs) enclosing the Weyl points in a doped WSM~\cite{Li_018,Sun_020}, involves the superposition of the Berry phases of the individual single-particle states, thus determining the Berry phase of the emerging Cooper pair. Therefore, the resulting chirality, i.e. the vorticity of the pair, will be characterized by the combination of those of the individual Weyl nodes. For spherically symmetric Fermi surfaces, it was shown~\cite{Li_018} that the pairing will correspond to monopole harmonic functions, in contrast to the spherical harmonics in conventional superconductivity. In order to experimentally detect those non-trivial vorticity effects, magneto-transport measurements can be applied to discriminate chiral versus non-chiral pairing states. This mechanism has been investigated by one of the authors~\cite{Munoz-2020,Muñoz_et_al_2024} in the context of the competition between the monopole pairing, characterized by the monopole harmonic functions $\mathcal{Y}_{q,j,m}(\theta,\varphi)$, with conventional spherical harmonic states $Y_{j,m}(\theta,\varphi)$. In this previous work, within a mean-field BCS theory, we showed that the monopole  and a conventional spherical harmonic phase may coexist with one another~\cite{Munoz-2020,Muñoz_et_al_2024}, while exclusion of one phase in favor of the other arises when the $\theta$-dependent form factors of the monopole harmonic $\mathcal{Y}_{|q|,j,|m|}(\theta,\varphi)$ and the spherical harmonic $Y_{j,m}(\theta,\varphi)$ are proportional to each other. We termed this mechanism as "topological repulsion"~\cite{Munoz-2020}.
In the present work, we extend the theoretical analysis~\cite{Munoz-2020,Muñoz_et_al_2024} to study the possible emergence of topological superconducting phases, and in particular the competition between conventional and monopole pairing, in the context of multi-Weyl semimetals. Furthermore, we analyze the critical temperatures for each of those phases, and the specific heat as a possible probe for the chirality index $\nu$ in multi-WSMs.
\section{The Model}
Let us start from the general form of the Hamiltonian representing a Weyl semimetal in the presence of particle-particle interactions
\begin{equation}        
    \hat{H}=\hat{H}_{0}+\hat{V}.
\end{equation}
Here, the non-interacting contribution is defined, in a generalized two-band model, by
\begin{equation}
\hat{H}_{0}=\sum_{\mathbf{k},\sigma,\sigma'}\hat{c}^\dagger_{\sigma,\mathbf{k}}\left[\mathbf{N}(\mathbf{k})\cdot\boldsymbol{\sigma}\right]_{\sigma\sigma'}\hat{c}_{\sigma',\mathbf{k}},
\label{eq_H0}
\end{equation}
for fermion operators $[ \hat{c}_{\sigma,\mathbf{k}},\hat{c}^\dagger_{\sigma',\mathbf{k}'}  ]_{+} =\delta_{\mathbf{k},\mathbf{k}'}\delta_{\sigma,\sigma'}$,
with $\boldsymbol{\sigma} = \left(\hat{\sigma}^x,\hat{\sigma}^y,\hat{\sigma}^z\right)$ the usual Pauli matrices.
The explicit form of the components of the vector $\mathbf{N}(\mathbf{k}) = \left( N_1(\mathbf{k}), N_2(\mathbf{k}), N_3(\mathbf{k}) \right)^T$ determines the topological properties of the effective two-band model, and the corresponding energy bands are trivially $E^{\pm}_{\mathbf{k}} = \pm | \mathbf{N}(\mathbf{k})|$. For instance, a double WSM can be realized by~\cite{Roy_PhysRevB_2017}
\begin{eqnarray}
N_1(\mathbf{k}) &=& t\left( \cos k_x - \cos k_y \right)\nonumber\\
N_2(\mathbf{k}) &=& t \sin k_y\sin k_x\nonumber\\
N_3(\mathbf{k}) &=& t_z\cos (k_z a) - m_z + t_0 \left\{ 6 + \cos(2 k_x )  +  \cos( 2 k_y)\right.\nonumber\\
&&\left.- 4 \cos k_x - 4\cos k_y \right\}
\end{eqnarray}

In addition, we shall assume a short-range interaction, involving a nearest-neighbors Coulomb repulsion $V_1>0$ and a contact, phonon-mediated effective attractive interaction $V_0<0$. When both terms are combined we have~\cite{Muñoz_et_al_2024}
\begin{equation}
    \hat{V}=V_0\sum_i\hat{n}_i\hat{n}_i+\frac{1}{2}V_1\sum_{\langle i,j\rangle}\hat{n}_i\hat{n}_j.
    \label{eq_V}
\end{equation}
Here $\hat{n}_i=\sum_{\sigma}\hat{c}^\dagger_\sigma(\mathbf{R}_i)\hat{c}_\sigma(\mathbf{R}_i)$ is the occupation number operator at the site $\mathbf{R}_i$ on the lattice in the local Wannier basis. By the appropriate Fourier transformation $\hat{c}_\sigma(\mathbf{R}_i)=\frac{1}{\sqrt{N}}\sum e^{i\mathbf{k}\cdot\mathbf{R}_i}\hat{c}_\sigma(\mathbf{k})$, the interaction term is expressed in the Bloch-momentum basis by
\begin{equation}
\hat{V}=\sum_{\mathbf{k},\mathbf{p},\mathbf{q},\sigma,\tau}V_\mathbf{q}\hat{c}^\dagger_{\sigma,\mathbf{k}+\mathbf{q}}\hat{c}_{\sigma,\mathbf{k}}\hat{c}^\dagger_{\tau,\mathbf{p}-\mathbf{q}}\hat{c}_{\tau,\mathbf{p}},
\label{eq:original_momentum_interaction}
\end{equation}
with a matrix element given by
\begin{equation}
    V_\mathbf{q}=V_0+V_1(\cos q_x+\cos q_y+\cos q_z).
\label{eq:lattice_potential}
\end{equation}

In order to capture the topological features of the multi-Weyl semimetal band structure, we define operators in the nodal Fermi surface representation, by expanding the momenta in the vicinity of each Weyl point as follows $\hat{\psi}_{\pm,\sigma}(\mathbf{k})=\hat{c}_\sigma(\mathbf{k}+\mathbf{K}_\pm)$ for $|\mathbf{k}|\ll|\mathbf{K}_\pm|$. In this nodal representation, a generic low-energy model for a multi-Weyl semimetal with topological charge $\nu\in\mathbb{N}$ is~\cite{Roy_PhysRevB_2017}
\begin{eqnarray}
\hat{H}_{0}&=&\sum_{\mathbf{k},a=\pm}\hat{\psi}^\dagger_a(\mathbf{k})\big\{v_F\left[\alpha_{\nu} k_{\perp}^{\nu} \cos(\nu\phi_{\mathbf{k}}) \hat{\sigma}^x\right. \nonumber\\
&& \left.+ \alpha_{\nu} k_{\perp}^{\nu} \sin(\nu\phi_{\mathbf{k}}) \hat{\sigma}^y 
 + a  k_z\hat{\sigma}^z\right] -\mu^{a}\big\}\hat{\psi}_a(\mathbf{k}),
\end{eqnarray}
where $a=\pm$ represents the sign of the nodal topological charge (chirality), and $\mu^a$ is the chemical potential at each node.

For the interaction operator, we take a similar approach, starting from the Bloch-momentum representation (\ref{eq:original_momentum_interaction}) and choosing the pairing mechanism of our interest, 
\begin{eqnarray}
\hat{V}&=&\sum_{\mathbf{k},\mathbf{q},\mathbf{p}}V^{ab,cd}_\mathbf{q}\hat{\psi}^\dagger_{a,\sigma}(\mathbf{k+q})\hat{\psi}^\dagger_{b,\tau}(\mathbf{p-q})\hat{\psi}_{c,\tau}(\mathbf{p})\hat{\psi}_{d,\sigma}(\mathbf{k}),\nonumber\\
\end{eqnarray}
where now we include all the interaction coefficients in a single matrix element $V^{ab,cd}_\mathbf{q}$. Explicitly, expanding up to second order in the momenta, we obtain~\cite{Muñoz_et_al_2024}
\begin{widetext}
\begin{eqnarray}
V^{\pm\pm,\pm\pm}_{\mathbf{k},\mathbf{q}}&\,=V^{\pm\mp,\mp\pm}_{\mathbf{k},\mathbf{q}}=V_0+V_1\bigg(3-\frac{(\mathbf{k}-\mathbf{q})^2}{2}\bigg)\nonumber\\
V^{\pm\mp,\pm\mp}_{\mathbf{k},\mathbf{q}}&=V_0+V_1\bigg\{2-\frac{|\mathbf{k}_\perp-\mathbf{q}_\perp|^2}{2}+\Big(1-\frac{(k_z-q_z)^2}{2}\Big)\cos2Q\mp(k_z-q_z)\sin 2Q\bigg\},
    \label{eq:Vcoeff}
\end{eqnarray}
\end{widetext}
with $\mathbf{k}_\perp=(k_x,k_y,0)$.  The proper treatment of these coefficients in the low-energy regime may lead to the emergence of different superconducting phases, according to their intra- or inter-node pairing structure~\cite{Munoz-2020,Muñoz_et_al_2024}.  In a standard BCS-like mean-field approximation, we write the interaction as
\begin{equation}
\hat{V}=\sum_{\mathbf{k},\sigma,\tau,a,b}\hat{\psi}^\dagger_{a,\sigma}(\mathbf{k})\Delta^{ab}_{\sigma\tau}(\mathbf{k})\hat{\psi}^\dagger_{b,\tau}(-\mathbf{k})+h.c.,
\end{equation}
where the pairing function is defined via the self-consistent relation
\begin{equation}
    \Delta^{ab}_{\sigma\tau}(\mathbf{k})=\sum_{\mathbf{q},c,d}V^{ab,cd}_{\mathbf{k},\mathbf{q}}\langle\hat{\psi}_{c,\tau}(-\mathbf{q})\hat{\psi}_{d,\sigma}(\mathbf{q})\rangle.
\end{equation}

Following the method presented in~\cite{Munoz-2020}, we introduce the Bogoliubov transformation $\hat{\alpha}^\dagger_a(\mathbf{k})=\sum_{\sigma}\zeta_{a,\sigma}(\mathbf{k})\hat{\psi}^\dagger_{a,\sigma}(\mathbf{k})$, where the spinor $\zeta_a(\mathbf{k})$ is the positive energy eigenvector of $\hat{H}_0$ at each Weyl node $a=\pm$.Thus, we have 
\begin{equation}
    \zeta_\pm (\mathbf{k})=\frac{1}{\sqrt{2\epsilon_{\mathbf{k}}(\epsilon_{\mathbf{k}}\pm k_{z})}}\begin{pmatrix}\epsilon_{\mathbf{k}}\pm k_{z}\\\alpha_{\nu} k_{\perp}^{\nu} e^{i\nu\phi_{\mathbf{k}}}\end{pmatrix},
\label{eq:bogoliubov_spinors}
\end{equation}
where was defined $\epsilon_{\mathbf{k}}\equiv\sqrt{k_{z}^{2}+\alpha^{2}_{\nu}k_{\perp}^{2\nu}}$. In this basis, the Hamiltonian is then reduced to 
\begin{equation}
\hat{H}=\sum_{\mathbf{k},a,b}\Big\{\xi_\mathbf{k}^a\hat{\alpha}^\dagger_a(\mathbf{k})\hat{\alpha}_a(\mathbf{k})+\Delta^{ab}(\mathbf{k})\hat{\alpha}^\dagger_a(\mathbf{k})\hat{\alpha}^\dagger_b(-\mathbf{k})+h.c.\Big\},
\end{equation}
where the single-particle contribution was diagonalized, leading to the dispersion relation
\begin{eqnarray}
\xi_{\textbf{k}}^{\pm} = v_{F} \sqrt{k_{z}^{2} + \alpha_{\nu}^{2} k_{\perp}^{2\nu}} - \mu^{\pm},
\end{eqnarray}
and the projected gap equation in the new basis becomes
\begin{equation}
    \Delta^{ab}(\mathbf{k})=\sum_{\mathbf{q},c,d}\Bar{V}^{ab,cd}_{\mathbf{k},\mathbf{q}}\langle\hat{\alpha}_c(-\mathbf{q})\hat{\alpha}_d(\mathbf{q})\rangle,
\label{eq:self_consistent_relation}
\end{equation}
with matrix elements $\Bar{V}^{ab,cd}_{\mathbf{k},\mathbf{q}}$ given by the corresponding transformation from the originals to this new basis.

Let us further define the Nambu spinor $\Psi^\dagger_\mathbf{k}=\big(\hat{\alpha}^\dagger_-(\mathbf{k}),\hat{\alpha}_-(-\mathbf{k}),\hat{\alpha}^\dagger_+(\mathbf{k}),\hat{\alpha}_+(-\mathbf{k})\big)$, in order to arrange the Hamiltonian in the Bogoliubov-de Gennes notation as a bilinear form~\cite{Munoz-2020, Muñoz_et_al_2024}
\begin{equation}
\hat{H}=\sum_\mathbf{k}\Psi^\dagger_\mathbf{k}\hat{H}_{\text{BdG}}(\mathbf{k})\Psi_\mathbf{k},
\end{equation}
where we defined the matrix
\begin{equation}
    \hat{H}_{BdG}(\textbf{k}) = \left(
    \begin{matrix}
        \xi_{\textbf{k}}^{-} & \Delta^{\text{intra}}_{\textbf{k}} & 0 & \Delta^{\text{inter}}_{\textbf{k}} \\
        (\Delta^{\text{intra}}_{\textbf{k}})^{*} & -\xi_{\textbf{k}}^{-} & (\Delta^{\text{inter}}_{\textbf{k}})^{*} & 0 \\
        0 & \Delta^{\text{inter}}_{\textbf{k}} & \xi_{\textbf{k}}^{+} & \Delta^{\text{intra}}_{\textbf{k}} \\
        (\Delta^{\text{inter}}_{\textbf{k}})^{*} & 0 & (\Delta^{\text{intra}}_{\textbf{k}})^{*} & -\xi_{\textbf{k}}^{+}
    \end{matrix} \right).
\end{equation}
For subsequent computations, it is convenient to express this matrix in a $\text{SU}(2)\otimes\text{SU}(2)$ basis composed by $\{\hat{\tau}_\alpha\otimes\hat{\eta}_\beta\}$, where $\hat{\tau}_\alpha$ span the subspace of the two Weyl nodes, while $\hat{\eta}_\beta$ represents the particle-hole degrees of freedom. Here, $\alpha,\beta=1,2,3$ are the Pauli matrix indexes, while $\alpha,\beta=0$ correspond to the two-dimensional representation of the identity. In this notation, we have
\begin{equation}
\begin{split}
    \hat{H}_{BdG}(\textbf{k}) &= \bar{\xi}_{\textbf{k}} \hat{\tau}_{0}\otimes\hat{\eta}_{3} + \frac{\delta\mu}{2} \hat{\tau}_{3}\otimes\hat{\eta}_{3} + \Re \Delta^{\text{intra}}_{\textbf{k}} \hat{\tau}_{0} \otimes \hat{\eta}_{1} \\
    - \Im \Delta^{\text{intra}}_{\textbf{k}}& \hat{\tau}_{0}\otimes\hat{\eta}_{2} + \Re \Delta^{\text{inter}}_{\textbf{k}} \hat{\tau}_{1}\otimes\hat{\eta}_{1} - \Im \Delta^{\text{inter}}_{\textbf{k}} \hat{\tau}_{1}\otimes\hat{\eta}_{2},
\end{split}
\end{equation}
where we defined the average chemical potential between the pair of nodes as $\bar{\mu} = (\mu_{+}+\mu_{-})/2$, and the difference by $\delta\mu = \mu_{+}-\mu_{-}$. Accordingly, the single quasiparticle dispersion at the average chemical potential is 
\begin{eqnarray}
\bar{\xi}_{\textbf{k}} = v_{F} \sqrt{k_{z}^{2} + \alpha^{2}_{\nu} k_{\perp}^{2\nu}} - \bar{\mu}.
\end{eqnarray}
\section{The Gap Equations}
\label{Section: Gap Eqns}
At finite temperature, the field theoretical representation of the partition function is
\begin{eqnarray}
Z = {\rm{Tr}}\, e^{-\beta \hat{H}} = \int \mathcal{D} \Psi^{\dagger}\mathcal{D}\Psi e^{-\int_{0}^{\beta}d\tau\sum_{\mathbf{k}}\Psi_{\mathbf{k}}^{\dagger} \left(  \frac{\partial}{\partial\tau} + \hat{H}_{BDG}\right)\Psi_{\mathbf{k}}}\nonumber\\
\end{eqnarray}
Therefore, the Green´s function matrix in momentum and Matsubara space is defined by
\begin{equation}
    \hat{\mathcal{G}}_{\textbf{k}}(\omega_{n}) = \left[-i\omega_{n} + \hat{H}_{BdG}(\textbf{k})\right]^{-1},
\end{equation}
where $\omega_{n} = (2n+1)\pi T$, $n\in\mathbb{Z}$ is the Matsubara frequency for Fermions.

The Green's function is composed by 4 blocks of $2\times 2$ matrices (details in appendix \ref{Appendix A: Green Function}), as follows
\begin{equation}
    \hat{\mathcal{G}}_{\textbf{k}}(\omega_{n}) = \left(
    \begin{matrix}
        \hat{\mathcal{G}}_{\textbf{k}}^{\text{intra},-} & \hat{\mathcal{G}}_{\textbf{k}}^{\text{inter}} \\
        (\hat{\mathcal{G}}_{\textbf{k}}^{\text{inter}})^{\dagger} & \hat{\mathcal{G}}_{\textbf{k}}^{\text{intra},+}
    \end{matrix}
    \right).
\end{equation}

Now, we formulate the BCS self-consistent equations, for both the intra-nodal and inter-nodal pairings, as follows
\begin{eqnarray}
    \Delta^{\text{intra}}_{\textbf{k}} &=& T\sum_{\textbf{k}',\omega_{n}} V_{\text{intra}}(\textbf{k},\textbf{k}') \langle \hat{\alpha}_{-}(\textbf{k}') \hat{\alpha}_{-}(-\textbf{k}') \rangle  \\
    \Delta^{\text{inter}}_{\textbf{k}} &=& T\sum_{\textbf{k}',\omega_{n}} V_{\text{inter}}(\textbf{k},\textbf{k}') \langle \hat{\alpha}_{-}(\textbf{k}') \hat{\alpha}_{+}(-\textbf{k}') \rangle.
\end{eqnarray}
The correlation functions involved in the definitions above can be directly obtained from the matrix elements of the Green's function, as follows
\begin{equation}
\begin{split}
    &\langle \hat{\alpha}_{-}(\textbf{k})  \hat{\alpha}_{-}(-\textbf{k}) \rangle = [\hat{\mathcal{G}} ^{\text{intra},-}_{\textbf{k}}]_{12} \\
    =& \frac{ \Delta^{\text{intra}}_{\textbf{k}} (E_{\textbf{k}}^{2} - \delta\mu \bar{\xi}_{\textbf{k}}) - 2 \Delta^{\text{inter}}_{\textbf{k}} B_{\textbf{k}} }{E_{\textbf{k}}^{4} - 4B_{\textbf{k}}^{2} - \delta\mu^{2}(\bar{\xi}_{\textbf{k}}^{2} + |\Delta^{\text{inter}}_{\textbf{k}}|^{2})}
\end{split}
\end{equation}
\begin{equation}
\begin{split}
    &\langle \hat{\alpha}_{-}(\textbf{k}) \hat{\alpha}_{+}(-\textbf{k}) \rangle = [\hat{\mathcal{G}} ^{\text{inter}}_{\textbf{k}}]_{12} \\
    &= \frac{ \Delta^{\text{inter}}_{\textbf{k}} (E_{\textbf{k}}^{2} - \frac{\delta\mu^{2}}{2} ) - 2 \Delta^{\text{intra}}_{\textbf{k}} B_{\textbf{k}} - i\omega_{n}\delta\mu \Delta^{\text{inter}}_{\textbf{k}} }{E_{\textbf{k}}^{4} - 4B_{\textbf{k}}^{2} - \delta\mu^{2}(\bar{\xi}_{\textbf{k}}^{2} + |\Delta^{\text{inter}}_{\textbf{k}}|^{2})}
\end{split}
\end{equation}
with
\begin{eqnarray}
    E_{\textbf{k}} &=& \sqrt{\bar{\xi}_{\textbf{k}}^{2} + |\Delta^{\text{intra}}_{\textbf{k}}|^{2} + |\Delta^{\text{inter}}_{\textbf{k}}|^{2} + \frac{\delta\mu^{2}}{4} + \omega_{n}^{2}} \\
    B_{\textbf{k}} &=& \Re\Delta^{\text{intra}}_{\textbf{k}} \Re\Delta^{\text{inter}}_{\textbf{k}} + \Im\Delta^{\text{intra}}_{\textbf{k}} \Im\Delta^{\text{inter}}_{\textbf{k}}
\end{eqnarray}
Then, we perform the sum over Matsubara frequencies $\omega_{n}$ in the gap equations (details in appendix \ref{Appendix B: Gap Eqns}), to obtain the self-consistent expressions
\begin{widetext}
\begin{equation}
\begin{split}
    \Delta^{\text{intra}}_{\textbf{k}} = \frac{1}{2} \sum_{\textbf{k}'} V_{\text{intra}}(\textbf{k},\textbf{k}') & \left[ \left( \Delta^{\text{intra}}_{\textbf{k}'} + \frac{ 2 \Delta^{\text{inter}}_{\textbf{k}'} B_{\textbf{k}'} + \delta\mu \bar{\xi}_{\textbf{k}'} \Delta^{\text{intra}}_{\textbf{k}'} }{\sqrt{4B_{\textbf{k}'}^{2} + \delta\mu^{2}(\bar{\xi}_{\textbf{k}'}^{2} + |\Delta^{\text{inter}}_{\textbf{k}'}|^{2})}} \right) \frac{\tanh(\beta \Gamma_{\textbf{k}'} / 2)}{\Gamma_{\textbf{k}'}} \right. \\ 
    & \left. + \left( \Delta^{\text{intra}}_{\textbf{k}'} - \frac{ 2 \Delta^{\text{inter}}_{\textbf{k}'} B_{\textbf{k}'} + \delta\mu \bar{\xi}_{\textbf{k}'} \Delta^{\text{intra}}_{\textbf{k}'} }{\sqrt{4B_{\textbf{k}'}^{2} + \delta\mu^{2}(\bar{\xi}_{\textbf{k}'}^{2} + |\Delta^{\text{inter}}_{\textbf{k}'}|^{2})}} \right) \frac{\tanh(\beta \gamma_{\textbf{k}'} / 2)}{\gamma_{\textbf{k}'}} \right]
\end{split}
\end{equation}
\begin{equation}
\begin{split}
    \Delta^{\text{inter}}_{\textbf{k}} = \frac{1}{2} \sum_{\textbf{k}'} V_{\text{inter}}(\textbf{k},\textbf{k}') & \left[ \left( \Delta^{\text{inter}}_{\textbf{k}'} + \frac{ 2 \Delta^{\text{intra}}_{\textbf{k}'} B_{\textbf{k}'} + \frac{\delta\mu^{2}}{2}  \Delta^{\text{inter}}_{\textbf{k}'} }{\sqrt{4B_{\textbf{k}'}^{2} + \delta\mu^{2}(\bar{\xi}_{\textbf{k}'}^{2} + |\Delta^{\text{inter}}_{\textbf{k}'}|^{2})}} \right) \frac{\tanh(\beta \Gamma_{\textbf{k}'} / 2)}{\Gamma_{\textbf{k}'}} \right. \\ 
    & \left. + \left( \Delta^{\text{inter}}_{\textbf{k}'} - \frac{ 2 \Delta^{\text{intra}}_{\textbf{k}'} B_{\textbf{k}'} + \frac{\delta\mu^{2}}{2}  \Delta^{\text{inter}}_{\textbf{k}'} }{\sqrt{4B_{\textbf{k}'}^{2} + \delta\mu^{2}(\bar{\xi}_{\textbf{k}'}^{2} + |\Delta^{\text{inter}}_{\textbf{k}'}|^{2})}} \right) \frac{\tanh(\beta \gamma_{\textbf{k}'} / 2)}{\gamma_{\textbf{k}'}} \right],
\end{split}
\end{equation}
\end{widetext}
where were defined
\begin{equation}
\begin{split}
    \Gamma_{\textbf{k}}^{2} &= \bar{\xi}_{\textbf{k}}^{2} + |\Delta^{\text{intra}}_{\textbf{k}}|^{2} + |\Delta^{\text{inter}}_{\textbf{k}}|^{2} + \frac{\delta\mu^{2}}{4} \\
    & + \sqrt{4B_{\textbf{k}}^{2} + \delta\mu^{2}(\bar{\xi}_{\textbf{k}}^{2} + |\Delta^{\text{inter}}_{\textbf{k}}|^{2})}, \\
\end{split}
\end{equation}
\begin{equation}
\begin{split}
    \gamma_{\textbf{k}}^{2} &= \bar{\xi}_{\textbf{k}}^{2} + |\Delta^{\text{intra}}_{\textbf{k}}|^{2} + |\Delta^{\text{inter}}_{\textbf{k}}|^{2} + \frac{\delta\mu^{2}}{4} \\
    & - \sqrt{4B_{\textbf{k}}^{2} + \delta\mu^{2}(\bar{\xi}_{\textbf{k}}^{2} + |\Delta^{\text{inter}}_{\textbf{k}}|^{2})}.
\end{split}
\end{equation}
\section{The Zero Temperature Limit}
\label{sec_T0}
We shall study the competition between both pairings, and therefore the corresponding superconducting phases in the zero temperature limit $T\rightarrow 0$. For small values of $|\Delta_{\mathbf{k}}^{\text{intra}}|$ and $|\Delta_{\mathbf{k}}^{\text{inter}}|$, and assuming that the chemical potential is nearly symmetrical at both nodes $|\delta\mu/\overline{\mu}| \ll 1$, the gap equations reduce to
\begin{equation}
    \Delta^{\eta}_{\textbf{k}} = \frac{1}{2} \sum_{\textbf{k}'} V_{\eta}(\textbf{k},\textbf{k}') \Delta^{\eta}_{\textbf{k}'} \left(\frac{1}{\Gamma_{\textbf{k}'}} + \frac{1}{\gamma_{\textbf{k}'}}\right),
    \label{eq_BCS_T0}
\end{equation}
where $\eta=\text{intra}/\text{inter}$. 

We shall assume the following angular dependence on the interaction potentials~\cite{Munoz-2020,Muñoz_et_al_2024}
\begin{equation}
    \begin{split}
        V_{\text{intra}}(\textbf{k},\textbf{k}') & =  V_{0}^{\text{intra}} Y_{l,m'} (\theta_{\textbf{k}}, \phi_{\textbf{k}}) Y_{l,m'}^{*} (\theta_{\textbf{k}'}, \phi_{\textbf{k}'}) \\
        V_{\text{inter}}(\textbf{k},\textbf{k}') & = V_{0}^{\text{inter}} \mathcal{Y}_{\nu,j,m} (\theta_{\textbf{k}}, \phi_{\textbf{k}}) \mathcal{Y}_{\nu,j,m}^{*} (\theta_{\textbf{k}'}, \phi_{\textbf{k}'}),
    \end{split}
\end{equation}
with $Y_{l,m} (\theta_{\textbf{k}},\phi_{\textbf{k}})$ and $\mathcal{Y}_{\nu,j,m}(\theta_{\textbf{k}},\phi_{\textbf{k}})$ the conventional spherical and monopole harmonics, respectively. These angular dependencies are consequently inherited by the pairing functions,
\begin{equation}
    \begin{split}
        \Delta^{\text{intra}}_{\textbf{k}} & =  \Delta_{l,m'} Y_{l,m'} (\theta_{\textbf{k}}, \phi_{\textbf{k}}) = \Delta_{l,m'} f_{\text{intra}}(\theta_{\textbf{k}}) e^{im' \phi_{\textbf{k}}} \\
        \Delta^{\text{inter}}_{\textbf{k}} & =  \bar{\Delta}_{\nu} \mathcal{Y}_{\nu,j,m} (\theta_{\textbf{k}}, \phi_{\textbf{k}}) = \bar{\Delta}_{\nu} f_{\text{inter}}(\theta_{\textbf{k}}) e^{i(m+\nu) \phi_{\textbf{k}}}.
    \end{split}
\end{equation}
Inserting these expressions into Eq.~\eqref{eq_BCS_T0}, we get
\begin{equation}
    \frac{1}{V_{0}^{\eta}} = \frac{1}{2} \sum_{\textbf{k}}  |f_{\eta}(\theta_{\textbf{k}})|^{2} \left(\frac{1}{\Gamma_{\textbf{k}}} + \frac{1}{\gamma_{\textbf{k}}}\right).
\end{equation}
Going to the continuum limit, and defining the density of states $\rho(\xi) = \int\frac{d^3 k}{(2\pi)^3}\delta(\xi - \xi_{\mathbf{k}})$ (details in Appendix~\ref{Appendix C: Continuum limit}), we obtain the expressions
\begin{eqnarray}
    \frac{1}{\lambda_{\eta}} &=& \frac{1}{2B_{\nu}} \int_{0}^{1}\frac{dx x}{\sqrt{1-x^{2\nu}}} \int_{0}^{2\pi} \frac{d\phi}{2\pi} |f_{\eta}(\theta_{x})|^{2}\nonumber\\
    &&\times\int_{-\omega_{D}}^{\omega_{D}} d\xi  \left(\frac{1}{\Gamma_{\textbf{k}}} + \frac{1}{\gamma_{\textbf{k}}}\right),
\end{eqnarray}
where the effective couplings were defined as $\lambda_{\eta} = \rho(0) V_{0}^{\eta}$, with $\rho(0)$ the density of states function at the Fermi level. The coefficient $B_{\nu} = \beta(\frac{1}{\nu}, \frac{1}{2})/(2\nu)$ defined in terms of the Beta function, and $\theta_{x}$ is defined according to the expression
\begin{equation}
    \tan\theta_{x} = \frac{cx}{\sqrt{1-x^{2\nu}}},
    \label{eq:tantx}
\end{equation}
with the constant $c=\alpha^{-1/\nu} (\mu/v_{F})^{1/\nu-1}$.

By further performing the integral in $\xi$, we end up with
\begin{widetext}
\begin{equation}
    \frac{1}{\lambda_{\eta}} = \frac{1}{B_{\nu}} \int_{0}^{1}\frac{dx x}{\sqrt{1-x^{2\nu}}} |f_{\eta}(\theta_{x})|^{2} \int_{0}^{2\pi} \frac{d\phi}{2\pi} \left[ 2\ln(2\omega_{D}) - \frac{1}{2} \sum_{s=\pm} \ln(A_{+} + 2sB_{\textbf{k}}) \right],
\end{equation}
\end{widetext}
where 
\begin{equation}
A_{\pm} = \Delta_{l,m'}^{2} f_{\text{intra}}^{2}(\theta_x) \pm \bar{\Delta}_{\nu}^{2} f_{\text{inter}}^{2}(\theta_x).
\end{equation}
Noting that 
\begin{eqnarray}
2B_{\textbf{k}} &=& 2\Delta_{l,m'} \bar{\Delta}_{\nu} f_{\text{intra}}(\theta_x) f_{\text{inter}}(\theta_x) \cos[(m'-m-\nu) \phi]\nonumber\\ 
&\equiv& B \cos(r\phi),
\end{eqnarray}
the integral in $\phi$ must be solved in different ways depending on the cases $m'=m+\nu$ or $m'\neq m+\nu$, respectively.

For Case 1: $m'=m+\nu$, we have
\begin{widetext}
\begin{equation}
    \frac{1}{2} \int_{0}^{2\pi} \frac{d\phi}{2\pi} \sum_{s=\pm} \ln (A_{+} + 2sB_{\textbf{k}}) = \frac{1}{2} \sum_{s=\pm} \ln (A_{+} + sB) = \frac{1}{2} \ln(A_{+}^{2} - B^{2}) = \ln(|A_{-}|),
\end{equation}
and then we obtain
\begin{equation}
    \frac{1}{\lambda_{\eta}} = \frac{1}{B_{\nu}} \int_{0}^{1}\frac{dx x}{\sqrt{1-x^{2\nu}}} |f_{\eta}(\theta_{x})|^{2}\ln\left[ \frac{4\omega_{D}^{2}}{\left|\Delta_{l,m'}^{2} f_{\text{intra}}^{2}(\theta_{x}) - \bar{\Delta}_{\nu}^{2} f_{\text{inter}}^{2}(\theta_{x})\right|} \right].
    \label{eq_case1}
\end{equation}
\end{widetext}
On the other hand, for Case 2: $m'\neq m+\nu$, we solve
\begin{widetext}
\begin{equation}
\begin{split}
    \int_{0}^{2\pi} \frac{d\phi}{2\pi}  \ln \left[A_{+} \pm B\cos(r\phi)\right] & = \sum_{n=1}^{r} \int_{2\pi(n-1)/r}^{2\pi n/r} \frac{d\phi}{2\pi}  \ln [A_{+} \pm B\cos(r\phi)] = \frac{1}{r} \sum_{n=1}^{r} \int_{2\pi(n-1)}^{2\pi n} \frac{d\phi'}{2\pi}  \ln (A_{+} \pm B\cos\phi') \\
    = & \int_{0}^{2\pi} \frac{d\phi'}{2\pi}  \ln (A_{+} \pm B\cos\phi') = \ln\left[\frac{1}{2} \left(A_{+} + \sqrt{A_{+}^{2} - B^{2}} \right)\right] = \ln\left[\frac{1}{2} \left(A_{+}+|A_{-}|\right)\right],
\end{split}
\end{equation}
and then we obtain
\begin{equation}
    \frac{1}{\lambda_{\eta}} = \frac{1}{B_{\nu}} \int_{0}^{1}\frac{dx x}{\sqrt{1-x^{2\nu}}} |f_{\eta}(\theta_{x})|^{2} \left\{ 2\ln(2\omega_{D}) - \ln\left[\frac{1}{2}(A_{+}+|A_{-}|)\right] \right\}.
    \label{eq_case2}
\end{equation}
\end{widetext}

\subsection{Topological Repulsion}
Let us now examine whether the mechanism that we call "topological repulsion"~\cite{Munoz-2020} arises in the case of multi-Weyl semimetals as well, according to our present model.
Consider Case 1: $m'=m+\nu$, corresponding to Eq.~\eqref{eq_case1}, and further assume that $f_{\text{intra}}(\theta_x) \approx f_{\text{inter}}(\theta_x)$. Then, we have
\begin{equation}
\begin{split}
    \frac{1}{\lambda_{\eta}} = & \frac{1}{B_{\nu}} \int_{0}^{1}\frac{dx x}{\sqrt{1-x^{2\nu}}} |f_{\text{intra}}(\theta_{x})|^{2} \\
    & \times \ln\left[ \frac{4\omega_{D}^{2}}{f_{\text{intra}}^{2}(\theta_{x}) \left|\Delta_{l,m'}^{2} - \bar{\Delta}_{\nu}^{2} \right|} \right].
\end{split}
\end{equation}
This equation is analogous to the one obtained in \cite{Munoz-2020}, but here we extend it to multi-Weyl semimetals. Since the right-hand side is the same for $\eta = {\text{intra}}$ and $\eta = {\text{inter}}$, we conclude that the phase boundary occurs precisely at $\lambda_{\text{intra}}=\lambda_{\text{inter}}$, thus leaving no room for coexistence in the phase diagram. This is precisely the topological repulsion mechanism already described in~\cite{Munoz-2020} for simple Weyl semimetals.

\subsection{s-wave vs Monopole}

\begin{figure*}[ht]
    \centering
    \begin{subfigure}[b]{0.32\textwidth}
        \centering
        \includegraphics[width=\textwidth]{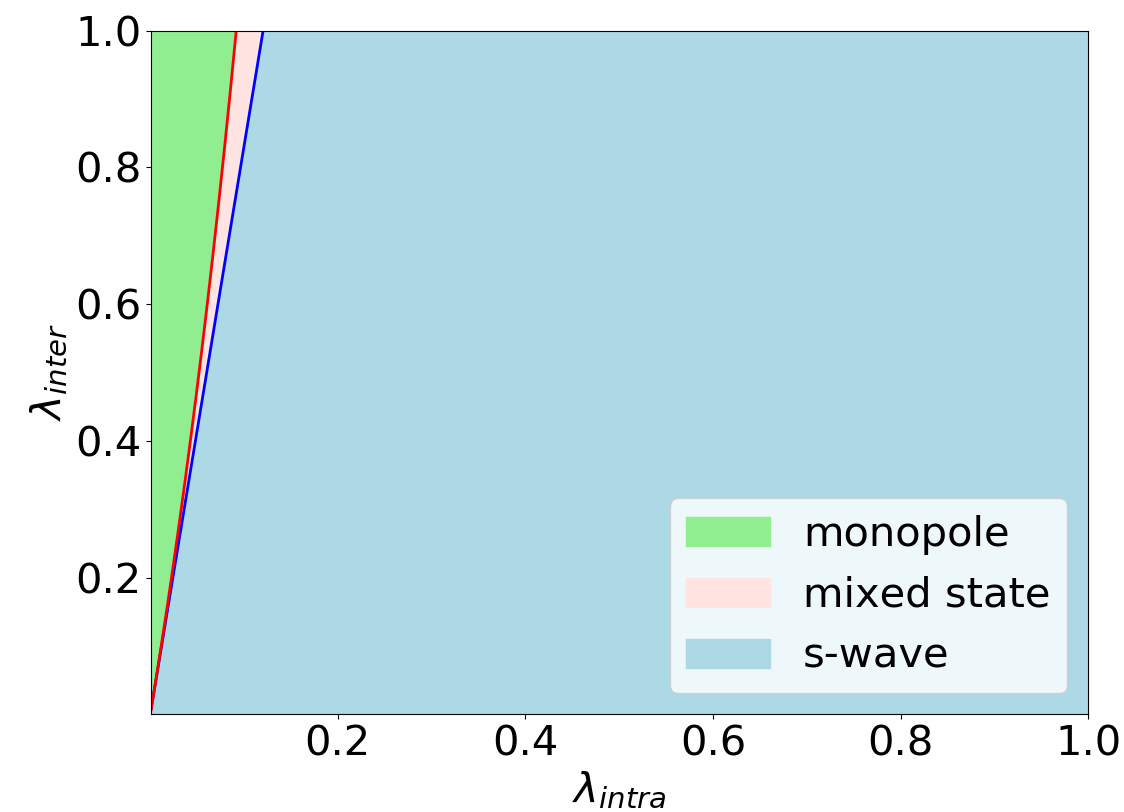}
        \caption{$\nu=1$, $c=0.1$}
        \label{s_wave_a}
    \end{subfigure}
    \begin{subfigure}[b]{0.32\textwidth}
        \centering
        \includegraphics[width=\textwidth]{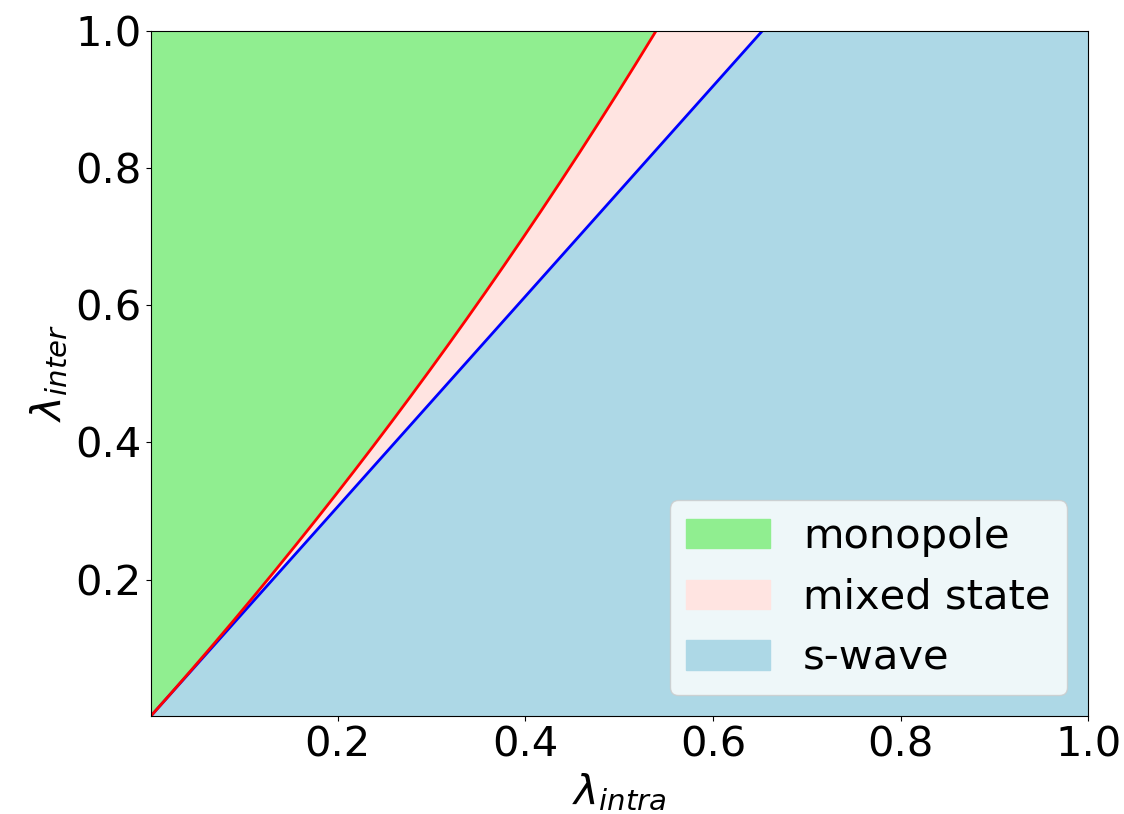}
        \caption{$\nu=1$, $c=1$}
        \label{s_wave_b}
    \end{subfigure}
    \begin{subfigure}[b]{0.32\textwidth}
        \centering
        \includegraphics[width=\textwidth]{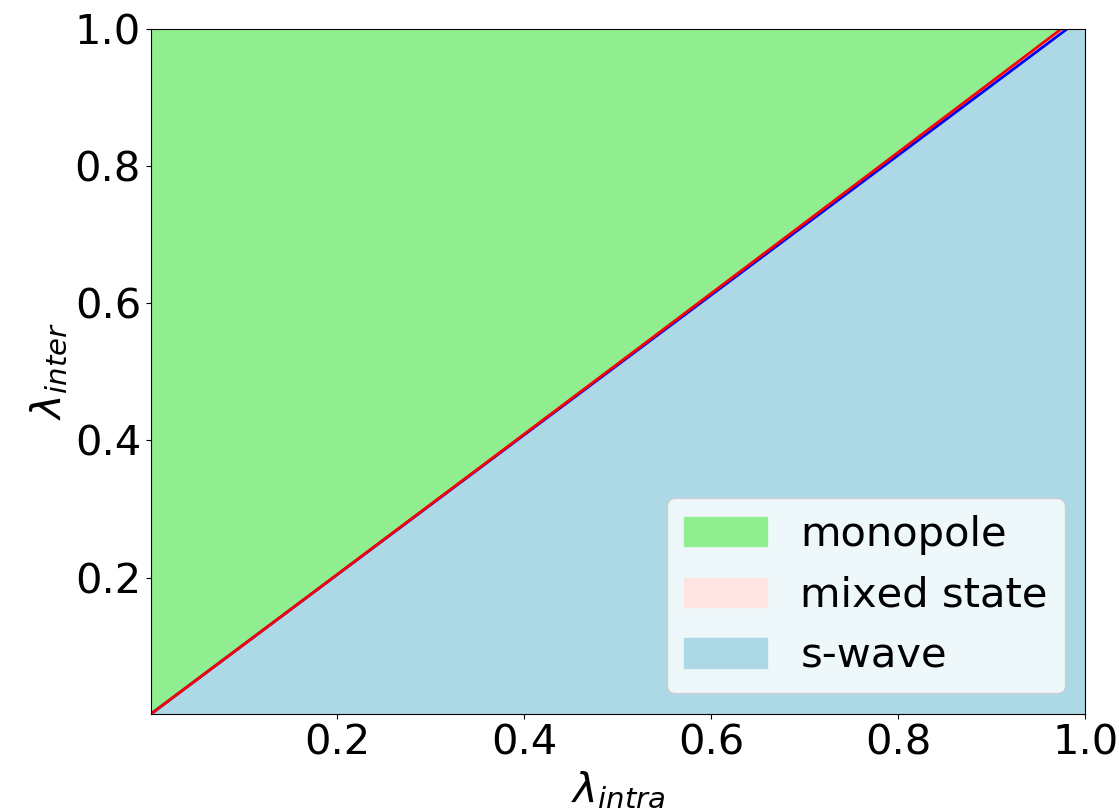}
        \caption{$\nu=1$, $c=10$}
        \label{s_wave_c}
    \end{subfigure}
    \begin{subfigure}[b]{0.32\textwidth}
        \centering
        \includegraphics[width=\textwidth]{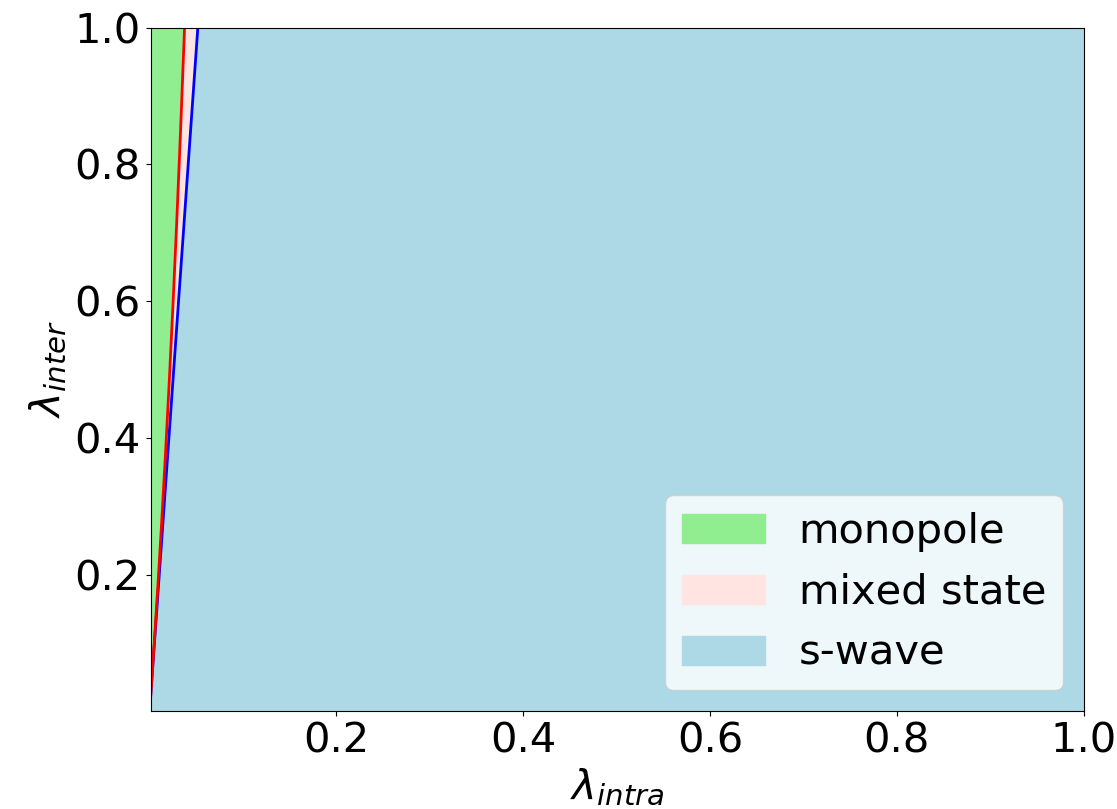}
        \caption{$\nu=2$, $c=0.1$}
        \label{s_wave_d}
    \end{subfigure}
    \begin{subfigure}[b]{0.32\textwidth}
        \centering
        \includegraphics[width=\textwidth]{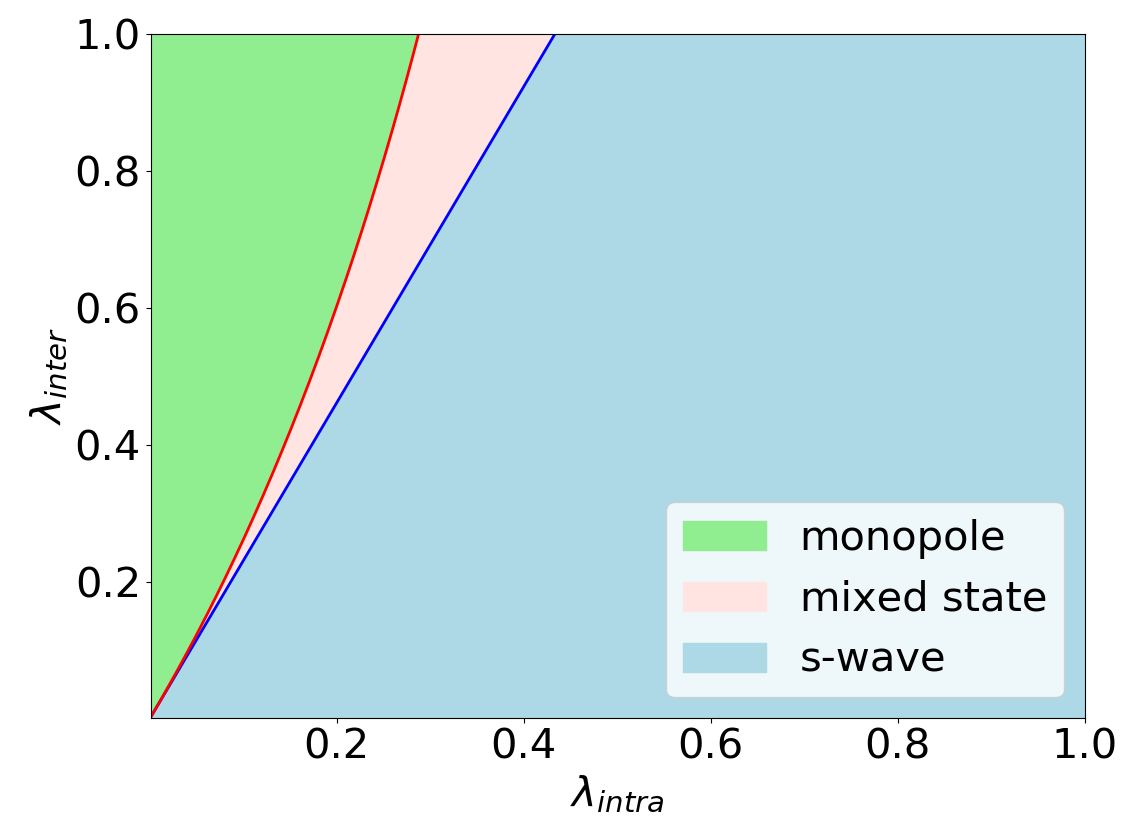}
        \caption{$\nu=2$, $c=1$}
        \label{s_wave_e}
    \end{subfigure}
    \begin{subfigure}[b]{0.32\textwidth}
        \centering
        \includegraphics[width=\textwidth]{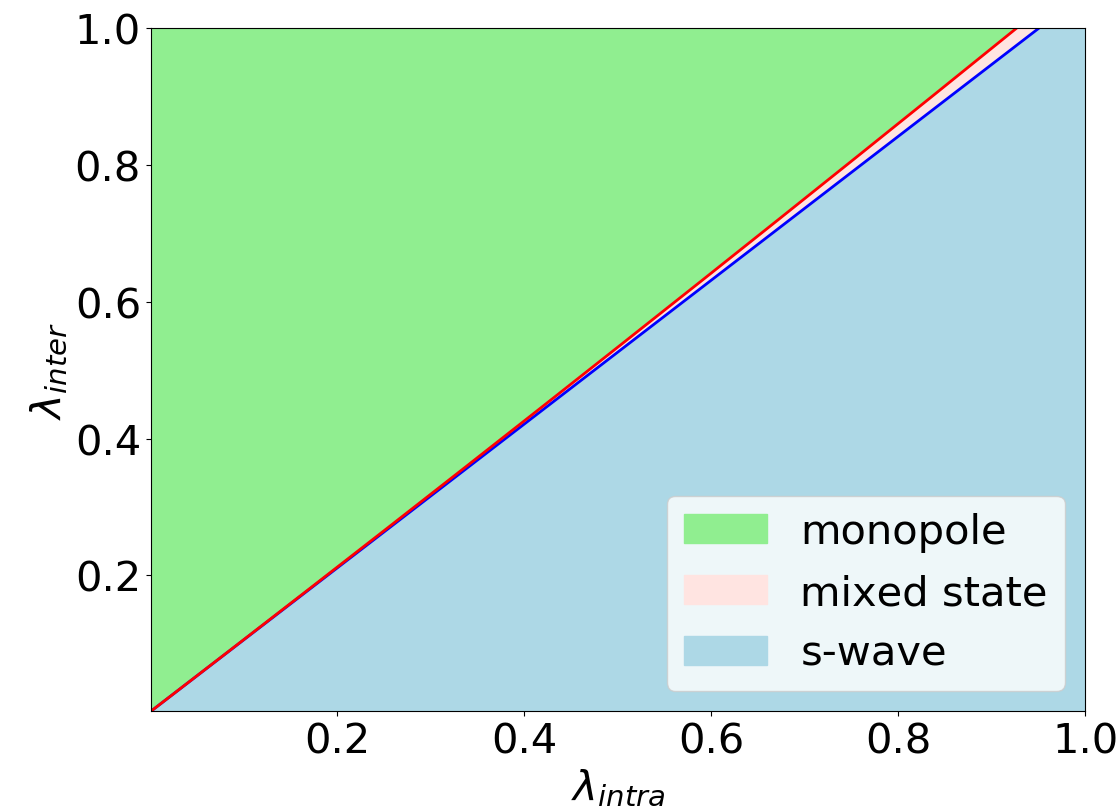}
        \caption{$\nu=2$, $c=10$}
        \label{s_wave_f}
    \end{subfigure}
    \begin{subfigure}[b]{0.32\textwidth}
        \centering
        \includegraphics[width=\textwidth]{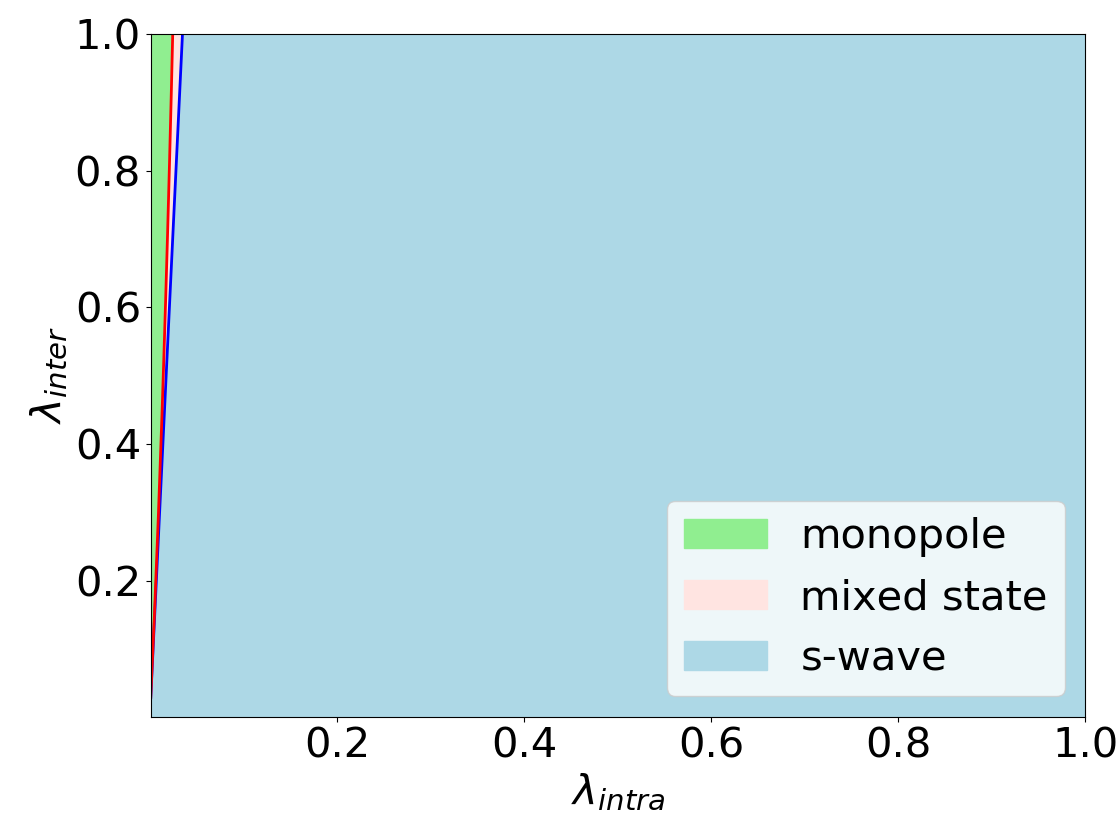}
        \caption{$\nu=3$, $c=0.1$}
        \label{s_wave_g}
    \end{subfigure}
    \begin{subfigure}[b]{0.32\textwidth}
        \centering
        \includegraphics[width=\textwidth]{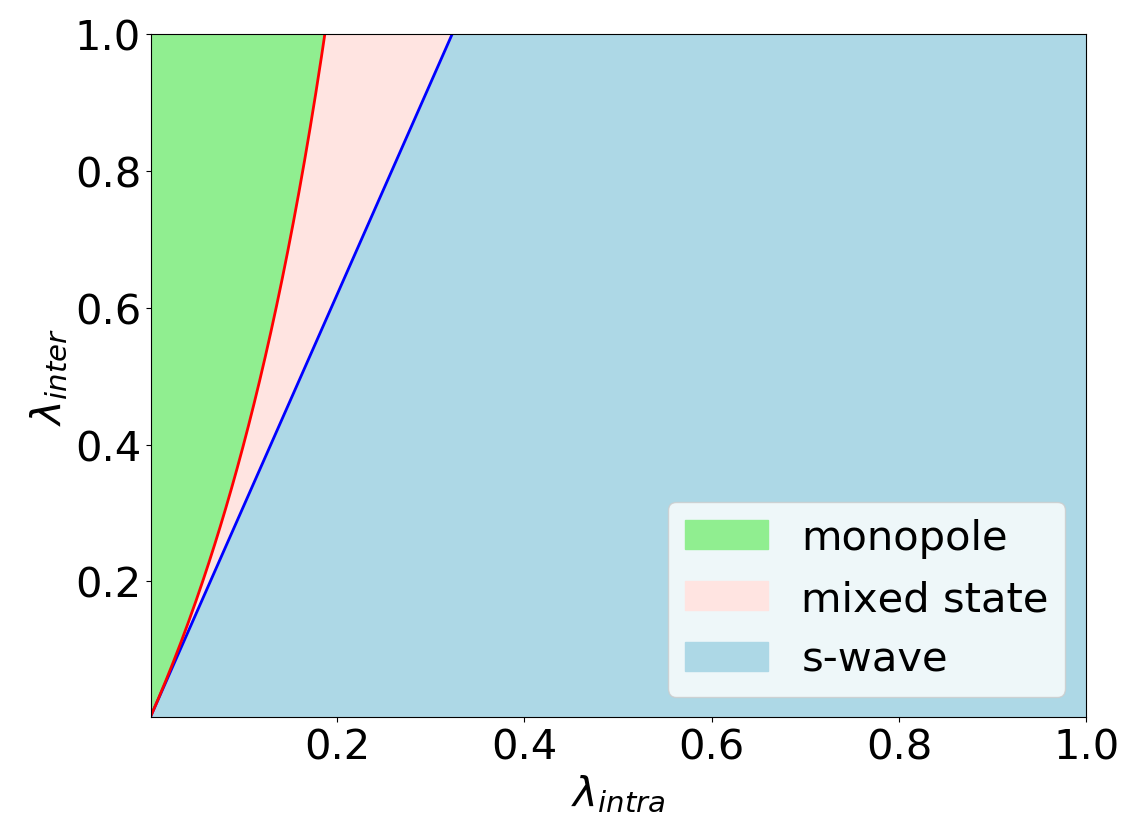}
        \caption{$\nu=3$, $c=1$}
        \label{s_wave_h}
    \end{subfigure}
    \begin{subfigure}[b]{0.32\textwidth}
        \centering
        \includegraphics[width=\textwidth]{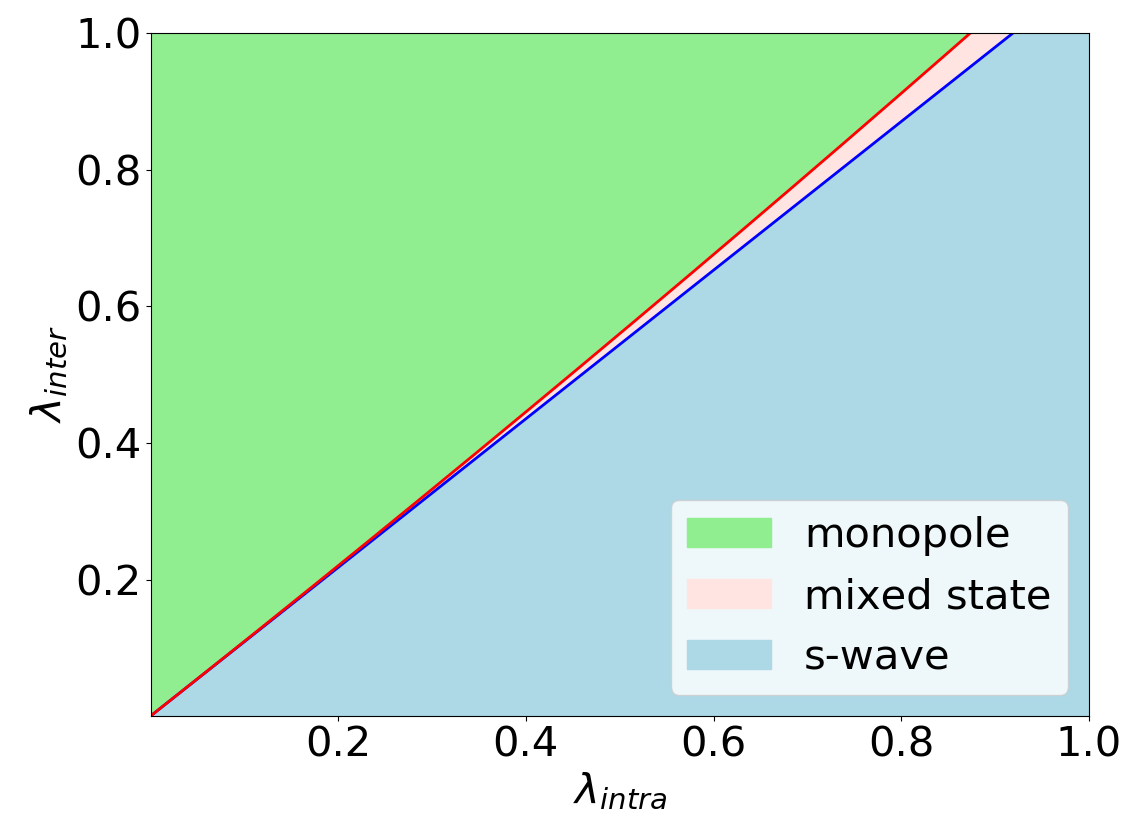}
        \caption{$\nu=3$, $c=10$}
        \label{s_wave_i}
    \end{subfigure}
        \caption{Phase diagram in the coupling space for {\bf{s-wave vs monopole}}, for chiralities $\nu=1,2,3$ and $c=0.1, 1, 10$ to illustrate the behavior when $c\rightarrow 0$, intermediate $c$ and $c\rightarrow \infty$, respectively. Blue line is Eq.~\eqref{Diagrama: curva s wave > monopolo} and red line is Eq.~\eqref{Diagrama: curva monopolo > s wave}, respectively.}
        \label{fig: s-wave vs monopole}
\end{figure*}

\begin{figure*}[ht]
    \centering
    \begin{subfigure}[b]{0.32\textwidth}
        \centering
        \includegraphics[width=\textwidth]{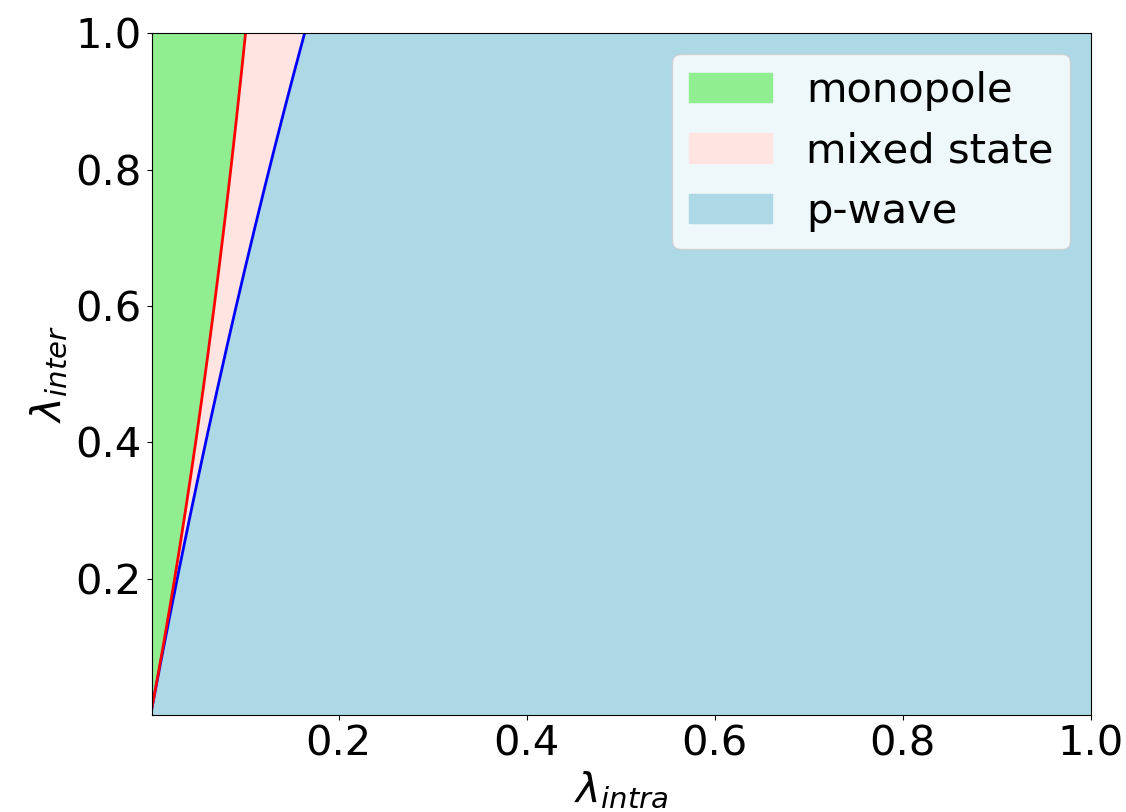}
        \caption{$\nu=1$, $c=0.1$}
        \label{p_wave_a}
    \end{subfigure}
    \begin{subfigure}[b]{0.32\textwidth}
        \centering
        \includegraphics[width=\textwidth]{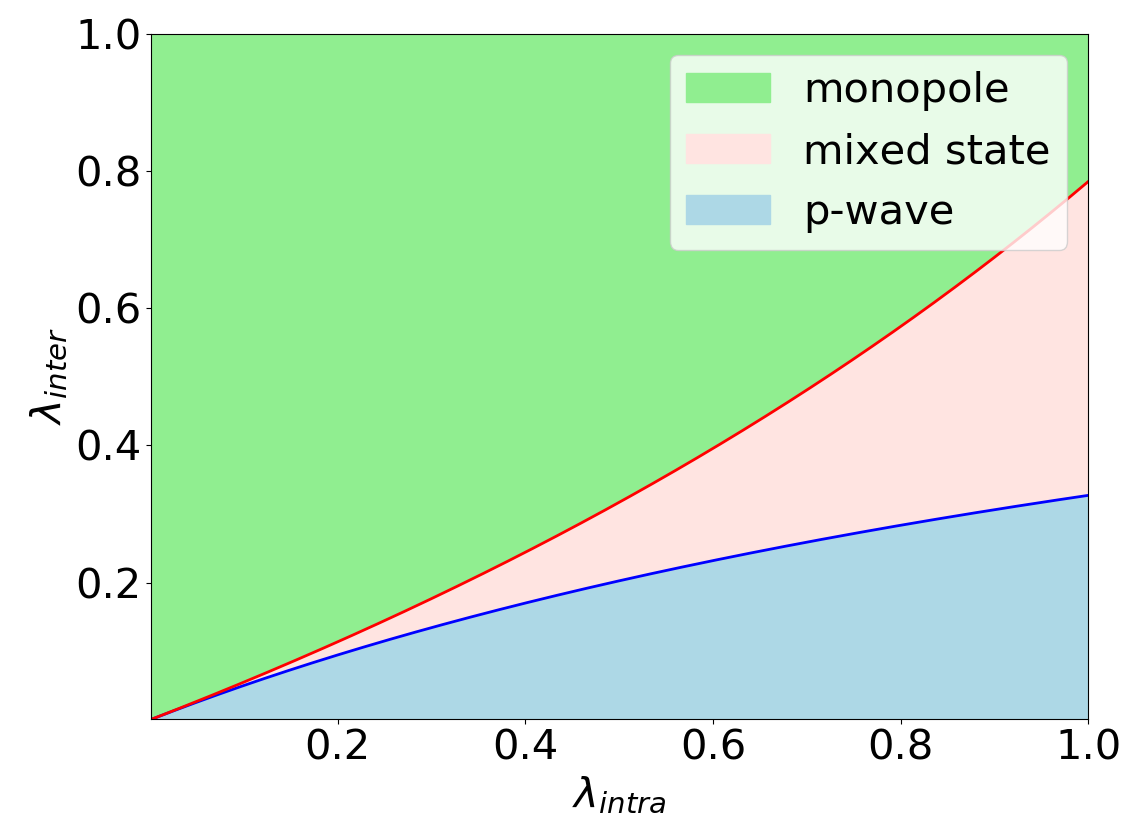}
        \caption{$\nu=1$, $c=1$}
        \label{p_wave_b}
    \end{subfigure}
    \begin{subfigure}[b]{0.32\textwidth}
        \centering
        \includegraphics[width=\textwidth]{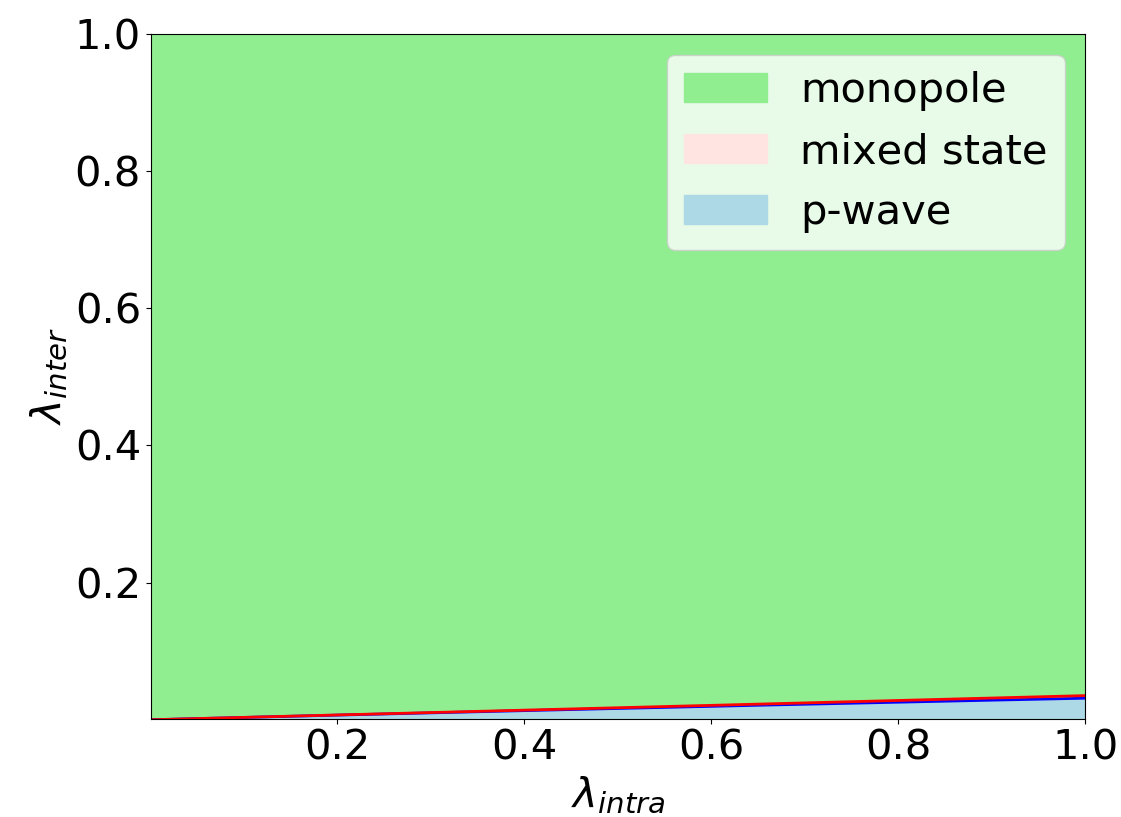}
        \caption{$\nu=1$, $c=7$}
        \label{p_wave_c}
    \end{subfigure}
    \begin{subfigure}[b]{0.32\textwidth}
        \centering
        \includegraphics[width=\textwidth]{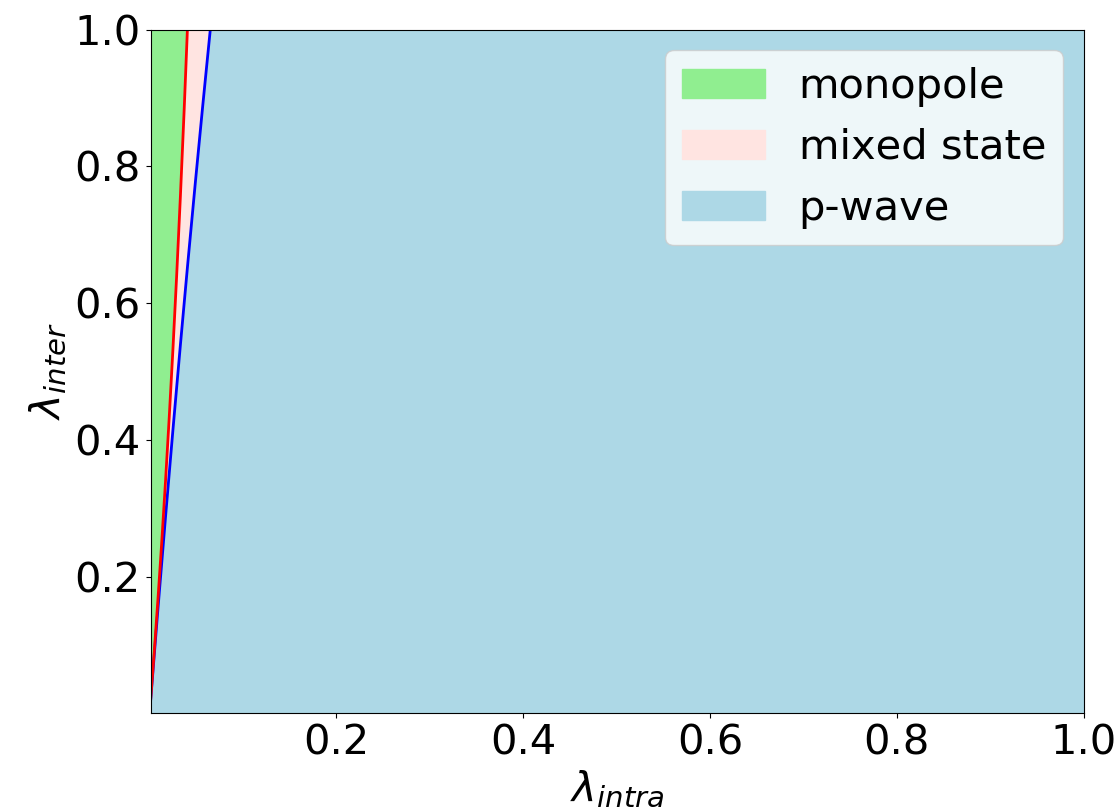}
        \caption{$\nu=2$, $c=0.1$}
        \label{p_wave_d}
    \end{subfigure}
    \begin{subfigure}[b]{0.32\textwidth}
        \centering
        \includegraphics[width=\textwidth]{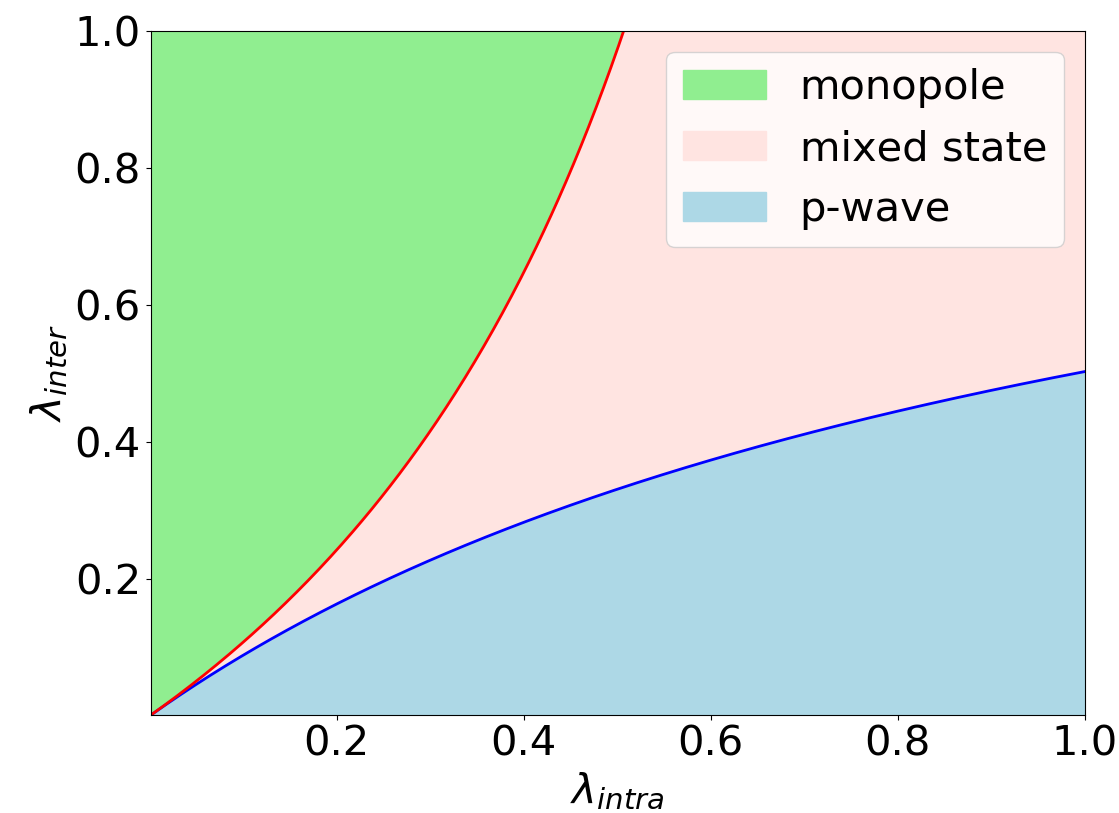}
        \caption{$\nu=2$, $c=1$}
        \label{p_wave_e}
    \end{subfigure}
    \begin{subfigure}[b]{0.32\textwidth}
        \centering
        \includegraphics[width=\textwidth]{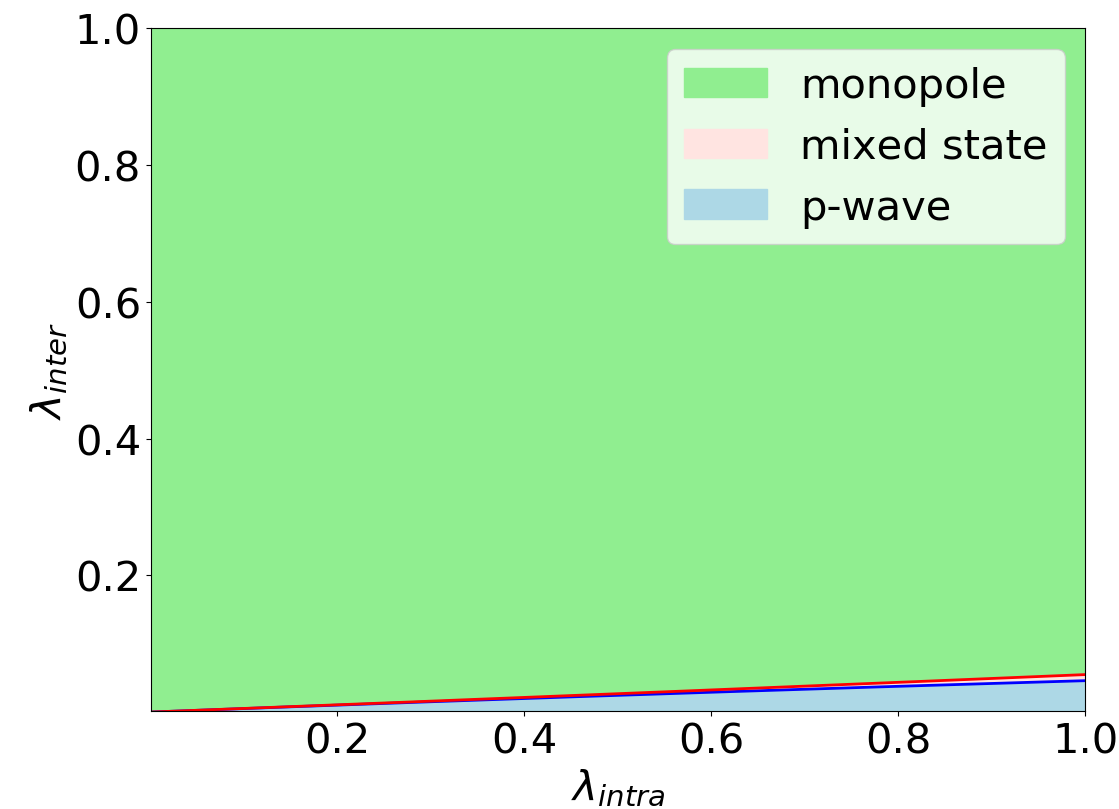}
        \caption{$\nu=2$, $c=7$}
        \label{p_wave_f}
    \end{subfigure}
    \begin{subfigure}[b]{0.32\textwidth}
        \centering
        \includegraphics[width=\textwidth]{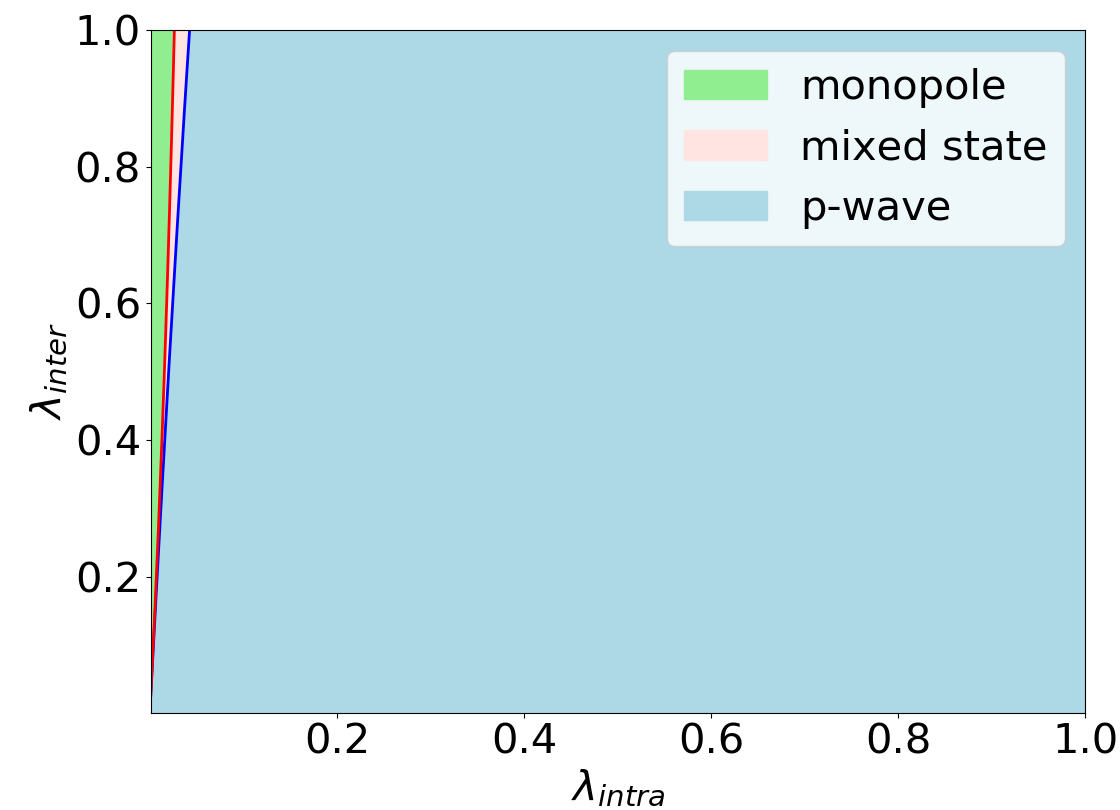}
        \caption{$\nu=3$, $c=0.1$}
        \label{p_wave_g}
    \end{subfigure}
    \begin{subfigure}[b]{0.32\textwidth}
        \centering
        \includegraphics[width=\textwidth]{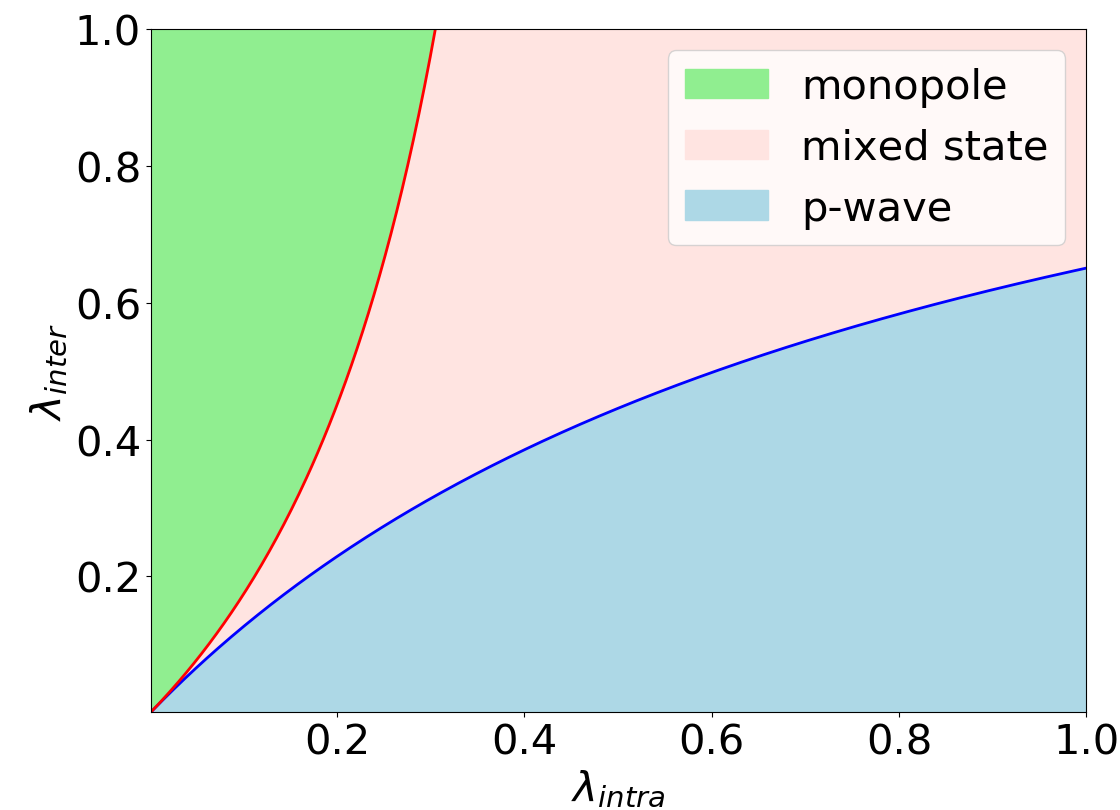}
        \caption{$\nu=3$, $c=1$}
        \label{p_wave_h}
    \end{subfigure}
    \begin{subfigure}[b]{0.32\textwidth}
        \centering
        \includegraphics[width=\textwidth]{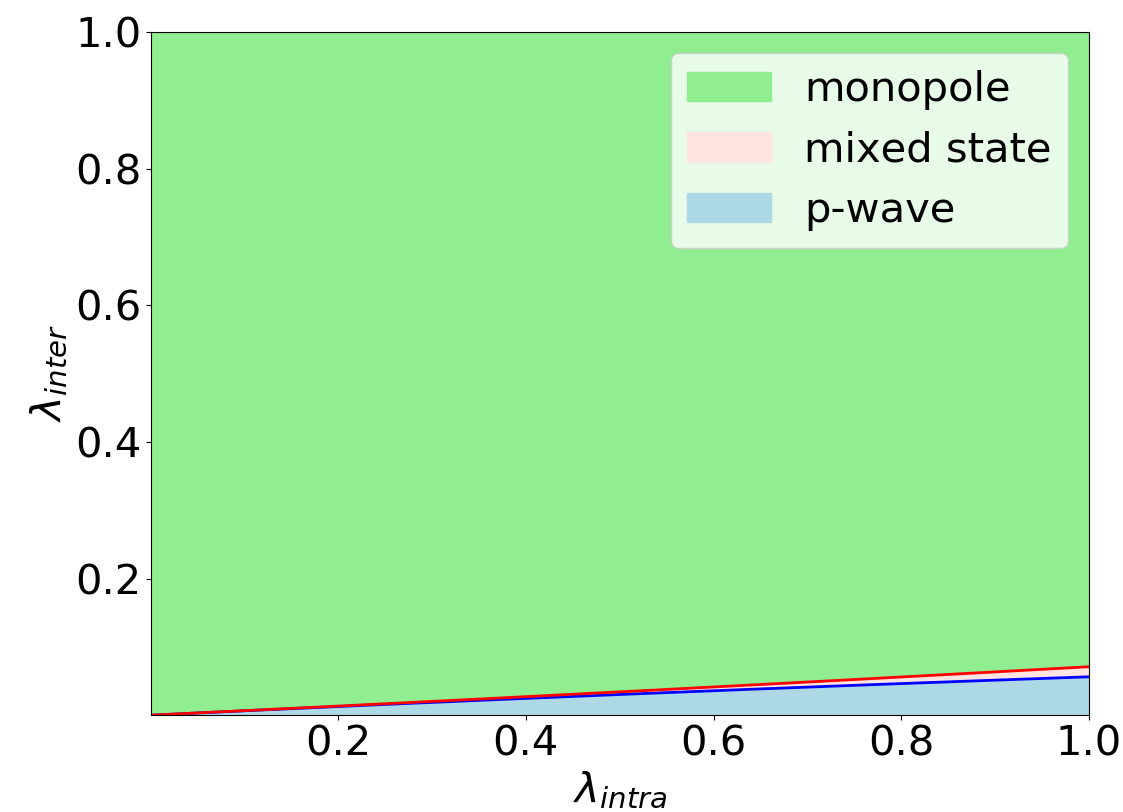}
        \caption{$\nu=3$, $c=7$}
        \label{p_wave_i}
    \end{subfigure}
    \caption{Phase diagram in the coupling space for $\boldsymbol{p_{z}}$-{\bf{wave vs monopole}}, for chiralities $\nu=1,2,3$ and $c=0.1, 1, 7$ to illustrate the behavior when $c\rightarrow 0$, intermediate $c$ and $c\rightarrow \infty$, respectively. Blue line is Eq.~\eqref{Diagrama: curva p wave > monopolo} and red line is Eq.~\eqref{Diagrama: curva monopolo > p wave}, respectively.}
    \label{fig: p-wave vs monopole}
\end{figure*}

Let us first analyze the competition between a conventional s-wave harmonic and a monopole superconducting phase. For this purpose, we consider the pairings $\Delta^{\text{intra}}_{\textbf{k}} =  \Delta_{s} Y_{0,0} (\theta_{\textbf{k}}, \phi_{\textbf{k}})$ and $\Delta^{\text{inter}}_{\textbf{k}} =  \bar{\Delta}_{\nu} \mathcal{Y}_{-\nu,\nu,0} (\theta_{\textbf{k}}, \phi_{\textbf{k}})$, corresponding to $f_{\text{intra}}(\theta) \sim 1$ and $f_{\text{inter}}(\theta) \sim \sin^{\nu}\theta$, respectively. Therefore, we are in Case 2, and hence applying Eq.~\eqref{eq_case2} we have
\begin{equation}
\begin{split}
    \frac{1}{\lambda_{\eta}} = \frac{1}{B_{\nu}} \int_{0}^{1}\frac{dx x}{\sqrt{1-x^{2\nu}}} |f_{\eta}(\theta_{x})|^{2} \left\{ 2\ln(2\omega_{D}) \vphantom{\frac{1}{2}} \right. \\
    \left. - \ln\left[\frac{1}{2}(\Delta_{s}^{2} + \bar{\Delta}_{\nu}^{2} \sin^{2\nu}\theta_{x} + |\Delta_{s}^{2} - \bar{\Delta}_{\nu}^{2} \sin^{2\nu}\theta_{x}|)\right] \right\}.
\end{split}
\end{equation}

We first analyze the situation $\Delta_{s} \gg \bar{\Delta}_{\nu}$, for which we obtain the following
\begin{eqnarray}
    \frac{1}{\lambda_{\text{intra}}} &=& \frac{1}{B_{\nu}} \int_{0}^{1}\frac{dx x}{\sqrt{1-x^{2\nu}}} \left[ 2\ln(2\omega_{D}) - \ln(\Delta_{s}^{2}) \right] \notag \\
    &=& 2\ln(2\omega_{D}) - \ln(\Delta_{s}^{2}),
    \label{eq_lintra1}
\end{eqnarray}
and
\begin{eqnarray}
    \frac{1}{\lambda_{\text{inter}}} &=& \frac{1}{B_{\nu}} \int_{0}^{1}\frac{dx x}{\sqrt{1-x^{2\nu}}} \sin^{2\nu}\theta_{x} \left[ 2\ln(2\omega_{D}) - \ln(\Delta_{s}^{2}) \right] \notag\\
    &=& I_{\nu}(c) \left[ 2\ln(2\omega_{D}) - \ln(\Delta_{s}^{2}) \right],
    \label{eq_linter1}
\end{eqnarray}
where we defined the coefficient;
\begin{eqnarray}
    I_{\nu}(c) = \frac{1}{B_{\nu}} \int_{0}^{1}\frac{dx x}{\sqrt{1-x^{2\nu}}} \sin^{2\nu}\theta_{x},
\end{eqnarray}
for $\theta_x$ defined in Eq.~\eqref{eq:tantx}.
Combining Eqs.~\eqref{eq_lintra1} and \eqref{eq_linter1}, we obtain
\begin{eqnarray}\label{Diagrama: curva s wave > monopolo}
    \frac{1}{\lambda_{\text{inter}}} = I_{\nu}(c) \frac{1}{\lambda_{\text{intra}}}.
\end{eqnarray}
This equation defines the boundary across which the s-wave dominates over the monopole in the phase diagram.

For the case $\Delta_{s}\ll \bar{\Delta}_{\nu}$ we define $\theta'$ so that $\sin^{\nu}\theta' = \Delta_{s} / \bar{\Delta}_{\nu}$, with $\theta' \rightarrow 0$. We also define $x'$ according to $\theta_{x=x'} = \theta'$, so that $x'\rightarrow 0$. With these definitions, Eq.~\eqref{eq_case2} reduces to the expression
\begin{eqnarray}
\begin{split}
    \frac{1}{\lambda_{\eta}} = \frac{1}{B_{\nu}} \int_{0}^{1}\frac{dx x |f_{\eta}(\theta_{x})|^{2}}{\sqrt{1-x^{2\nu}}} \left\{ 2\ln(2\omega_{D}) \vphantom{\frac{1}{2}}\right. \\
    \left. - \ln\left[\frac{\Delta_{s}^{2}}{2} \left(1 +  \frac{\sin^{2\nu}\theta_{x}}{\sin^{2\nu}\theta'} + \left| 1 - \frac{\sin^{2\nu}\theta_{x}}{\sin^{2\nu}\theta'} \right| \right) \right] \right\}.
\end{split}
\end{eqnarray}
To compute the integral in $x$, we split the domain into intervals $x\in[0,x']\cup [x',1]$. Therefore, we have
\begin{equation}\label{eqn general: s-wave < monopole}
\begin{split}
    \frac{1}{\lambda_{\eta}} = & \frac{1}{B_{\nu}} \int_{0}^{1} \frac{dx x |f_{\eta}(\theta_{x})|^{2}}{\sqrt{1-x^{2\nu}}} \left[2\ln(2\omega_{D}) - \ln(\Delta_{s}^{2})\right] \\
    & + \frac{1}{B_{\nu}} \int_{x'}^{1} \frac{dx x |f_{\eta}(\theta_{x})|^{2}}{\sqrt{1-x^{2\nu}}} \ln\left( \sin^{2\nu}\theta' \right) \\
    & - \frac{1}{B_{\nu}} \int_{x'}^{1} \frac{dx x |f_{\eta}(\theta_{x})|^{2}}{\sqrt{1-x^{2\nu}}} \ln\left( \sin^{2\nu}\theta_{x} \right).
\end{split}
\end{equation}
Furthermore, we introduce the additional splitting $\int_{x'}^{1} = \int_{0}^{1} - \int_{0}^{x'}$ only in the last integral. Since $x' \rightarrow 0$, the integral $\int_{0}^{x'}$ can be accurately calculated with the mean value theorem, substituting the factor $\ln(\sin^{2\nu}\theta_{x}) \approx \ln(\sin^{2\nu}\theta')$ to join it with the second integral in (\ref{eqn general: s-wave < monopole}). Thus, after this procedure, we obtain
\begin{eqnarray}
    \frac{1}{\lambda_{\eta}} &=& \frac{1}{B_{\nu}} \int_{0}^{1} \frac{dx x |f_{\eta}(\theta_{x})|^{2}}{\sqrt{1-x^{2\nu}}} \left[2\ln(2\omega_{D}) + \ln \left( \frac{\sin^{2\nu}\theta'}{\Delta_{s}^{2}} \right) \right] \notag\\
    && - \frac{1}{B_{\nu}} \int_{0}^{1} \frac{dx x |f_{\eta}(\theta_{x})|^{2}}{\sqrt{1-x^{2\nu}}} \ln\left( \sin^{2\nu}\theta_{x} \right).
\end{eqnarray}
Expressing this equation for $\eta = {\text{inter}}$ and $\eta = {\text{intra}}$, respectively, we obtain
\begin{eqnarray}
    \frac{1}{\lambda_{\text{intra}}} &=& 2\ln(2\omega_{D}) + \ln \left( \frac{\sin^{2\nu}\theta'}{\Delta_{s}^{2}} \right) + I_{s\nu}(c) \\
    \frac{1}{\lambda_{\text{inter}}} &=& I_{\nu}(c)\left[2\ln(2\omega_{D}) + \ln \left( \frac{\sin^{2\nu}\theta'}{\Delta_{s}^{2}} \right) \right] + I_{\nu\nu} (c) \notag,
\end{eqnarray}
where we defined
\begin{eqnarray}
    I_{s\nu}(c) &=& -\frac{1}{B_{\nu}} \int_{0}^{1} \frac{dx x}{\sqrt{1-x^{2\nu}}} \ln\left( \sin^{2\nu}\theta_{x} \right) \\
    I_{\nu\nu}(c) &=& -\frac{1}{B_{\nu}} \int_{0}^{1} \frac{dx x \sin^{2\nu}\theta_{x}}{\sqrt{1-x^{2\nu}}} \ln\left( \sin^{2\nu}\theta_{x} \right).
\end{eqnarray}
Combining both equations, we finally obtain the following result
\begin{equation}\label{Diagrama: curva monopolo > s wave}
    \frac{1}{\lambda_{\text{inter}}} = I_{\nu}(c)\left[\frac{1}{\lambda_{\text{intra}}} - I_{s\nu}(c)\right] + I_{\nu\nu} (c).
\end{equation}
This equation determines the boundary across which the monopole predominates over the s-wave.

From (\ref{Diagrama: curva s wave > monopolo}) and (\ref{Diagrama: curva monopolo > s wave}), we can construct the phase diagram $\lambda_{\text{inter}}$ versus $\lambda_{\text{intra}}$ for any value of $\nu$ and $c$. Some explicit examples are shown in Fig.~\ref{fig: s-wave vs monopole}.

\subsection{p-wave vs Monopole}

Similarly to the previous case, we shall now study the coexistence of a conventional spherical harmonic $p_{z}$ wave with the superconducting monopole. Therefore, we assume for the pairings $\Delta^{\text{intra}}_{\textbf{k}} =  \Delta_{p_{z}} Y_{1,0} (\theta_{\textbf{k}}, \phi_{\textbf{k}})$ and $\Delta^{\text{inter}}_{\textbf{k}} =  \bar{\Delta}_{\nu} \mathcal{Y}_{-\nu,\nu,0} (\theta_{\textbf{k}}, \phi_{\textbf{k}})$, with which $f_{\text{intra}}(\theta) \sim \cos\theta$ and $f_{\text{inter}}(\theta) \sim \sin^{\nu}\theta$, respectively.

Defining $\theta'$ so that $\Delta_{p_{z}}^{2}/\bar{\Delta}_{\nu}^{2} = \sin^{2\nu} \theta'/\cos^{2} \theta'$, we have after Eq.~\eqref{eq_case2}
\begin{equation}
\begin{split}
    \frac{1}{\lambda_{\eta}} = \frac{1}{B_{\nu}} \int_{0}^{1}\frac{dx x|f_{\eta}(\theta_{x})|^{2}}{\sqrt{1-x^{2\nu}}} \left\{ 2\ln(2\omega_{D}) \vphantom{\frac{1}{2}} \right. \\
    - \left. \ln\left[\frac{\Delta_{p_{z}}^{2} \cos^{2}\theta_{x}}{2} \left(1 +  g(\theta_{x}) + \left| 1 - g(\theta_{x}) \right| \right) \right] \right\},
\end{split}
\end{equation}
where we defined the function
\begin{equation}
    g(\theta_{x}) = \left(\frac{\sin^{2\nu}\theta_{x}}{\cos^{2}\theta_{x}}\right) \left(\frac{\cos^{2}\theta'}{\sin^{2\nu}\theta'}\right).
\end{equation}
In addition, we define $x'$ such that $\theta_{x=x'} \equiv \theta'$. By splitting the integration region into subintervals $x\in[0,x']\cup [x',1]$, we obtain
\begin{eqnarray}
    \frac{1}{\lambda_{\eta}} &=& \frac{1}{B_{\nu}} \int_{0}^{1}\frac{dx x|f_{\eta}(\theta_{x})|^{2}}{\sqrt{1-x^{2\nu}}} \left[ 2\ln(2\omega_{D}) - \ln\left(\Delta_{p_{z}}^{2} \cos^{2}\theta_{x} \right) \right] \notag\\
    && -\frac{1}{B_{\nu}} \int_{x'}^{1}\frac{dx x|f_{\eta}(\theta_{x})|^{2}}{\sqrt{1-x^{2\nu}}} \ln\left[g(\theta_{x})\right].
\end{eqnarray}

For $\Delta_{p_{z}} \gg \bar{\Delta}_{\nu}$, $x' \rightarrow 1$, the last integral vanishes and we obtain
\begin{eqnarray}\label{Diagrama: curva p wave > monopolo}
    \frac{1}{\lambda_{\text{inter}}} = \frac{I_{\nu}(c)}{I_{p}(c)} \left[\frac{1}{\lambda_{\text{intra}}} - I_{pp}(c)\right] + I_{\nu p}(c)
\end{eqnarray}
with
\begin{eqnarray}
    I_{p}(c) &=& \frac{1}{B_{\nu}} \int_{0}^{1}\frac{dx x}{\sqrt{1-x^{2\nu}}} \cos^{2}\theta_{x} \\
    I_{pp}(c) &=& -\frac{1}{B_{\nu}} \int_{0}^{1}\frac{dx x}{\sqrt{1-x^{2\nu}}} \cos^{2}\theta_{x}, \ln\left( \cos^{2}\theta_{x} \right), \\
    I_{\nu p}(c) &=& -\frac{1}{B_{\nu}} \int_{0}^{1}\frac{dx x}{\sqrt{1-x^{2\nu}}} \sin^{2\nu}\theta_{x} \ln\left( \cos^{2}\theta_{x} \right).
\end{eqnarray}

The case $\Delta_{p_{z}} \ll \bar{\Delta}_{\nu}$ is analogous to that of the s wave, thus reducing to the expression
\begin{equation}\label{Diagrama: curva monopolo > p wave}
    \frac{1}{\lambda_{\text{inter}}} = \frac{I_{\nu}(c)}{I_{p}(c)} \left[\frac{1}{\lambda_{\text{intra}}} - I_{p\nu}(c)\right] + I_{\nu\nu}(c),
\end{equation}
along with the definition
\begin{equation}
    I_{p\nu}(c) = -\frac{1}{B_{\nu}} \int_{0}^{1}\frac{dx x}{\sqrt{1-x^{2\nu}}} \cos^{2}\theta_{x} \ln\left( \sin^{2\nu}\theta_{x} \right).
\end{equation}

In Figure~\ref{fig: p-wave vs monopole} we show the $p_{z}$-wave vs monopole phase diagram for different values of $\nu$ and $c$.

\section{Critical Behavior}
\label{Sec:Critical}
\begin{figure*}[ht]
    \centering
    \begin{subfigure}[b]{0.32\textwidth}
        \centering
        \includegraphics[width=\textwidth]{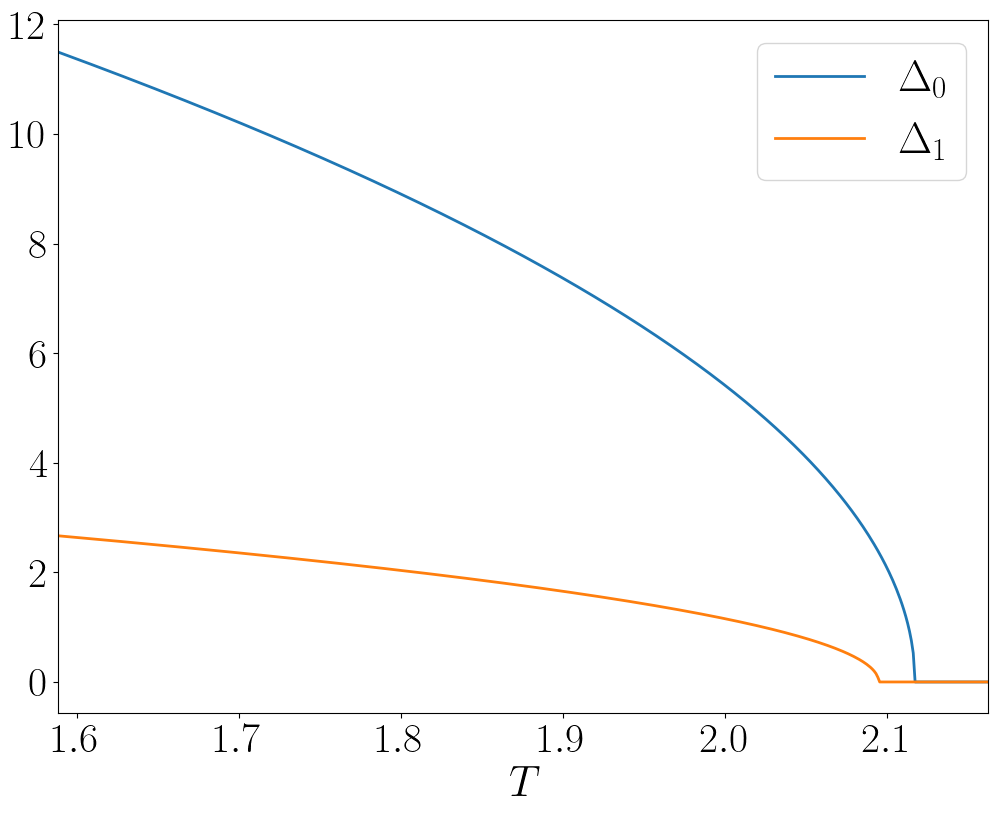}
        \caption{$\nu=1$, $c=1$}
        \label{s_wave_CB_a}
    \end{subfigure}
    \begin{subfigure}[b]{0.32\textwidth}
        \centering
        \includegraphics[width=\textwidth]{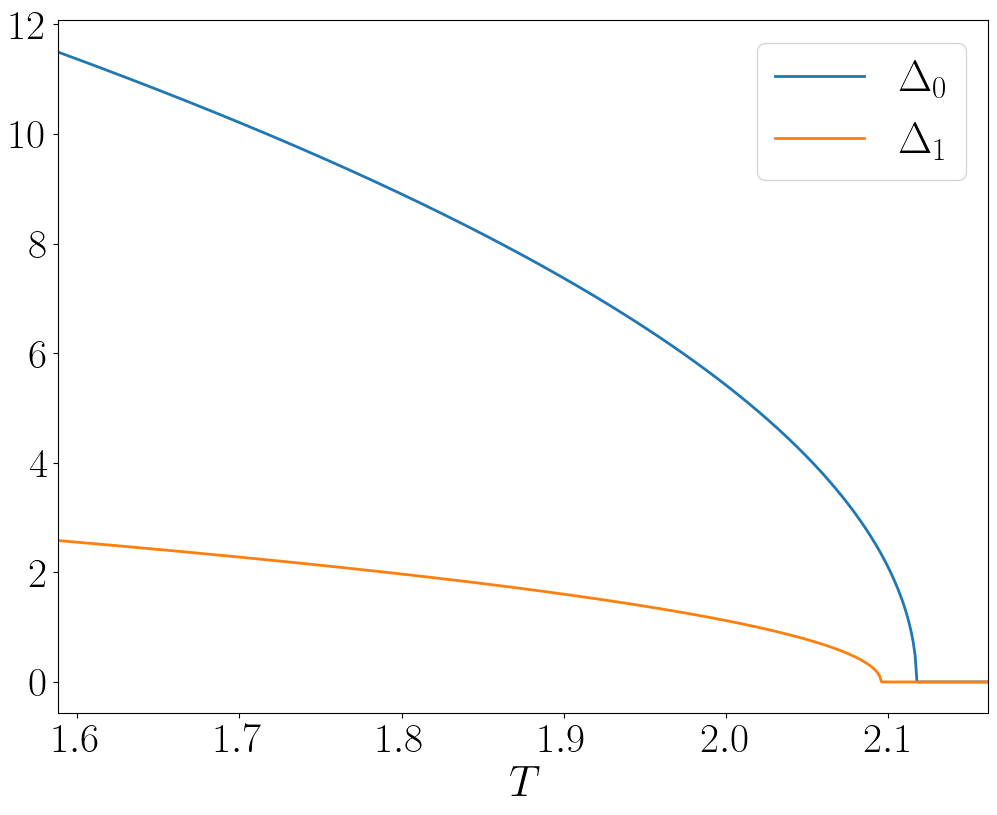}
        \caption{$\nu=1$, $c=10$}
        \label{s_wave_CB_b}
    \end{subfigure}
    \begin{subfigure}[b]{0.32\textwidth}
        \centering
        \includegraphics[width=\textwidth]{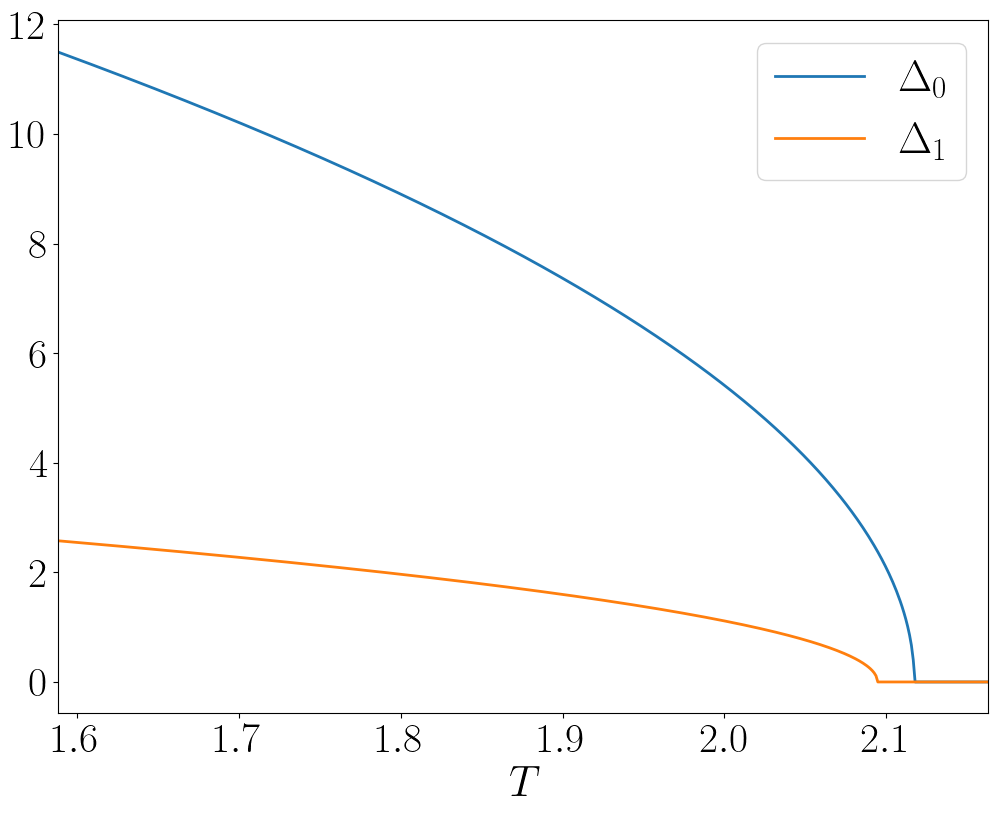}
        \caption{$\nu=1$, $c=100$}
        \label{s_wave_CB_c}
    \end{subfigure}
    \begin{subfigure}[b]{0.32\textwidth}
        \centering
        \includegraphics[width=\textwidth]{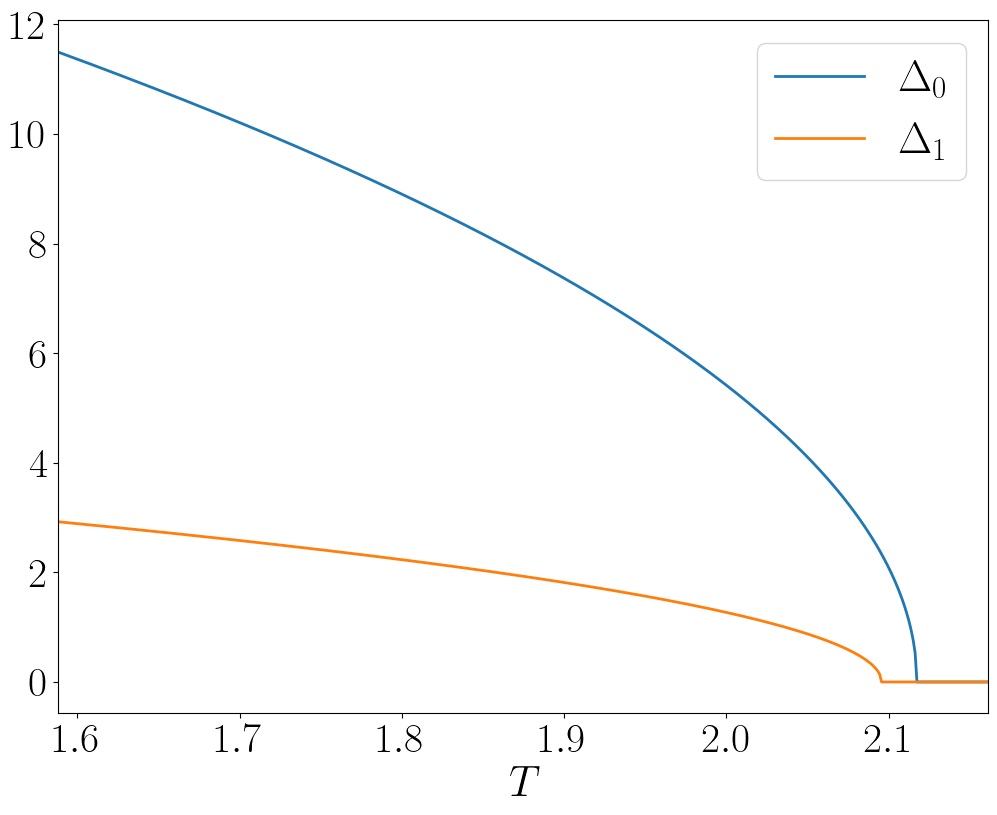}
        \caption{$\nu=2$, $c=1$}
        \label{s_wave_CB_d}
    \end{subfigure}
    \begin{subfigure}[b]{0.32\textwidth}
        \centering
        \includegraphics[width=\textwidth]{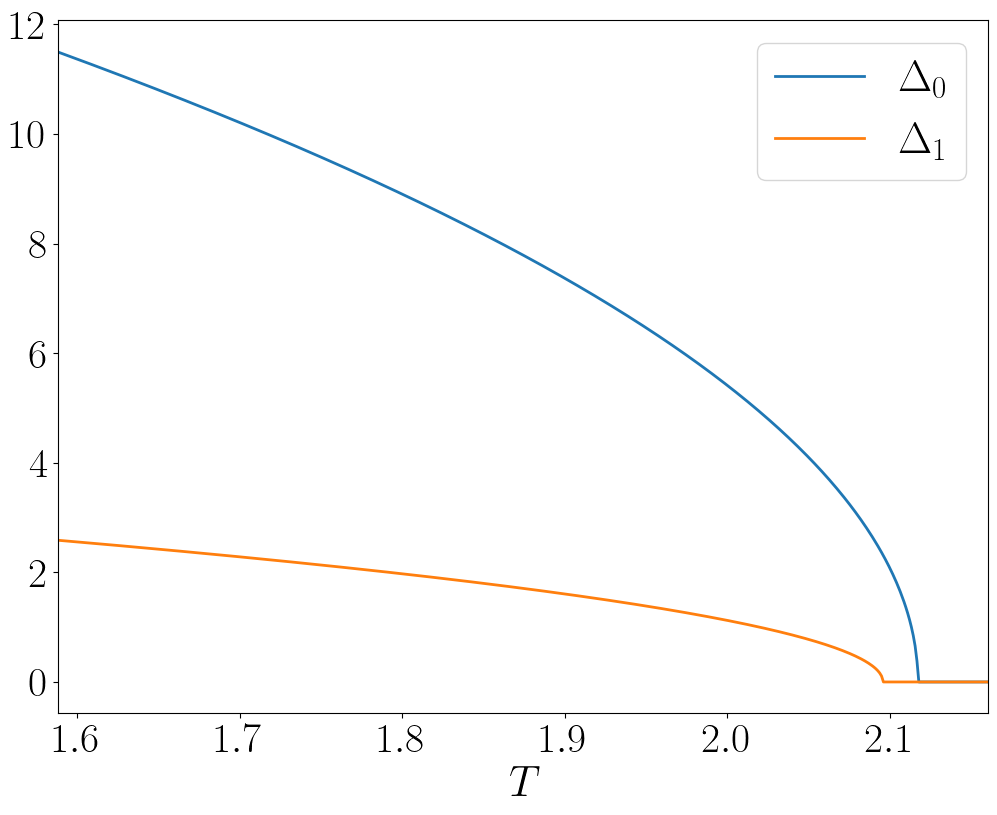}
        \caption{$\nu=2$, $c=10$}
        \label{s_wave_CB_e}
    \end{subfigure}
    \begin{subfigure}[b]{0.32\textwidth}
        \centering
        \includegraphics[width=\textwidth]{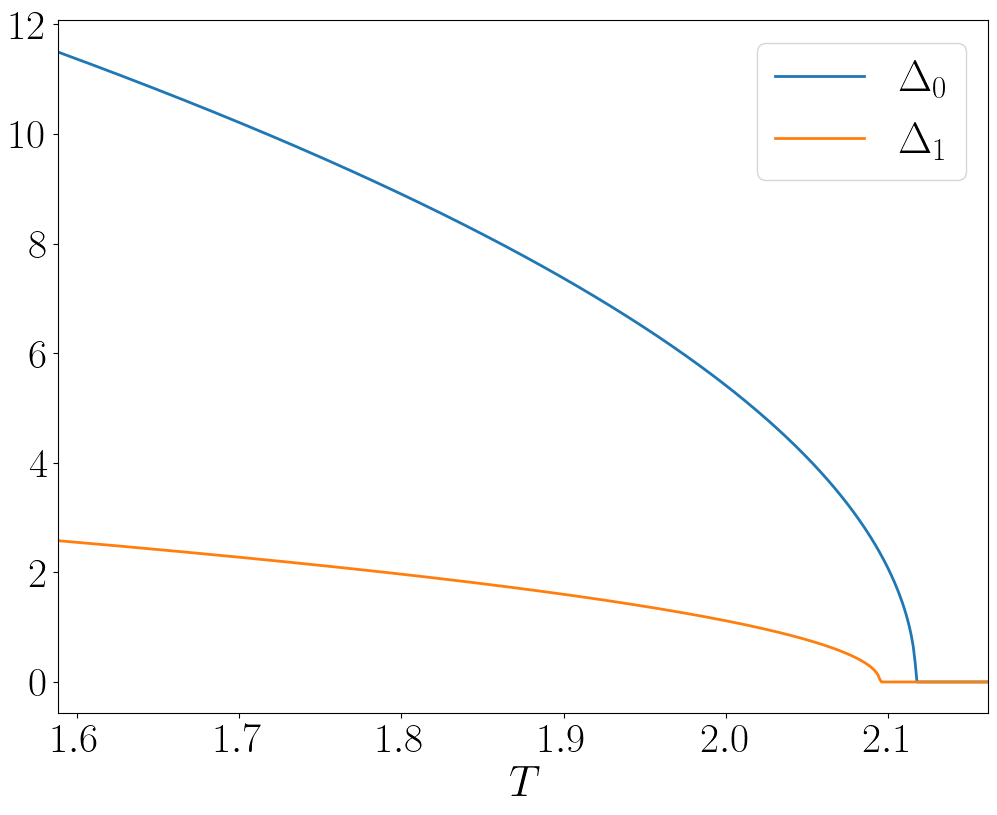}
        \caption{$\nu=2$, $c=100$}
        \label{s_wave_CB_f}
    \end{subfigure}
    \begin{subfigure}[b]{0.32\textwidth}
        \centering
        \includegraphics[width=\textwidth]{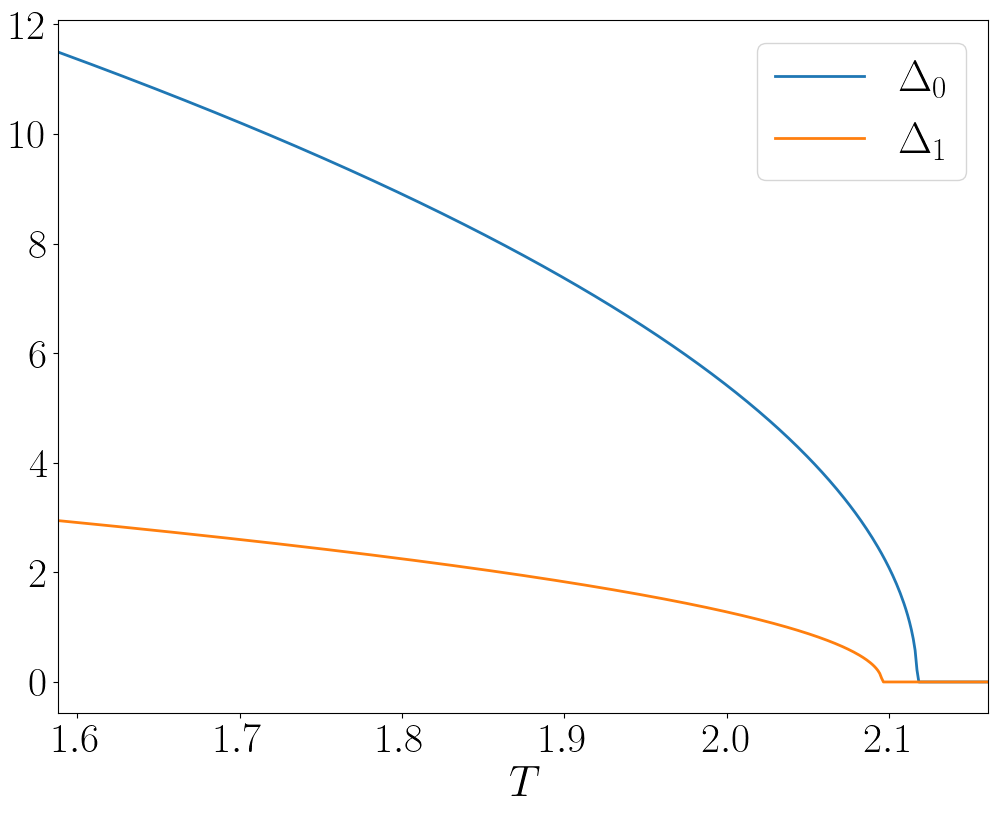}
        \caption{$\nu=3$, $c=1.5$}
        \label{s_wave_CB_g}
    \end{subfigure}
    \begin{subfigure}[b]{0.32\textwidth}
        \centering
        \includegraphics[width=\textwidth]{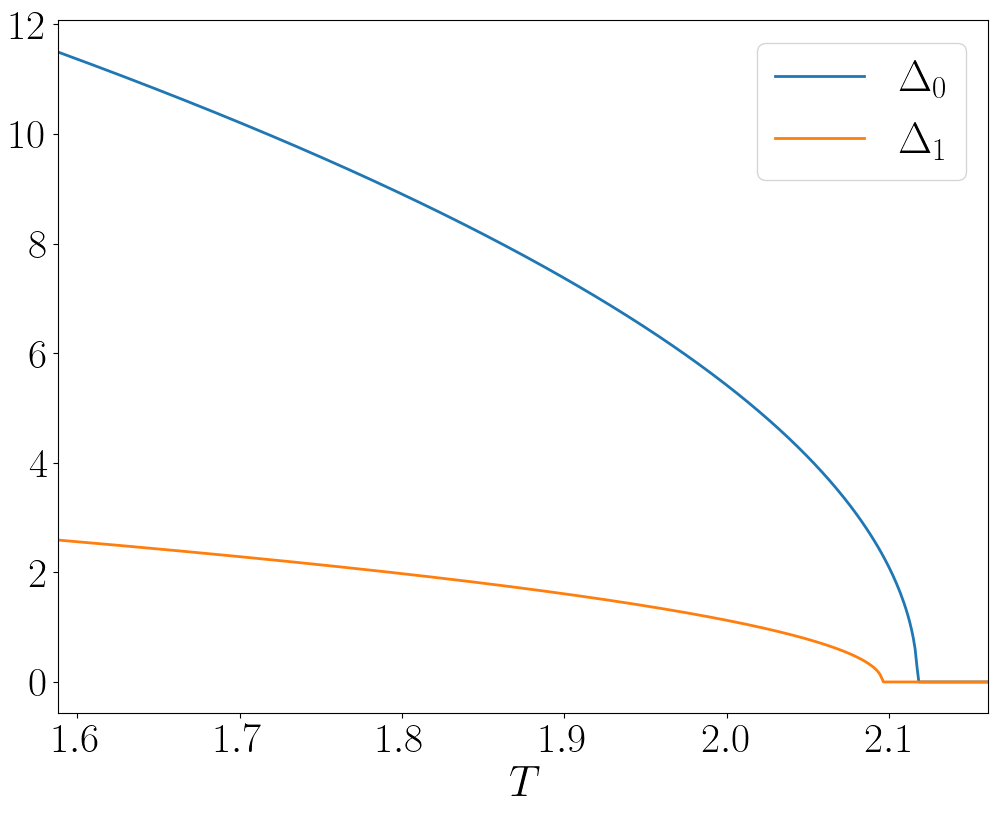}
        \caption{$\nu=3$, $c=10$}
        \label{s_wave_CB_h}
    \end{subfigure}
    \begin{subfigure}[b]{0.32\textwidth}
        \centering
        \includegraphics[width=\textwidth]{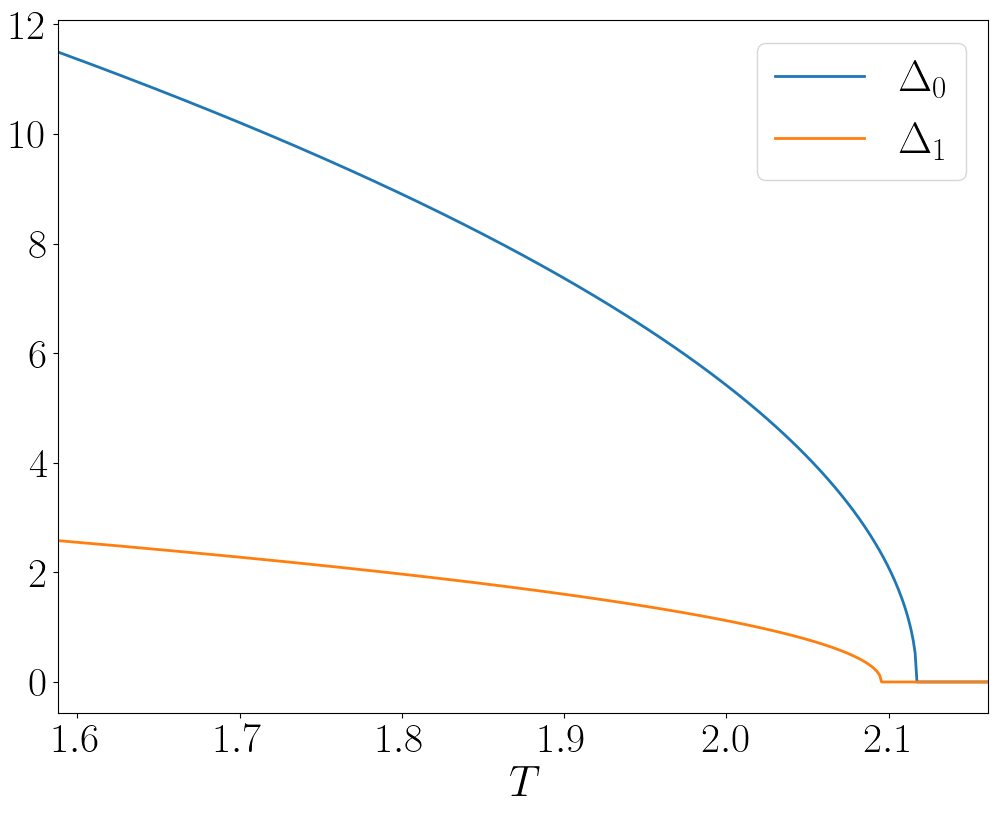}
        \caption{$\nu=3$, $c=100$}
        \label{s_wave_CB_i}
    \end{subfigure}
    \caption{Pairings $\Delta_{0}$ (s-wave in the intra-nodal channel) and $\Delta_{1}$ (monopole in the inter-nodal channel), both in units of $10^{-3}\hbar\omega_D$, as function of $T$ (in units of $10^{-3}\hbar\omega_D/k_B$) near the critical temperatures.}
    \label{fig: pairings s-wave and monopole}
\end{figure*}

\begin{figure*}[ht]
    \centering
    \begin{subfigure}[b]{0.32\textwidth}
        \centering
        \includegraphics[width=\textwidth]{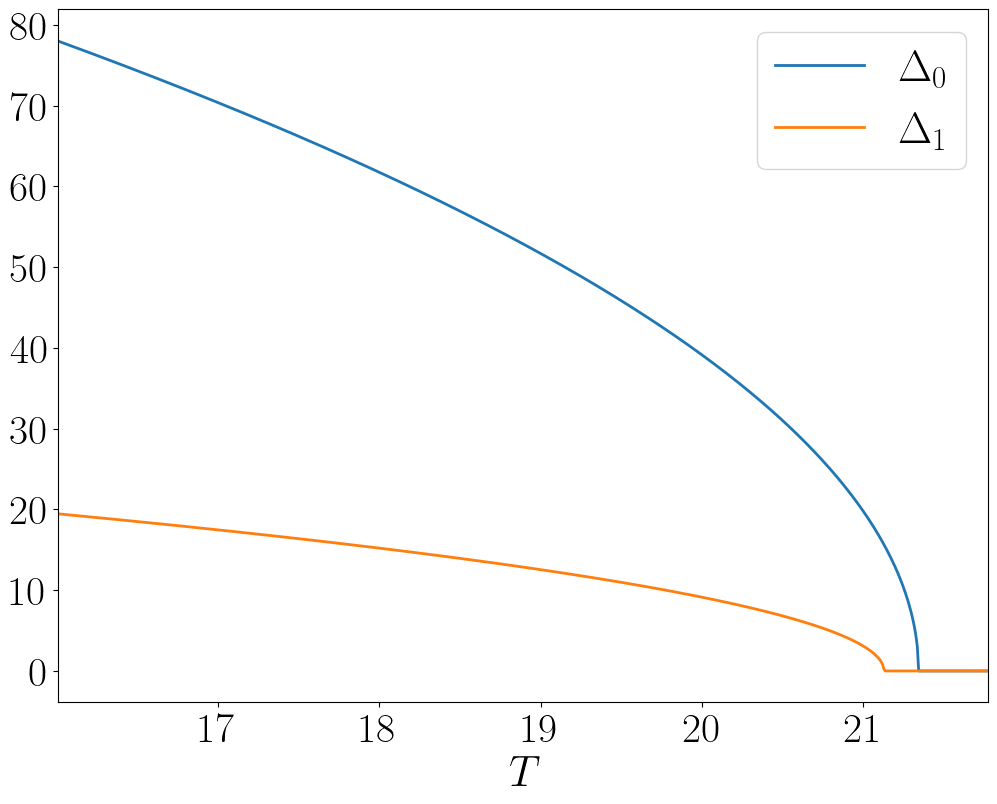}
        \caption{$\nu=1$, $c=0.5$}
        \label{p_wave_CB_a}
    \end{subfigure}
    \begin{subfigure}[b]{0.32\textwidth}
        \centering
        \includegraphics[width=\textwidth]{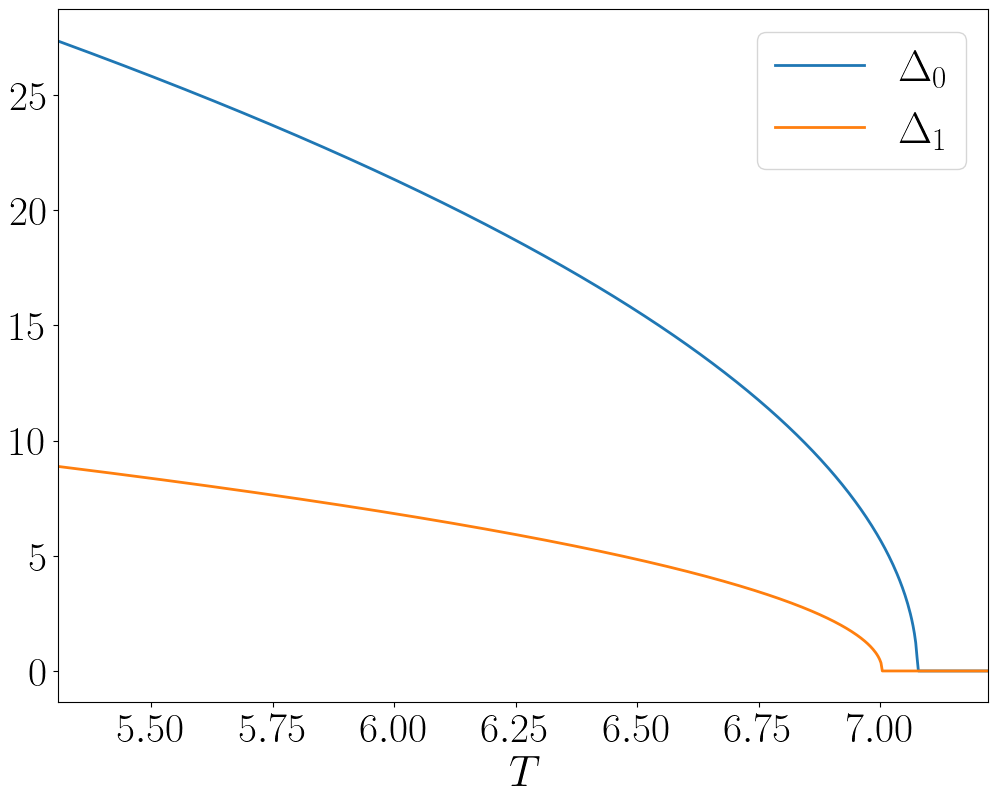}
        \caption{$\nu=1$, $c=0.75$}
        \label{p_wave_CB_b}
    \end{subfigure}
    \begin{subfigure}[b]{0.32\textwidth}
        \centering
        \includegraphics[width=\textwidth]{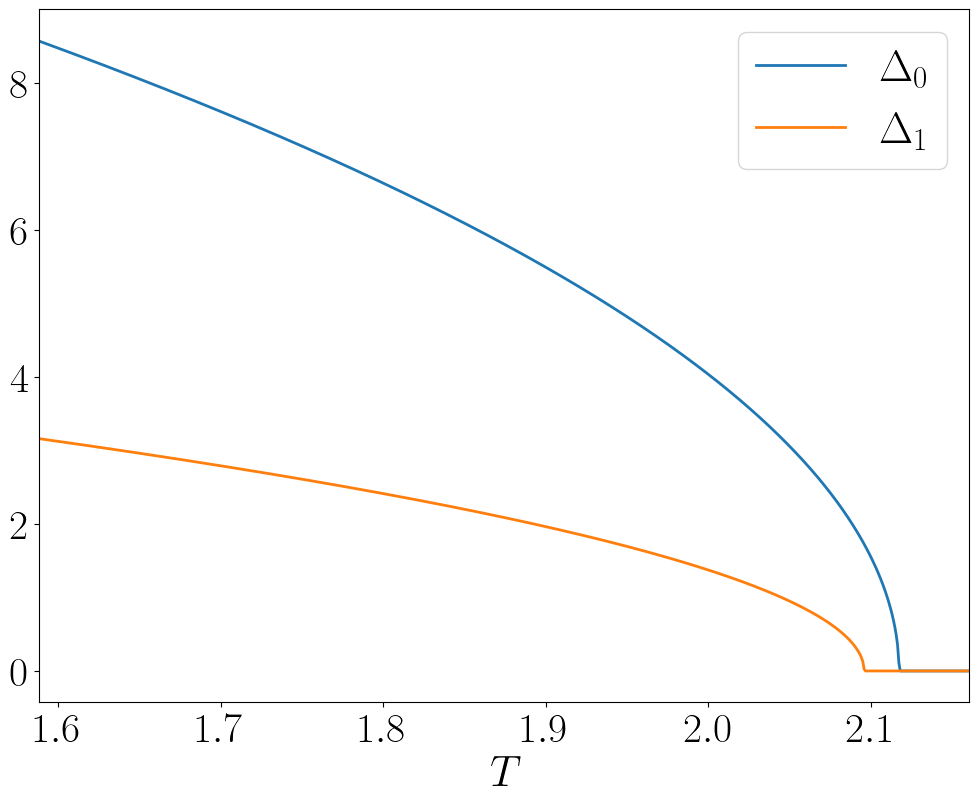}
        \caption{$\nu=1$, $c=1$}
        \label{p_wave_CB_c}
    \end{subfigure}
    \begin{subfigure}[b]{0.32\textwidth}
        \centering
        \includegraphics[width=\textwidth]{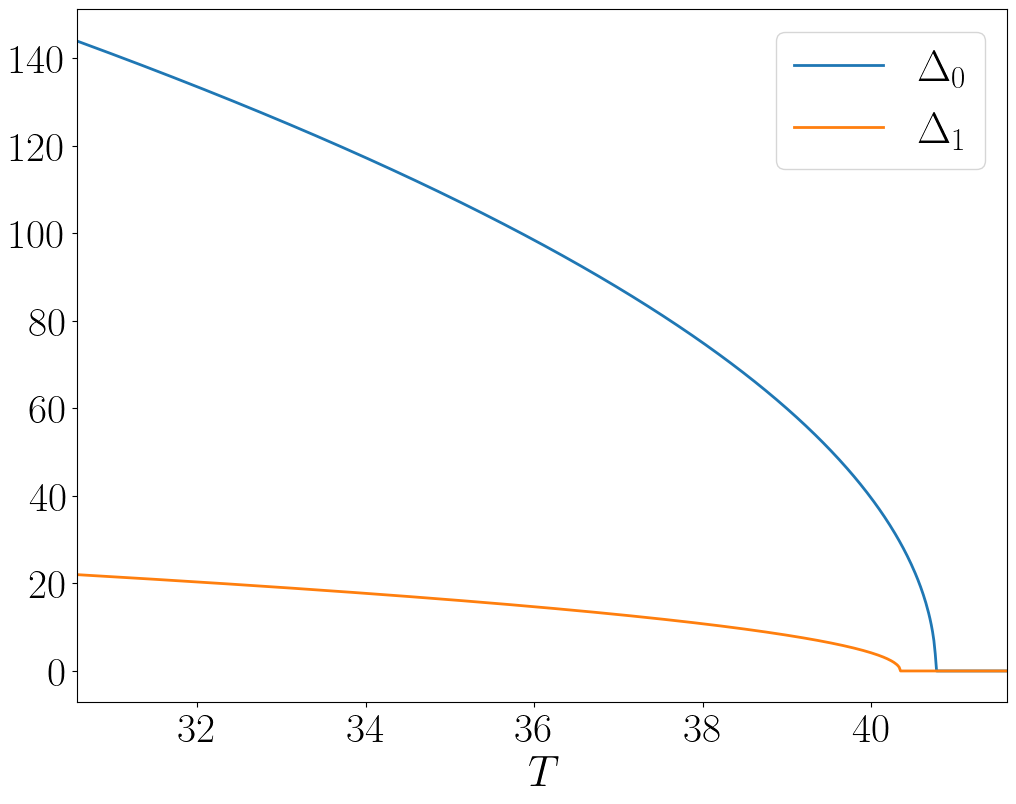}
        \caption{$\nu=2$, $c=0.5$}
        \label{p_wave_CB_d}
    \end{subfigure}
    \begin{subfigure}[b]{0.32\textwidth}
        \centering
        \includegraphics[width=\textwidth]{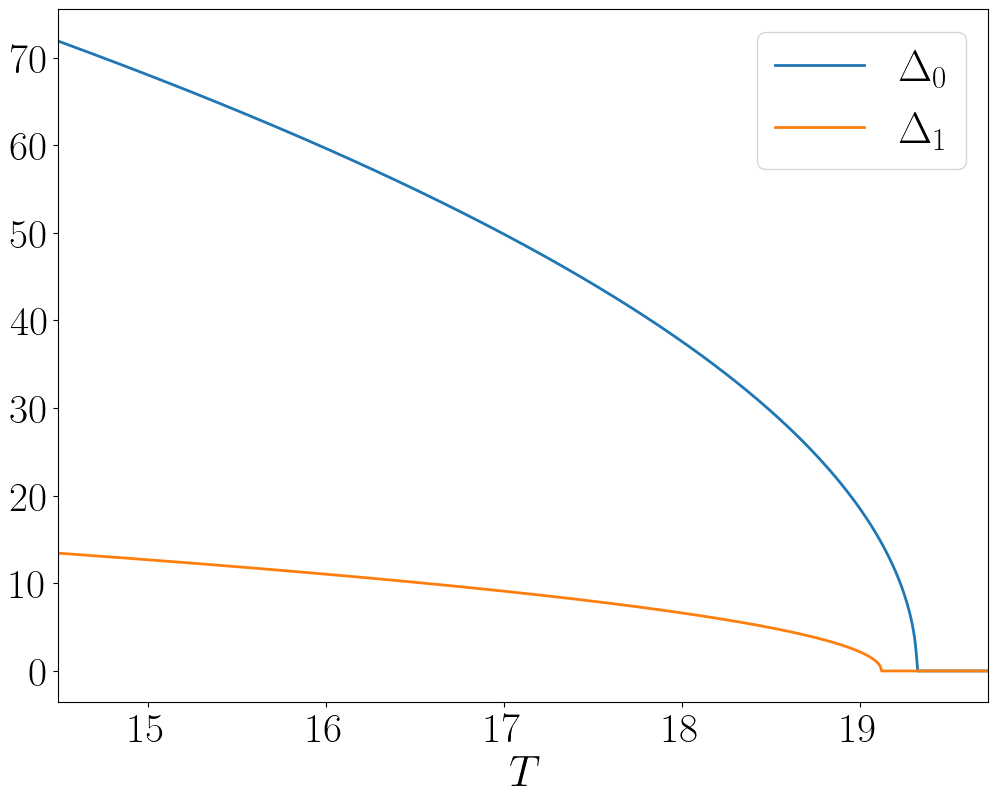}
        \caption{$\nu=2$, $c=0.75$}
        \label{p_wave_CB_e}
    \end{subfigure}
    \begin{subfigure}[b]{0.32\textwidth}
        \centering
        \includegraphics[width=\textwidth]{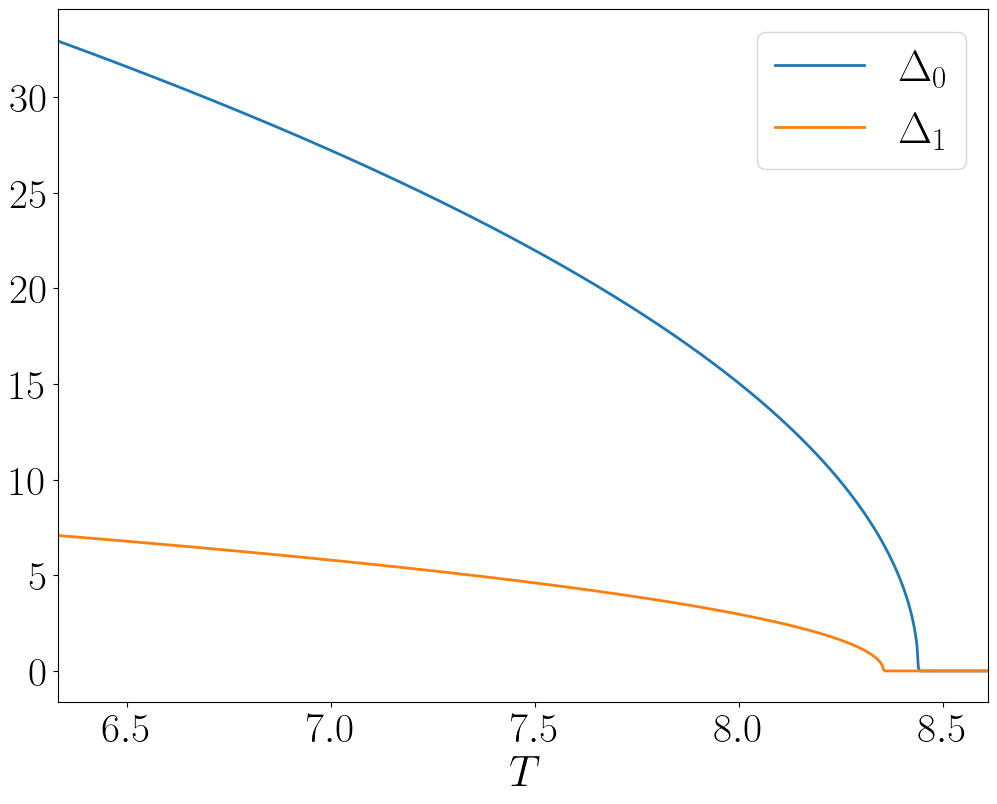}
        \caption{$\nu=2$, $c=1$}
        \label{p_wave_CB_f}
    \end{subfigure}
    \begin{subfigure}[b]{0.32\textwidth}
        \centering
        \includegraphics[width=\textwidth]{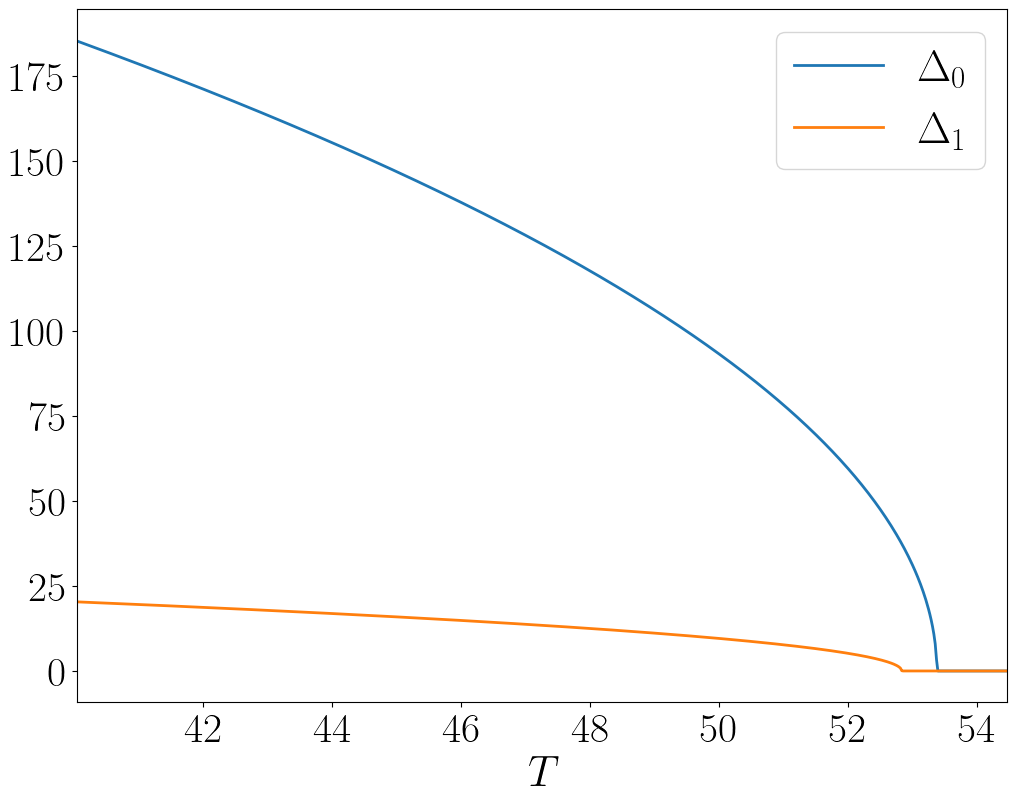}
        \caption{$\nu=3$, $c=0.5$}
        \label{p_wave_CB_g}
    \end{subfigure}
    \begin{subfigure}[b]{0.32\textwidth}
        \centering
        \includegraphics[width=\textwidth]{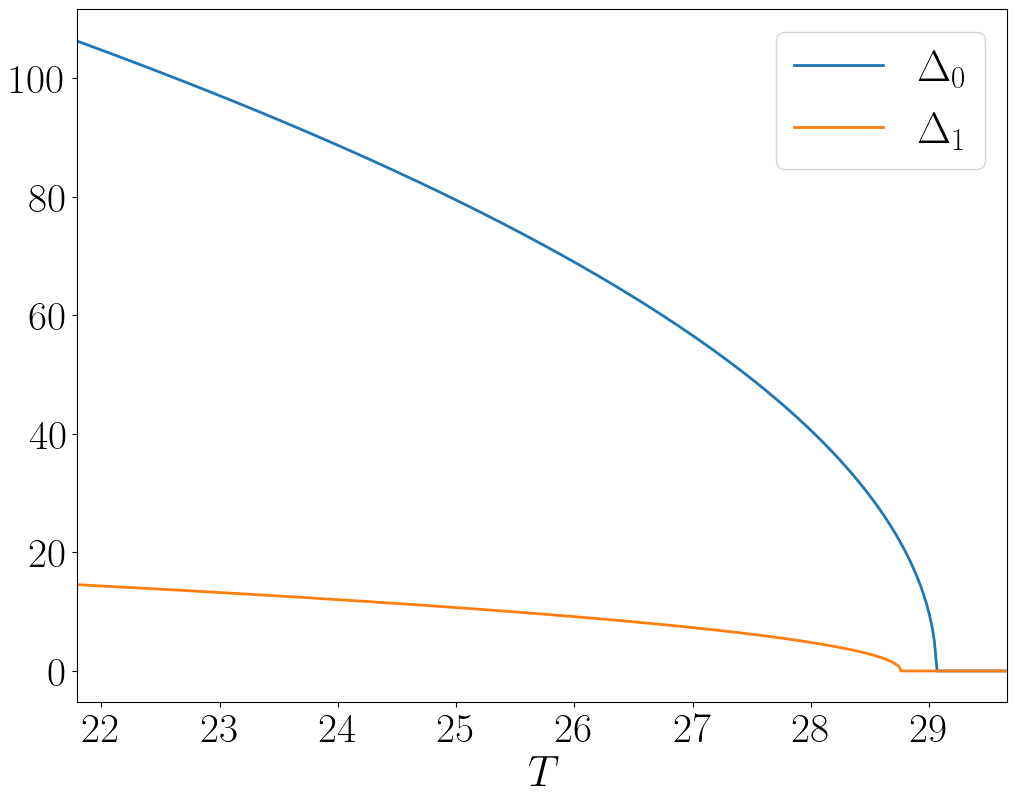}
        \caption{$\nu=3$, $c=0.75$}
        \label{p_wave_CB_h}
    \end{subfigure}
    \begin{subfigure}[b]{0.32\textwidth}
        \centering
        \includegraphics[width=\textwidth]{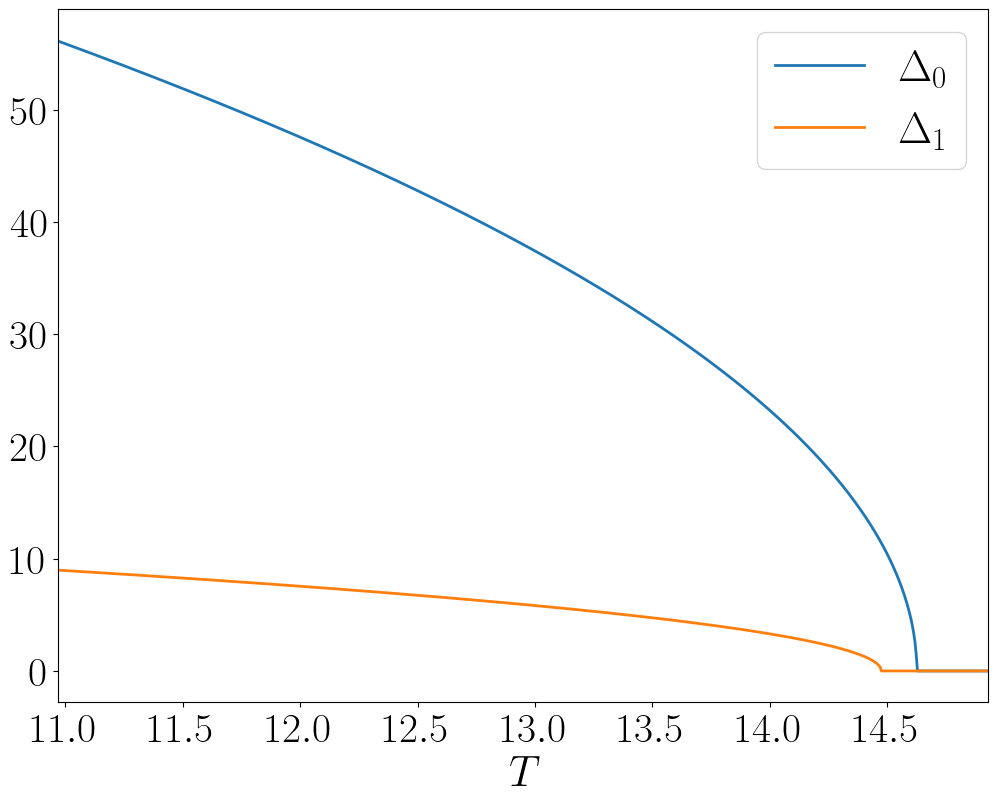}
        \caption{$\nu=3$, $c=1$}
        \label{p_wave_CB_i}
    \end{subfigure}
    \caption{Pairings $\Delta_{0}$ ($p_{z}$-wave in the intra-nodal channel) and $\Delta_{1}$ (monopole in the inter-nodal channel), both in units of $10^{-3}\hbar\omega_D$}, as function of $T$ (in units of $10^{-3}\hbar\omega_D/k_B$) near the critical temperatures.
    \label{fig: pairings p-wave and monopole}
\end{figure*}
In this section, we shall analyze the critical temperatures and the corresponding critical exponents that determine the two superconducting phases, i.e. conventional and monopole. 
Proceeding along the same simplification steps presented in Section~\ref{sec_T0}, we can write the system of BCS  Equations~\eqref{eq_BCS_T0} at finite temperature in the form
\begin{eqnarray}
\frac{\Delta_{\eta}}{\lambda_{\eta}} &=& C_{\eta}\Delta_{\eta} + D_{\eta}\Delta_{\overline{\eta}}
\label{eq_BCS_T}
\end{eqnarray}
where for notational convenience we use the indexes $\eta = 0 \equiv$~"intra", $\eta = 1 \equiv$~"inter", and we define $\overline{\eta} = 1 - \eta$. Consequently, we shall index the order parameters as  $\Delta_{l,m'} \equiv \Delta_0$ and $\bar{\Delta}_{\nu} \equiv \Delta_1$, respectively.
In this more convenient and compact notation, we define the coefficients
\begin{eqnarray}
C_{\eta}(\Delta_0,\Delta_1) &=& \frac{1}{B_{\nu}}\int_0^1  \frac{x f_{\eta}^{2}(\theta_x) dx}{\sqrt{1 - x^{2\nu}}}\int_{0}^{2\pi}\frac{d\phi}{2\pi}\nonumber\\
&&\times\int_{0}^{\omega_D}d\xi
\left[ \mathcal{T}_{\beta}\left({\Gamma}\right)+\mathcal{T}_{\beta}\left({\gamma}\right) \right],
\label{eq_Ccoeff}
\end{eqnarray}
along with the function
\begin{equation}
\mathcal{T}_{\beta}(z) \equiv \frac{\tanh(\beta z/2)}{z}.
\end{equation}
Similarly, for $r \equiv m' - m - \nu$, we define the  coefficients
\begin{eqnarray}
D_{\eta}(\Delta_0,\Delta_1) &=& \frac{1}{B_{\nu}}\int_0^1  \frac{x f_{\eta}(\theta_x) f_{\overline{\eta}}(\theta_x)dx}{\sqrt{1 - x^{2\nu}}}\int_{0}^{2\pi}\frac{d\phi}{2\pi}e^{\pm i r\phi}\nonumber\\
&&\times\int_{0}^{\omega_D}d\xi
\left[ \mathcal{T}_{\beta}\left({\Gamma}\right)-\mathcal{T}_{\beta}\left({\gamma}\right) \right].
\label{eq_Dcoeff}
\end{eqnarray}
where we defined the functions (ignoring any asymmetry in the chemical potential $\delta\mu = 0$)
\begin{eqnarray}
\Gamma &=& \left[\xi^2 + \Delta_{0}^2 f_{intra}^2(\theta_x) + \Delta_{1}^2f_{inter}^2 (\theta_x)\right.\\
&&\left.+ 2\Delta_{0}\Delta_{1}f_{intra}(\theta_x)f_{inter}(\theta_x)\cos(r\phi)\right]^{1/2}\nonumber\\
\gamma &=& \left[\xi^2 + \Delta_{0}^2 f_{intra}^2(\theta_x) + \Delta_{1}^2f_{inter}^2 (\theta_x)\right.\nonumber\\
&&\left.- 2\Delta_{0}\Delta_{1}f_{intra}(\theta_x)f_{inter}(\theta_x)\cos(r\phi)\right]^{1/2}\nonumber
\end{eqnarray}
Let us start by considering two critical temperatures $T^c_{0}$, $T^c_{1}$, each of them associated to the vanishing conditions of the corresponding order parameter: $\Delta_{\eta}(T\rightarrow T^{c-}_{\eta}) = 0$ for $\eta = 0,1$~(intra/inter). 

Without loss of generality, from now on we choose the index $\eta$ such that it corresponds to the highest critical temperature, i.e. 
\begin{equation}
T_{\eta}^{c} = {\rm{Max}}\left\{ T_{0}^{c},T_{1}^{c}\right\},
\end{equation}
and hence $T_{\eta}^{c} > T_{\overline{\eta}}^{c}$, for $\overline{\eta} = 1-\eta$. Under this prescription, let us first consider the region $T \lesssim T_{\eta}^{c}$. Then, while $\Delta_{\overline{\eta}} \equiv 0$ in this temperature zone, we may cancel the factor $\Delta_{\eta}$ on both sides, and afterward take the limit $\lim_{T\rightarrow T_{\eta}^{c-}}\Delta_{\eta}(T) = 0$ in Eq.~\eqref{eq_BCS_T}, to obtain
\begin{eqnarray}
\frac{1}{\lambda_{\eta}} &=& \lim_{\Delta_{\eta}\rightarrow 0}C_{\eta}\left(\Delta_{\eta},\Delta_{\overline{\eta}}=0\right) = \Lambda_{\eta\overline{\eta}}^{20} \int_0^{\frac{\omega_D}{2 T_{\eta}^{c}}}dz\frac{\tanh(z)}{z}\nonumber\\
&\simeq& \Lambda_{\eta\overline{\eta}}^{20}\ln\left( \frac{2}{\pi}e^{\gamma_E}\frac{\omega_D}{T_{\eta}^c} \right),
\label{eq_TC1}
\end{eqnarray}
where $\gamma_E$ is the Euler-Mascheroni constant, and we defined the angular integrals
\begin{eqnarray}
\Lambda_{\eta\overline{\eta}}^{n m} \equiv \frac{2}{B_{\nu}}\int_0^1 \frac{dx\, x f_{\eta}^{n}(\theta_x) f_{\overline{\eta}}^{m}(\theta_x)}{\sqrt{1 - x^{2\nu}}}.
\label{eq:Lambdaeta}
\end{eqnarray}
Solving explicitly from Eq.~\eqref{eq_TC1}, we obtain the highest critical temperature as
\begin{equation}
T_{\eta}^c = \frac{2 \omega_D}{\pi}e^{\gamma_E} e^{-\frac{1}{\lambda_{\eta}\Lambda_{\eta\overline{\eta}}^{20}}}.
\label{eq:Tceta}
\end{equation}
Based on this first result, we then conclude that the highest critical temperature $T_{\eta}^c$ will depend, in terms of the microscopic parameters of the model, upon de condition ${\rm{Max}}\left\{ \lambda_1 \Lambda_{10}^{20},\lambda_0\Lambda_{10}^{02} \right\}$.

Let us now consider the interval between the two critical temperatures, i.e. $T_{\overline{\eta}}^c < T < T_{\eta}^c$. Therefore, the condition $\Delta_{\overline{\eta}}(T) \equiv 0$ still applies, but $\Delta_{\eta}(T) \ne 0$. We can thus expand the functions in the integrand of Eq.~\eqref{eq_Ccoeff} up to second-order in $\Delta_{\eta}(T)$, as follows
\begin{eqnarray}
\frac{\mathcal{T}_{\beta}(\Gamma) + \mathcal{T}_{\beta}(\gamma)}{2} \simeq \frac{\tanh\left(\frac{\beta\xi}{2}\right)}{\xi}+  \frac{\Delta_{\eta}^2 f_{\eta}^2(\theta_x)}{16}\psi_{\beta}\left(\frac{\beta\xi}{2}\right),\nonumber\\
\end{eqnarray}
where we defined the function
\begin{equation}
\psi_{\beta}(z) = \frac{\beta^3}{z^3}\left[z\, \rm{sech}^{2}(z) - \tanh(z)\right].
\label{eq_psi}
\end{equation}
Substituting these relations into Eq.~\eqref{eq_BCS_T}, we obtain (for $T_{\overline{\eta}}^c < T < T_{\eta}^c$)
\begin{eqnarray}
\frac{1}{\lambda_{\eta}} &=& C_{\eta}\left(\Delta_{\eta},\Delta_{\overline{\eta}}=0\right)\nonumber\\
&=& \Lambda_{\eta\overline{\eta}}^{20}\ln\left( \frac{2}{\pi}e^{\gamma_E}\frac{\omega_D}{T} \right) - \alpha\Lambda_{\eta\overline{\eta}}^{40}\frac{\Delta_{\eta}^2(T)}{T^2},
\label{eq_lambdaeta}
\end{eqnarray}
with $\alpha = 7\zeta(3)/(8 \pi^2) \sim 0.10657$ and $\zeta(3) \sim 1.202$ the Riemann zeta function.
Substituting the definition of the highest critical temperature from Eq.~\eqref{eq_TC1} into Eq.\eqref{eq_lambdaeta}, we obtain the temperature dependence of the corresponding order parameter
\begin{eqnarray}
\Delta_{\eta}^2 (T) = \frac{\Lambda_{\eta\overline{\eta}}^{20}}{\Lambda_{\eta\overline{\eta}}^{40}}\frac{T^2}{\alpha}\ln\left(  \frac{T_{\eta}^c}{T}\right)\sim \kappa_{\eta}\left| T_{\eta}^c - T \right|,
\label{eq_DeltaT}
\end{eqnarray}
with
$\kappa_{\eta} = \frac{\Lambda_{\eta\overline{\eta}}^{20}}{\Lambda_{\eta\overline{\eta}}^{40}}\frac{T^c_{\eta}}{\alpha}$. This clearly shows that the corresponding critical exponent  is $1/2$, just as in the standard BCS theory.

In order to investigate the lowest critical temperature $T_{\overline{\eta}}^{c}$, let us now consider the second equation in the system Eq.~\eqref{eq_BCS_T} corresponding to $\overline{\eta}$, and take the limit $T \rightarrow T_{\overline{\eta}}^{c-}$, where $\Delta_{\overline{\eta}}\rightarrow 0$, but $\Delta_{\eta}\ne 0$, 
\begin{eqnarray}
\frac{1}{\lambda_{\overline{\eta}}} &=& C_{\overline{\eta}}(\Delta_{\eta},\Delta_{\overline{\eta}}=0) + \lim_{\Delta_{\overline{\eta}}\rightarrow 0}D_{\overline{\eta}}\left(\Delta_{\eta},\Delta_{\overline{\eta}}\right)\cdot\frac{\Delta_{\eta}}{\Delta_{\overline{\eta}}}
\label{eq_lambdabareta1}
\end{eqnarray}
The limit in the second term above is obtained by first using the expansion
\begin{eqnarray}
\mathcal{T}_{\beta}(\Gamma) - \mathcal{T}_{\beta}(\gamma) \simeq \frac{\Delta_{\eta}\Delta_{\overline{\eta}}}{4}f_{\eta}(\theta_x)f_{\overline{\eta}}(\theta_x)\cos(r\phi)\psi_{\beta}\left( \frac{\beta\xi}{2} \right)\nonumber\\
\end{eqnarray}
in Eq.~\eqref{eq_Dcoeff},
and by further applying the identity
\begin{equation}
\int_0^{2\pi}\frac{d\phi}{2\pi}e^{\pm ir\phi}\cos(r\phi) = \frac{1 + \delta_{r,0}}{2}.
\end{equation}
Therefore, we obtain 
\begin{eqnarray}
\lim_{\Delta_{\overline{\eta}}\rightarrow 0}D_{\overline{\eta}}\left(\Delta_{\eta},\Delta_{\overline{\eta}}\right)\cdot\frac{\Delta_{\eta}}{\Delta_{\overline{\eta}}} = -\alpha\left( 1 + \delta_{r,0} \right)\Lambda_{\eta\overline{\eta}}^{22}\frac{\Delta_{\eta}^2(T_{\overline{\eta}}^c)}{T_{\overline{\eta}}^{c 2}}.
\end{eqnarray}
Substituting this result into Eq.~\eqref{eq_lambdabareta1}, and assuming that the two critical temperatures are close enough so we can still use Eq.~\eqref{eq_DeltaT} for $\Delta_{\eta}(T)$, we obtain the expression
\begin{eqnarray}
\frac{1}{\Lambda_{\eta\overline{\eta}}^{02}\lambda_{\overline{\eta}}}=\ln\left( \frac{2}{\pi}e^{\gamma_E}\frac{\omega_D}{T_{\overline{\eta}}^c} \right)- \left(2 + \delta_{r,0}\right)\frac{\Lambda_{\eta\overline{\eta}}^{22}\Lambda_{\eta\overline{\eta}}^{20}}{\Lambda_{\eta\overline{\eta}}^{02}\Lambda_{\eta\overline{\eta}}^{40}}  \ln\left( \frac{T_{\eta}^c}{T_{\overline{\eta}}^c} \right)\nonumber\\
\label{eq_TC2}
\end{eqnarray}
Subtracting Eq.~\eqref{eq_TC1} from Eq.~\eqref{eq_TC2}, after some elementary algebra we obtain
\begin{eqnarray}
\frac{1}{\lambda_{\overline{\eta}}\Lambda_{\eta\overline{\eta}}^{02}}-\frac{1}{\lambda_{\eta}\Lambda_{\eta\overline{\eta}}^{20}}  = \frac{1}{\sigma_r} \ln\left( \frac{T_{\eta}^c}{T_{\overline{\eta}}^c} \right),
\label{eq_ratio1}
\end{eqnarray}
where we defined the parameter
\begin{equation}
1/\sigma_r = 1 - \left(2 + \delta_{r,0}\right)\frac{\Lambda_{\eta\overline{\eta}}^{20}\Lambda_{\eta\overline{\eta}}^{22}}{\Lambda_{\eta\overline{\eta}}^{02}\Lambda_{\eta\overline{\eta}}^{40}}.  
\label{eq_sigmar}
\end{equation}
From Eq.~\eqref{eq_ratio1}, we obtain the expression
\begin{eqnarray}
T_{\overline{\eta}}^c = T_{\eta}^c e^{-\sigma_r\left(\frac{1}{\lambda_{\overline{\eta}}\Lambda_{\eta\overline{\eta}}^{02}}-\frac{1}{\lambda_{\eta}\Lambda_{\eta\overline{\eta}}^{20}}  \right)}.
\end{eqnarray}
This result implies a necessary condition for the existence of the lowest temperature superconducting phase $\Delta_{\overline{\eta}}(T)\ne 0$, i.e. that $\sigma_r\left(\frac{1}{\lambda_{\overline{\eta}}\Lambda_{\eta\overline{\eta}}^{02}}-\frac{1}{\lambda_{\eta}\Lambda_{\eta\overline{\eta}}^{20}}  \right) > 0$.

Finally, in the low temperature region $T < T_{\overline{\eta}}^c$, both order parameters acquire small finite values, such that applying a similar expansion procedure as described before, we find that the coefficients in the system Eq.~\eqref{eq_BCS_T} are given (up to second order) by
\begin{eqnarray}
C_{\eta} &\simeq&  \Lambda_{\eta\overline{\eta}}^{20}\ln\left( \frac{2}{\pi}e^{\gamma_E}\frac{\omega_D}{T} \right) - \frac{\alpha}{T^2}\left(\Lambda_{\eta\overline{\eta}}^{40}\Delta_{\eta}^2 + \Lambda_{\eta\overline{\eta}}^{22}\Delta_{\overline{\eta}}^2\right)\nonumber\\
C_{\overline{\eta}} &\simeq&  \Lambda_{\eta\overline{\eta}}^{02}\ln\left( \frac{2}{\pi}e^{\gamma_E}\frac{\omega_D}{T} \right) - \frac{\alpha}{T^2}\left(\Lambda_{\eta\overline{\eta}}^{04}\Delta_{\overline{\eta}}^2 + \Lambda_{\eta\overline{\eta}}^{22}\Delta_{\eta}^2\right)\nonumber\\
D_{\eta} = D_{\overline{\eta}} &\simeq& -\frac{\alpha(1 + \delta_{r,0})}{T^2}\Lambda_{\eta\overline{\eta}}^{22}\Delta_{\eta}\Delta_{\overline{\eta}}
\end{eqnarray}
Inserting these expressions into the system Eq.~\eqref{eq_BCS_T}, canceling common factors, and substituting the coupling constants in terms of both critical temperatures, we end up with the linear system
\begin{eqnarray}
\frac{\Lambda^{40}_{\eta\overline{\eta}}}{\Lambda^{20}_{\eta\overline{\eta}}}\Delta_{\eta}^2 + \frac{\Lambda^{22}_{\eta\overline{\eta}}}{\Lambda^{20}_{\eta\overline{\eta}}}\left( 2 + \delta_{r,0} \right)\Delta_{\overline{\eta}}^2 &=& -\frac{T^2}{\alpha}\ln\left(\frac{T}{T_{\eta}^c}  \right)\\
\frac{\Lambda^{04}_{\eta\overline{\eta}}}{\Lambda^{02}_{\eta\overline{\eta}}}\Delta_{\overline{\eta}}^2 + \frac{\Lambda^{22}_{\eta\overline{\eta}}}{\Lambda^{02}_{\eta\overline{\eta}}}\left( 2 + \delta_{r,0} \right)\Delta_{\eta}^2 &=& -\frac{T^2}{\alpha}\ln\left[\left(\frac{T}{T_{\eta}^c}  \right)\left(\frac{T_{\eta}^c}{T_{\overline{\eta}}^c}  \right)^{\frac{1}{\sigma_r}}\right]\nonumber
\end{eqnarray}
By explicitly solving this linear system for the order parameters, we finally obtain, in the region close to the lower critical temperature $0 < T \lesssim T_{\overline{\eta}}^c$
\begin{eqnarray}
\Delta_{\eta}^{2}(T) &=& q_r \frac{T^2}{\alpha} \ln\left( \frac{T_{\eta}^c}{T_{\overline{\eta}^c}} \right)
- p_r \frac{T^2}{\alpha} \ln\left( \frac{T}{T_{\eta}^c}  \right)\nonumber\\
\Delta_{\overline{\eta}}^{2}(T) &=&- \overline{p}_r \frac{T^2}{\alpha} \ln\left( \frac{T}{T_{\overline{\eta}}^c}\right) \simeq \kappa_{\overline{\eta}}|T - T_{\overline{\eta}}^c|.
\end{eqnarray}
Here, we define the coefficients
\begin{eqnarray}
q_r &\equiv& \frac{\Lambda_{\eta\overline{\eta}}^{22} \Lambda_{\eta\overline{\eta}}^{02}\left( 2 + \delta_{r,0}  \right)}{\sigma_r\left[ \Lambda_{\eta\overline{\eta}}^{40}\Lambda_{\eta\overline{\eta}}^{04} - \left( 2 + \delta_{r,0}  \right)^2\left(\Lambda_{\eta\overline{\eta}}^{22}\right)^2\right]}\nonumber\\
p_r &\equiv& \frac{ \Lambda_{\eta\overline{\eta}}^{04} \Lambda_{\eta\overline{\eta}}^{20}   -\left(  2 + \delta_{r,0}\right)\Lambda_{\eta\overline{\eta}}^{22} \Lambda_{\eta\overline{\eta}}^{02} }{\left[ \Lambda_{\eta\overline{\eta}}^{40}\Lambda_{\eta\overline{\eta}}^{04} - \left( 2 + \delta_{r,0}  \right)^2\left(\Lambda_{\eta\overline{\eta}}^{22}\right)^2\right]}\nonumber\\
\overline{p}_r &\equiv& \frac{ \Lambda_{\eta\overline{\eta}}^{40} \Lambda_{\eta\overline{\eta}}^{02}   -\left(  2 + \delta_{r,0}\right)\Lambda_{\eta\overline{\eta}}^{22} \Lambda_{\eta\overline{\eta}}^{20} }{\left[ \Lambda_{\eta\overline{\eta}}^{40}\Lambda_{\eta\overline{\eta}}^{04} - \left( 2 + \delta_{r,0}  \right)^2\left(\Lambda_{\eta\overline{\eta}}^{22}\right)^2\right]}\nonumber\\
\kappa_{\overline{\eta}} &\equiv& \frac{\overline{p}_r T_{\overline{\eta}}^c}{\alpha}
\end{eqnarray}

We present explicit numerical examples of the temperature dependence of both pairings in Fig.~\ref{fig: pairings s-wave and monopole} for the s-wave and in Fig.~\ref{fig: pairings p-wave and monopole} for the p-wave, respectively. As can be observed in both figures, the critical behavior of these order parameters is qualitatively similar, as follows from the previous mathematical analysis in the vicinity of the critical temperatures. Clearly, in agreement with Eq.~\eqref{eq:Tceta}, the numerical value of each critical temperature $T_0^c$ and $T_1^c$ is established by the angular dependence of the corresponding order parameters, and by the topological charge $\nu$, via the angular integrals Eq.~\eqref{eq:Lambdaeta}. In particular, for the two examples considered here, i.e., the s-wave in Fig.~\ref{fig: pairings s-wave and monopole} and the p-wave in Fig.~\ref{fig: pairings p-wave and monopole}, the critical temperature associated with the monopole (inter-nodal) is slightly lower than the corresponding to conventional spherical harmonic (intra-nodal) $T_1^c < T_0^1$, for all three topological charges $\nu = 1,\,2,\,3$. On the other hand, we see that for the s-wave in Fig.~\ref{fig: pairings s-wave and monopole}, the order parameters become practically independent of the parameter $c$, for relatively large values $c > 1$. In contrast, for the p-wave in Fig.~\ref{fig: pairings p-wave and monopole} we see that there is a strong dependence on this parameter in the region $c < 1$.

\section{Specific Heat}
\begin{figure}[ht]
    \centering
    \includegraphics[width=1.0\linewidth]{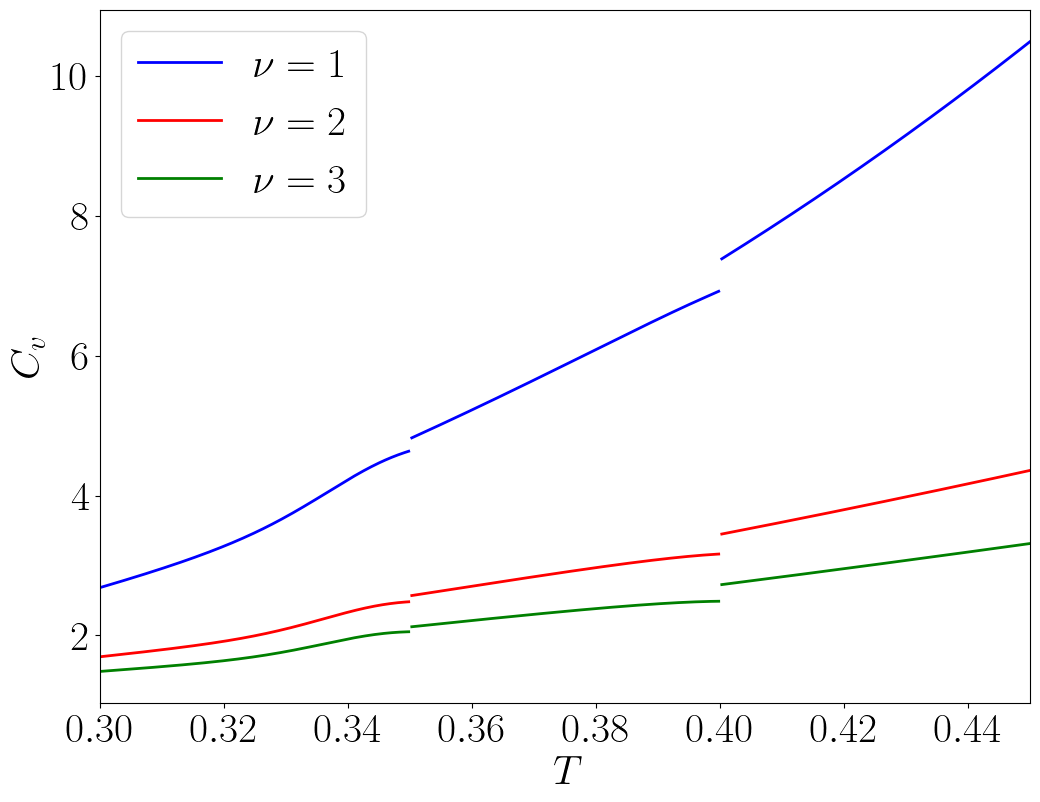}
    \caption{Total specific heaet $C_{v} = C_{v}^{\text{norm}} + C_{v}^{SC}$ (in units of $k_B\omega_{D}/\left[v_{f}\alpha_{\nu}^{2/(\nu-1)}\right]$)}, as function of temperature (in units of $\hbar v_f \alpha_{\nu}^{1/(1-\nu)}/k_B$), for $\nu=1,2,3$. Notice that $\alpha_{\nu} has $.
    \label{fig:specific heat}
\end{figure}
In cases where both superconducting phases coexist, we expect two superconducting phase transitions to exist at the corresponding critical temperatures, which are likely to be different. As a possible experimental probe for these transitions, we seek to calculate the specific heat of the system as a function of temperature, since we expect a discontinuous jump in the specific heat at each critical point.

For this purpose, we first calculate the internal energy of the system from the grand canonical partition function, according to the general expression
\begin{equation}
    \langle\hat{H}\rangle = -\frac{\partial}{\partial\beta} \ln \mathcal{Z} + \frac{\mu}{\beta} \frac{\partial}{\partial \mu} \ln\mathcal{Z}.
\end{equation}
The contribution to the partition function associated with the superconducting phases is calculated from the finite-temperature Green's function of our effective model in the Matsubara frequency space.
\begin{eqnarray}
    \ln\mathcal{Z}^{SC} &=& \ln\left( \det[\hat{\mathcal{G}}^{-1}_{\textbf{k}}(\omega_{n})] \right) = - \text{Tr} \left[ \ln \hat{\mathcal{G}}_{\textbf{k}}(\omega_{n}) \right] \notag\\
    &=& \sum_{\textbf{k}, \omega_{n}} \ln\left[(\omega_{n}^{2} + \Gamma^{2}_{\textbf{k}}) (\omega_{n}^{2} + \gamma^{2}_{\textbf{k}})\right] \\
    &=& \sum_{\textbf{k}} \ln\left[(1 + \cosh(\beta \Gamma_{\textbf{k}})) (1 + \cosh(\beta \gamma_{\textbf{k}}))\right]. \notag
\end{eqnarray}

Then, the specific heat associated with each of the superconducting transitions is calculated as
\begin{equation}
    C_{v}^{SC} = \beta^{2} \frac{\partial^{2}}{\partial\beta^{2}} \ln\mathcal{Z}^{SC} - \beta^{2}\frac{\partial}{\partial\beta} \left(\frac{\mu}{\beta} \frac{\partial}{\partial \mu} \ln\mathcal{Z}^{SC}\right).
\end{equation}
Here we must consider the temperature dependence of the gap functions to calculate the derivatives. In agreement with~\cite{Muñoz_et_al_2024}, we have
\begin{equation}
    |\Delta^{\eta}| \sim |T-T_{c}^{\eta}|^{1/2},
\end{equation}
for both $\eta=\text{intra/inter}$, $T<T_{c}^{\eta}$ and for a simple WSM with isotropic dispersion relation. Here we will use the same relation, because the development carried out in \cite{Muñoz_et_al_2024} can be generalized to multi-WSM obtaining the same critical exponents.

On the other hand, we calculate the specific heat component due to fermions in the normal WSM phase, with energy dispersion $\xi_{\textbf{k}} = v_{f} \sqrt{k_{z}^{2} + \alpha^{2} k_{\perp}^{2\nu}}$, as follows
\begin{equation}
    C_{v}^{\text{norm}} = \beta^{2} \int \frac{d^{3}k}{(2\pi)^{3}} \frac{(\xi_{\textbf{k}} - \mu)^{2} e^{-\beta(\xi_{\textbf{k}} - \mu)}}{(1 + e^{-\beta(\xi_{\textbf{k}} - \mu)})^{2}}.
\end{equation}
This integral is solved by means of the change of variables
\begin{eqnarray}
    k_{x} &=& \left(\frac{r}{\alpha_{\nu}}\sin\theta\right)^{1/\nu} \cos\phi, \notag\\
    k_{y} &=& \left(\frac{r}{\alpha_{\nu}}\sin\theta\right)^{1/\nu} \sin\phi, \\
    k_{z} &=& r \cos\theta \notag,
\end{eqnarray}
with which we have $\xi_{\textbf{k}}=v_{f} r$ and the Jacobian
\begin{equation}
    \left| \frac{\partial(k_{x},k_{y},k_{z})}{\partial(r,\theta,\phi)} \right| = \frac{1}{\nu} \left(\frac{r}{\alpha_{\nu}}\right)^{2/\nu} \sin^{2/\nu-1}\theta.
\end{equation}
Then, in the limit $\mu\rightarrow 0$ we obtain
\begin{equation}
    C_{v}^{\text{norm}} = \frac{\beta(\frac{1}{\nu},\frac{1}{2})}{(2\pi)^{2}\nu \alpha_{\nu}^{2/\nu}} \left(\frac{T}{v_{f}}\right)^{\frac{2}{\nu}+1} \int_{0}^{\infty}dx \frac{x^{2+2/\nu} e^{-x}}{(1 + e^{-x})^{2}}.
\end{equation}

In Figure \ref{fig:specific heat} we plot the total specific heat $C_{v}^{\text{norm}} + C_{v}^{SC}$. As expected, the specific heat displays discontinuities precisely at the critical temperatures corresponding to each monopole and conventional superconducting phases, respectively. In addition, as seen in Fig.~\ref{fig:specific heat}, the curves display different slopes depending on the chiral index $\nu = 1,2,3$, which may offer an experimental probe to characterize these topological materials by measuring their specific heat.
\section{Discussion and Conclusions}

In this work, we presented a theoretical model to explore the possible emergence and competition of conventional and monopole superconducting phases in multi-WSMs, generalizing the previous analysis presented in~\cite{Munoz-2020,Muñoz_et_al_2024}. We explicitly showed that the order parameters associated to intra-nodal and inter-nodal pairing, exhibit conventional spherical harmonics $\Delta^{\text{intra}}_{\textbf{k}} =  \Delta_{l,m'} Y_{l,m'} (\theta_{\textbf{k}}, \phi_{\textbf{k}}) = \Delta_{l,m'} f_{\text{intra}}(\theta_{\textbf{k}}) e^{im' \phi_{\textbf{k}}}$, and monopole harmonics $\Delta^{\text{inter}}_{\textbf{k}} =  \bar{\Delta}_{\nu} \mathcal{Y}_{\nu,j,m} (\theta_{\textbf{k}}, \phi_{\textbf{k}}) = \bar{\Delta}_{\nu} f_{\text{inter}}(\theta_{\textbf{k}}) e^{i(m+\nu) \phi_{\textbf{k}}}$ angular dependencies, respectively.
According to our model, we have shown that, depending on the nature of the $\theta$-component of the spherical harmonics of the order parameters defined above, it is possible to observe both a conventional and a monopole superconducting phases, separated by a region of coexistence or mixed state. However, we also showed that if the azimuthal indexes of the conventional ($m'$) and monopole ($m$) satisfy the relation $m' = m + \nu$, and furthermore the $f_{intra}(\theta) \simeq f_{inter}(\theta)$, then topological repulsion takes place and hence the coexistence region between both phases disappears, in agreement with~\cite{Munoz-2020}.

We further computed the specific heat as a function of temperature, and we showed that according to our model this property displays discontinuities at the critical temperatures corresponding to each monopole and conventional superconducting phases, respectively. As discussed in Section~\ref{Section: Gap Eqns}, the critical temperatures for both possible superconducting phases depend explicitly on the topological charge $\nu$, as well as on the microscopic parameter $c$.
Moreover, we also showed that the slope of the specific heat curve as a function of temperature strongly depends on the chirality index $\nu = 1,2,3$. Both features suggest that, according to our model, the specific heat offers a possible experimental probe to characterize the topological properties of these quantum materials.

\section*{Acknowledgments}
E. Mu\~noz acknowledges financial support from ANID Fondecyt grant No 1230440. A. Tapia acknowledges financial support from ANID-Subdirección de Capital
Humano/Magíster Nacional/2024 - 22240909.


\begin{thebibliography}{41}%
\makeatletter
\providecommand \@ifxundefined [1]{%
 \@ifx{#1\undefined}
}%
\providecommand \@ifnum [1]{%
 \ifnum #1\expandafter \@firstoftwo
 \else \expandafter \@secondoftwo
 \fi
}%
\providecommand \@ifx [1]{%
 \ifx #1\expandafter \@firstoftwo
 \else \expandafter \@secondoftwo
 \fi
}%
\providecommand \natexlab [1]{#1}%
\providecommand \enquote  [1]{``#1''}%
\providecommand \bibnamefont  [1]{#1}%
\providecommand \bibfnamefont [1]{#1}%
\providecommand \citenamefont [1]{#1}%
\providecommand \href@noop [0]{\@secondoftwo}%
\providecommand \href [0]{\begingroup \@sanitize@url \@href}%
\providecommand \@href[1]{\@@startlink{#1}\@@href}%
\providecommand \@@href[1]{\endgroup#1\@@endlink}%
\providecommand \@sanitize@url [0]{\catcode `\\12\catcode `\$12\catcode `\&12\catcode `\#12\catcode `\^12\catcode `\_12\catcode `\%12\relax}%
\providecommand \@@startlink[1]{}%
\providecommand \@@endlink[0]{}%
\providecommand \url  [0]{\begingroup\@sanitize@url \@url }%
\providecommand \@url [1]{\endgroup\@href {#1}{\urlprefix }}%
\providecommand \urlprefix  [0]{URL }%
\providecommand \Eprint [0]{\href }%
\providecommand \doibase [0]{http://dx.doi.org/}%
\providecommand \selectlanguage [0]{\@gobble}%
\providecommand \bibinfo  [0]{\@secondoftwo}%
\providecommand \bibfield  [0]{\@secondoftwo}%
\providecommand \translation [1]{[#1]}%
\providecommand \BibitemOpen [0]{}%
\providecommand \bibitemStop [0]{}%
\providecommand \bibitemNoStop [0]{.\EOS\space}%
\providecommand \EOS [0]{\spacefactor3000\relax}%
\providecommand \BibitemShut  [1]{\csname bibitem#1\endcsname}%
\let\auto@bib@innerbib\@empty
\bibitem [{\citenamefont {Murakami}(2007)}]{Murakami_2007}%
  \BibitemOpen
  \bibfield  {author} {\bibinfo {author} {\bibfnamefont {Shuichi}\ \bibnamefont {Murakami}},\ }\bibfield  {title} {\enquote {\bibinfo {title} {Phase transition between the quantum spin hall and insulator phases in 3d: emergence of a topological gapless phase},}\ }\href {\doibase 10.1088/1367-2630/9/9/356} {\bibfield  {journal} {\bibinfo  {journal} {New Journal of Physics}\ }\textbf {\bibinfo {volume} {9}},\ \bibinfo {pages} {356} (\bibinfo {year} {2007})}\BibitemShut {NoStop}%
\bibitem [{\citenamefont {Burkov}\ and\ \citenamefont {Balents}(2011)}]{Burkov_Balents_2011}%
  \BibitemOpen
  \bibfield  {author} {\bibinfo {author} {\bibfnamefont {A.~A.}\ \bibnamefont {Burkov}}\ and\ \bibinfo {author} {\bibfnamefont {Leon}\ \bibnamefont {Balents}},\ }\bibfield  {title} {\enquote {\bibinfo {title} {Weyl semimetal in a topological insulator multilayer},}\ }\href {\doibase 10.1103/PhysRevLett.107.127205} {\bibfield  {journal} {\bibinfo  {journal} {Phys. Rev. Lett.}\ }\textbf {\bibinfo {volume} {107}},\ \bibinfo {pages} {127205} (\bibinfo {year} {2011})}\BibitemShut {NoStop}%
\bibitem [{\citenamefont {Wan}\ \emph {et~al.}(2011)\citenamefont {Wan}, \citenamefont {Turner}, \citenamefont {Vishwanath},\ and\ \citenamefont {Savrasov}}]{Wan_et_al_2011}%
  \BibitemOpen
  \bibfield  {author} {\bibinfo {author} {\bibfnamefont {Xiangang}\ \bibnamefont {Wan}}, \bibinfo {author} {\bibfnamefont {Ari~M.}\ \bibnamefont {Turner}}, \bibinfo {author} {\bibfnamefont {Ashvin}\ \bibnamefont {Vishwanath}}, \ and\ \bibinfo {author} {\bibfnamefont {Sergey~Y.}\ \bibnamefont {Savrasov}},\ }\bibfield  {title} {\enquote {\bibinfo {title} {Topological semimetal and fermi-arc surface states in the electronic structure of pyrochlore iridates},}\ }\href {\doibase 10.1103/PhysRevB.83.205101} {\bibfield  {journal} {\bibinfo  {journal} {Phys. Rev. B}\ }\textbf {\bibinfo {volume} {83}},\ \bibinfo {pages} {205101} (\bibinfo {year} {2011})}\BibitemShut {NoStop}%
\bibitem [{\citenamefont {T.O.~Wehling}\ and\ \citenamefont {Balatsky}(2014)}]{Wehling_014}%
  \BibitemOpen
  \bibfield  {author} {\bibinfo {author} {\bibfnamefont {A.M. Black-Schaffer}\ \bibnamefont {T.O.~Wehling}}\ and\ \bibinfo {author} {\bibfnamefont {A.V.}\ \bibnamefont {Balatsky}},\ }\bibfield  {title} {\enquote {\bibinfo {title} {Dirac materials},}\ }\href {\doibase 10.1080/00018732.2014.927109} {\bibfield  {journal} {\bibinfo  {journal} {Advances in Physics}\ }\textbf {\bibinfo {volume} {63}},\ \bibinfo {pages} {1--76} (\bibinfo {year} {2014})},\ \Eprint {http://arxiv.org/abs/https://doi.org/10.1080/00018732.2014.927109} {https://doi.org/10.1080/00018732.2014.927109} \BibitemShut {NoStop}%
\bibitem [{\citenamefont {Chiu}\ \emph {et~al.}(2016)\citenamefont {Chiu}, \citenamefont {Teo}, \citenamefont {Schnyder},\ and\ \citenamefont {Ryu}}]{Chiu_016}%
  \BibitemOpen
  \bibfield  {author} {\bibinfo {author} {\bibfnamefont {C.-K.}\ \bibnamefont {Chiu}}, \bibinfo {author} {\bibfnamefont {J.~C.~Y.}\ \bibnamefont {Teo}}, \bibinfo {author} {\bibfnamefont {A.~P.}\ \bibnamefont {Schnyder}}, \ and\ \bibinfo {author} {\bibfnamefont {S.}~\bibnamefont {Ryu}},\ }\bibfield  {title} {\enquote {\bibinfo {title} {Classification of topological quantum matter with symmetries},}\ }\href@noop {} {\bibfield  {journal} {\bibinfo  {journal} {Rev. Mod. Phys.}\ }\textbf {\bibinfo {volume} {88}},\ \bibinfo {pages} {035005} (\bibinfo {year} {2016})}\BibitemShut {NoStop}%
\bibitem [{\citenamefont {Bradlyn}\ \emph {et~al.}(2016)\citenamefont {Bradlyn}, \citenamefont {Cano}, \citenamefont {Wang}, \citenamefont {Vergniory}, \citenamefont {Felser}, \citenamefont {Cava},\ and\ \citenamefont {Bernevig}}]{Bradlyn_016}%
  \BibitemOpen
  \bibfield  {author} {\bibinfo {author} {\bibfnamefont {Barry}\ \bibnamefont {Bradlyn}}, \bibinfo {author} {\bibfnamefont {Jennifer}\ \bibnamefont {Cano}}, \bibinfo {author} {\bibfnamefont {Zhijun}\ \bibnamefont {Wang}}, \bibinfo {author} {\bibfnamefont {M.~G.}\ \bibnamefont {Vergniory}}, \bibinfo {author} {\bibfnamefont {C.}~\bibnamefont {Felser}}, \bibinfo {author} {\bibfnamefont {R.~J.}\ \bibnamefont {Cava}}, \ and\ \bibinfo {author} {\bibfnamefont {B.~Andrei}\ \bibnamefont {Bernevig}},\ }\bibfield  {title} {\enquote {\bibinfo {title} {Beyond dirac and weyl fermions: Unconventional quasiparticles in conventional crystals},}\ }\href {\doibase 10.1126/science.aaf5037} {\bibfield  {journal} {\bibinfo  {journal} {Science}\ }\textbf {\bibinfo {volume} {353}},\ \bibinfo {pages} {aaf5037} (\bibinfo {year} {2016})},\ \Eprint {http://arxiv.org/abs/https://www.science.org/doi/pdf/10.1126/science.aaf5037} {https://www.science.org/doi/pdf/10.1126/science.aaf5037} \BibitemShut {NoStop}%
\bibitem [{\citenamefont {Armitage}\ \emph {et~al.}(2018)\citenamefont {Armitage}, \citenamefont {Mele},\ and\ \citenamefont {Vishwanath}}]{Armitage_018}%
  \BibitemOpen
  \bibfield  {author} {\bibinfo {author} {\bibfnamefont {N.~P.}\ \bibnamefont {Armitage}}, \bibinfo {author} {\bibfnamefont {E.~J.}\ \bibnamefont {Mele}}, \ and\ \bibinfo {author} {\bibfnamefont {Ashvin}\ \bibnamefont {Vishwanath}},\ }\bibfield  {title} {\enquote {\bibinfo {title} {Weyl and dirac semimetals in three-dimensional solids},}\ }\href {\doibase 10.1103/RevModPhys.90.015001} {\bibfield  {journal} {\bibinfo  {journal} {Rev. Mod. Phys.}\ }\textbf {\bibinfo {volume} {90}},\ \bibinfo {pages} {015001} (\bibinfo {year} {2018})}\BibitemShut {NoStop}%
\bibitem [{\citenamefont {Burkov}(2018)}]{Burkov_018}%
  \BibitemOpen
  \bibfield  {author} {\bibinfo {author} {\bibfnamefont {A.A.}\ \bibnamefont {Burkov}},\ }\bibfield  {title} {\enquote {\bibinfo {title} {Weyl metals},}\ }\href {\doibase https://doi.org/10.1146/annurev-conmatphys-033117-054129} {\bibfield  {journal} {\bibinfo  {journal} {Annual Review of Condensed Matter Physics}\ }\textbf {\bibinfo {volume} {9}},\ \bibinfo {pages} {359--378} (\bibinfo {year} {2018})}\BibitemShut {NoStop}%
\bibitem [{\citenamefont {Xu}\ \emph {et~al.}(2015{\natexlab{a}})\citenamefont {Xu}, \citenamefont {Belopolski}, \citenamefont {Alidoust}, \citenamefont {Neupane}, \citenamefont {Bian}, \citenamefont {Zhang}, \citenamefont {Sankar}, \citenamefont {Chang}, \citenamefont {Yuan}, \citenamefont {Lee}, \citenamefont {Huang}, \citenamefont {Zheng}, \citenamefont {Ma}, \citenamefont {Sanchez}, \citenamefont {Wang}, \citenamefont {Bansil}, \citenamefont {Chou}, \citenamefont {Shibayev}, \citenamefont {Lin}, \citenamefont {Jia},\ and\ \citenamefont {Hasan}}]{Xu_015}%
  \BibitemOpen
  \bibfield  {author} {\bibinfo {author} {\bibfnamefont {Su-Yang}\ \bibnamefont {Xu}}, \bibinfo {author} {\bibfnamefont {Ilya}\ \bibnamefont {Belopolski}}, \bibinfo {author} {\bibfnamefont {Nasser}\ \bibnamefont {Alidoust}}, \bibinfo {author} {\bibfnamefont {Madhab}\ \bibnamefont {Neupane}}, \bibinfo {author} {\bibfnamefont {Guang}\ \bibnamefont {Bian}}, \bibinfo {author} {\bibfnamefont {Chenglong}\ \bibnamefont {Zhang}}, \bibinfo {author} {\bibfnamefont {Raman}\ \bibnamefont {Sankar}}, \bibinfo {author} {\bibfnamefont {Guoqing}\ \bibnamefont {Chang}}, \bibinfo {author} {\bibfnamefont {Zhujun}\ \bibnamefont {Yuan}}, \bibinfo {author} {\bibfnamefont {Chi-Cheng}\ \bibnamefont {Lee}}, \bibinfo {author} {\bibfnamefont {Shin-Ming}\ \bibnamefont {Huang}}, \bibinfo {author} {\bibfnamefont {Hao}\ \bibnamefont {Zheng}}, \bibinfo {author} {\bibfnamefont {Jie}\ \bibnamefont {Ma}}, \bibinfo {author} {\bibfnamefont {Daniel~S.}\ \bibnamefont {Sanchez}}, \bibinfo {author} {\bibfnamefont {BaoKai}\ \bibnamefont {Wang}},
  \bibinfo {author} {\bibfnamefont {Arun}\ \bibnamefont {Bansil}}, \bibinfo {author} {\bibfnamefont {Fangcheng}\ \bibnamefont {Chou}}, \bibinfo {author} {\bibfnamefont {Pavel~P.}\ \bibnamefont {Shibayev}}, \bibinfo {author} {\bibfnamefont {Hsin}\ \bibnamefont {Lin}}, \bibinfo {author} {\bibfnamefont {Shuang}\ \bibnamefont {Jia}}, \ and\ \bibinfo {author} {\bibfnamefont {M.~Zahid}\ \bibnamefont {Hasan}},\ }\bibfield  {title} {\enquote {\bibinfo {title} {Discovery of a weyl fermion semimetal and topological fermi arcs},}\ }\href {\doibase 10.1126/science.aaa9297} {\bibfield  {journal} {\bibinfo  {journal} {Science}\ }\textbf {\bibinfo {volume} {349}},\ \bibinfo {pages} {613--617} (\bibinfo {year} {2015}{\natexlab{a}})},\ \Eprint {http://arxiv.org/abs/https://www.science.org/doi/pdf/10.1126/science.aaa9297} {https://www.science.org/doi/pdf/10.1126/science.aaa9297} \BibitemShut {NoStop}%
\bibitem [{\citenamefont {Grassano}\ \emph {et~al.}(2018)\citenamefont {Grassano}, \citenamefont {Pulci}, \citenamefont {Mosca~Conte},\ and\ \citenamefont {Bechstedt}}]{Grassano_018}%
  \BibitemOpen
  \bibfield  {author} {\bibinfo {author} {\bibfnamefont {Davide}\ \bibnamefont {Grassano}}, \bibinfo {author} {\bibfnamefont {Olivia}\ \bibnamefont {Pulci}}, \bibinfo {author} {\bibfnamefont {Adriano}\ \bibnamefont {Mosca~Conte}}, \ and\ \bibinfo {author} {\bibfnamefont {Friedhelm}\ \bibnamefont {Bechstedt}},\ }\bibfield  {title} {\enquote {\bibinfo {title} {Validity of weyl fermion picture for transition metals monopnictides taas, tap, nbas, and nbp from ab initio studies},}\ }\href {\doibase 10.1038/s41598-018-21465-z} {\bibfield  {journal} {\bibinfo  {journal} {Scientific Reports}\ }\textbf {\bibinfo {volume} {8}},\ \bibinfo {pages} {3534} (\bibinfo {year} {2018})}\BibitemShut {NoStop}%
\bibitem [{\citenamefont {Zhang}\ \emph {et~al.}(2016)\citenamefont {Zhang}, \citenamefont {Xu}, \citenamefont {Belopolski}, \citenamefont {Yuan}, \citenamefont {Lin}, \citenamefont {Tong}, \citenamefont {Bian}, \citenamefont {Alidoust}, \citenamefont {Lee}, \citenamefont {Huang}, \citenamefont {Chang}, \citenamefont {Chang}, \citenamefont {Hsu}, \citenamefont {Jeng}, \citenamefont {Neupane}, \citenamefont {Sanchez}, \citenamefont {Zheng}, \citenamefont {Wang}, \citenamefont {Lin}, \citenamefont {Zhang}, \citenamefont {Lu}, \citenamefont {Shen}, \citenamefont {Neupert}, \citenamefont {Zahid~Hasan},\ and\ \citenamefont {Jia}}]{Zhang_016}%
  \BibitemOpen
  \bibfield  {author} {\bibinfo {author} {\bibfnamefont {Cheng-Long}\ \bibnamefont {Zhang}}, \bibinfo {author} {\bibfnamefont {Su-Yang}\ \bibnamefont {Xu}}, \bibinfo {author} {\bibfnamefont {Ilya}\ \bibnamefont {Belopolski}}, \bibinfo {author} {\bibfnamefont {Zhujun}\ \bibnamefont {Yuan}}, \bibinfo {author} {\bibfnamefont {Ziquan}\ \bibnamefont {Lin}}, \bibinfo {author} {\bibfnamefont {Bingbing}\ \bibnamefont {Tong}}, \bibinfo {author} {\bibfnamefont {Guang}\ \bibnamefont {Bian}}, \bibinfo {author} {\bibfnamefont {Nasser}\ \bibnamefont {Alidoust}}, \bibinfo {author} {\bibfnamefont {Chi-Cheng}\ \bibnamefont {Lee}}, \bibinfo {author} {\bibfnamefont {Shin-Ming}\ \bibnamefont {Huang}}, \bibinfo {author} {\bibfnamefont {Tay-Rong}\ \bibnamefont {Chang}}, \bibinfo {author} {\bibfnamefont {Guoqing}\ \bibnamefont {Chang}}, \bibinfo {author} {\bibfnamefont {Chuang-Han}\ \bibnamefont {Hsu}}, \bibinfo {author} {\bibfnamefont {Horng-Tay}\ \bibnamefont {Jeng}}, \bibinfo {author} {\bibfnamefont {Madhab}\ \bibnamefont
  {Neupane}}, \bibinfo {author} {\bibfnamefont {Daniel~S.}\ \bibnamefont {Sanchez}}, \bibinfo {author} {\bibfnamefont {Hao}\ \bibnamefont {Zheng}}, \bibinfo {author} {\bibfnamefont {Junfeng}\ \bibnamefont {Wang}}, \bibinfo {author} {\bibfnamefont {Hsin}\ \bibnamefont {Lin}}, \bibinfo {author} {\bibfnamefont {Chi}\ \bibnamefont {Zhang}}, \bibinfo {author} {\bibfnamefont {Hai-Zhou}\ \bibnamefont {Lu}}, \bibinfo {author} {\bibfnamefont {Shun-Qing}\ \bibnamefont {Shen}}, \bibinfo {author} {\bibfnamefont {Titus}\ \bibnamefont {Neupert}}, \bibinfo {author} {\bibfnamefont {M.}~\bibnamefont {Zahid~Hasan}}, \ and\ \bibinfo {author} {\bibfnamefont {Shuang}\ \bibnamefont {Jia}},\ }\bibfield  {title} {\enquote {\bibinfo {title} {Signatures of the adler--bell--jackiw chiral anomaly in a weyl fermion semimetal},}\ }\href {\doibase 10.1038/ncomms10735} {\bibfield  {journal} {\bibinfo  {journal} {Nature Communications}\ }\textbf {\bibinfo {volume} {7}},\ \bibinfo {pages} {10735} (\bibinfo {year} {2016})}\BibitemShut
  {NoStop}%
\bibitem [{\citenamefont {Arnold}\ \emph {et~al.}(2016)\citenamefont {Arnold}, \citenamefont {Shekhar}, \citenamefont {Wu}, \citenamefont {Sun}, \citenamefont {dos Reis}, \citenamefont {Kumar}, \citenamefont {Naumann}, \citenamefont {Ajeesh}, \citenamefont {Schmidt}, \citenamefont {Grushin}, \citenamefont {Bardarson}, \citenamefont {Baenitz}, \citenamefont {Sokolov}, \citenamefont {Borrmann}, \citenamefont {Nicklas}, \citenamefont {Felser}, \citenamefont {Hassinger},\ and\ \citenamefont {Yan}}]{Arnold_016}%
  \BibitemOpen
  \bibfield  {author} {\bibinfo {author} {\bibfnamefont {Frank}\ \bibnamefont {Arnold}}, \bibinfo {author} {\bibfnamefont {Chandra}\ \bibnamefont {Shekhar}}, \bibinfo {author} {\bibfnamefont {Shu-Chun}\ \bibnamefont {Wu}}, \bibinfo {author} {\bibfnamefont {Yan}\ \bibnamefont {Sun}}, \bibinfo {author} {\bibfnamefont {Ricardo~Donizeth}\ \bibnamefont {dos Reis}}, \bibinfo {author} {\bibfnamefont {Nitesh}\ \bibnamefont {Kumar}}, \bibinfo {author} {\bibfnamefont {Marcel}\ \bibnamefont {Naumann}}, \bibinfo {author} {\bibfnamefont {Mukkattu~O.}\ \bibnamefont {Ajeesh}}, \bibinfo {author} {\bibfnamefont {Marcus}\ \bibnamefont {Schmidt}}, \bibinfo {author} {\bibfnamefont {Adolfo~G.}\ \bibnamefont {Grushin}}, \bibinfo {author} {\bibfnamefont {Jens~H.}\ \bibnamefont {Bardarson}}, \bibinfo {author} {\bibfnamefont {Michael}\ \bibnamefont {Baenitz}}, \bibinfo {author} {\bibfnamefont {Dmitry}\ \bibnamefont {Sokolov}}, \bibinfo {author} {\bibfnamefont {Horst}\ \bibnamefont {Borrmann}}, \bibinfo {author} {\bibfnamefont
  {Michael}\ \bibnamefont {Nicklas}}, \bibinfo {author} {\bibfnamefont {Claudia}\ \bibnamefont {Felser}}, \bibinfo {author} {\bibfnamefont {Elena}\ \bibnamefont {Hassinger}}, \ and\ \bibinfo {author} {\bibfnamefont {Binghai}\ \bibnamefont {Yan}},\ }\bibfield  {title} {\enquote {\bibinfo {title} {Negative magnetoresistance without well-defined chirality in the weyl semimetal tap},}\ }\href {\doibase 10.1038/ncomms11615} {\bibfield  {journal} {\bibinfo  {journal} {Nature Communications}\ }\textbf {\bibinfo {volume} {7}},\ \bibinfo {pages} {11615} (\bibinfo {year} {2016})}\BibitemShut {NoStop}%
\bibitem [{\citenamefont {Shekhar}\ \emph {et~al.}(2015)\citenamefont {Shekhar}, \citenamefont {Nayak}, \citenamefont {Sun}, \citenamefont {Schmidt}, \citenamefont {Nicklas}, \citenamefont {Leermakers}, \citenamefont {Zeitler}, \citenamefont {Skourski}, \citenamefont {Wosnitza}, \citenamefont {Liu}, \citenamefont {Chen}, \citenamefont {Schnelle}, \citenamefont {Borrmann}, \citenamefont {Grin}, \citenamefont {Felser},\ and\ \citenamefont {Yan}}]{Shekhar_015}%
  \BibitemOpen
  \bibfield  {author} {\bibinfo {author} {\bibfnamefont {Chandra}\ \bibnamefont {Shekhar}}, \bibinfo {author} {\bibfnamefont {Ajaya~K.}\ \bibnamefont {Nayak}}, \bibinfo {author} {\bibfnamefont {Yan}\ \bibnamefont {Sun}}, \bibinfo {author} {\bibfnamefont {Marcus}\ \bibnamefont {Schmidt}}, \bibinfo {author} {\bibfnamefont {Michael}\ \bibnamefont {Nicklas}}, \bibinfo {author} {\bibfnamefont {Inge}\ \bibnamefont {Leermakers}}, \bibinfo {author} {\bibfnamefont {Uli}\ \bibnamefont {Zeitler}}, \bibinfo {author} {\bibfnamefont {Yurii}\ \bibnamefont {Skourski}}, \bibinfo {author} {\bibfnamefont {Jochen}\ \bibnamefont {Wosnitza}}, \bibinfo {author} {\bibfnamefont {Zhongkai}\ \bibnamefont {Liu}}, \bibinfo {author} {\bibfnamefont {Yulin}\ \bibnamefont {Chen}}, \bibinfo {author} {\bibfnamefont {Walter}\ \bibnamefont {Schnelle}}, \bibinfo {author} {\bibfnamefont {Horst}\ \bibnamefont {Borrmann}}, \bibinfo {author} {\bibfnamefont {Yuri}\ \bibnamefont {Grin}}, \bibinfo {author} {\bibfnamefont {Claudia}\ \bibnamefont
  {Felser}}, \ and\ \bibinfo {author} {\bibfnamefont {Binghai}\ \bibnamefont {Yan}},\ }\bibfield  {title} {\enquote {\bibinfo {title} {Extremely large magnetoresistance and ultrahigh mobility in the topological weyl semimetal candidate nbp},}\ }\href {\doibase 10.1038/nphys3372} {\bibfield  {journal} {\bibinfo  {journal} {Nature Physics}\ }\textbf {\bibinfo {volume} {11}},\ \bibinfo {pages} {645--649} (\bibinfo {year} {2015})}\BibitemShut {NoStop}%
\bibitem [{\citenamefont {Lv}\ \emph {et~al.}(2015)\citenamefont {Lv}, \citenamefont {Weng}, \citenamefont {Fu}, \citenamefont {Wang}, \citenamefont {Miao}, \citenamefont {Ma}, \citenamefont {Richard}, \citenamefont {Huang}, \citenamefont {Zhao}, \citenamefont {Chen}, \citenamefont {Fang}, \citenamefont {Dai}, \citenamefont {Qian},\ and\ \citenamefont {Ding}}]{Lv_et_al_2015(TaAs)}%
  \BibitemOpen
  \bibfield  {author} {\bibinfo {author} {\bibfnamefont {B.~Q.}\ \bibnamefont {Lv}}, \bibinfo {author} {\bibfnamefont {H.~M.}\ \bibnamefont {Weng}}, \bibinfo {author} {\bibfnamefont {B.~B.}\ \bibnamefont {Fu}}, \bibinfo {author} {\bibfnamefont {X.~P.}\ \bibnamefont {Wang}}, \bibinfo {author} {\bibfnamefont {H.}~\bibnamefont {Miao}}, \bibinfo {author} {\bibfnamefont {J.}~\bibnamefont {Ma}}, \bibinfo {author} {\bibfnamefont {P.}~\bibnamefont {Richard}}, \bibinfo {author} {\bibfnamefont {X.~C.}\ \bibnamefont {Huang}}, \bibinfo {author} {\bibfnamefont {L.~X.}\ \bibnamefont {Zhao}}, \bibinfo {author} {\bibfnamefont {G.~F.}\ \bibnamefont {Chen}}, \bibinfo {author} {\bibfnamefont {Z.}~\bibnamefont {Fang}}, \bibinfo {author} {\bibfnamefont {X.}~\bibnamefont {Dai}}, \bibinfo {author} {\bibfnamefont {T.}~\bibnamefont {Qian}}, \ and\ \bibinfo {author} {\bibfnamefont {H.}~\bibnamefont {Ding}},\ }\bibfield  {title} {\enquote {\bibinfo {title} {Experimental discovery of weyl semimetal taas},}\ }\href {\doibase
  10.1103/PhysRevX.5.031013} {\bibfield  {journal} {\bibinfo  {journal} {Phys. Rev. X}\ }\textbf {\bibinfo {volume} {5}},\ \bibinfo {pages} {031013} (\bibinfo {year} {2015})}\BibitemShut {NoStop}%
\bibitem [{\citenamefont {Xu}\ \emph {et~al.}(2015{\natexlab{b}})\citenamefont {Xu}, \citenamefont {Belopolski}, \citenamefont {Alidoust}, \citenamefont {Neupane}, \citenamefont {Bian}, \citenamefont {Zhang}, \citenamefont {Sankar}, \citenamefont {Chang}, \citenamefont {Yuan}, \citenamefont {Lee}, \citenamefont {Huang}, \citenamefont {Zheng}, \citenamefont {Ma}, \citenamefont {Sanchez}, \citenamefont {Wang}, \citenamefont {Bansil}, \citenamefont {Chou}, \citenamefont {Shibayev}, \citenamefont {Lin}, \citenamefont {Jia},\ and\ \citenamefont {Hasan}}]{Xu_et_al_2015(TaAs)}%
  \BibitemOpen
  \bibfield  {author} {\bibinfo {author} {\bibfnamefont {Su-Yang}\ \bibnamefont {Xu}}, \bibinfo {author} {\bibfnamefont {Ilya}\ \bibnamefont {Belopolski}}, \bibinfo {author} {\bibfnamefont {Nasser}\ \bibnamefont {Alidoust}}, \bibinfo {author} {\bibfnamefont {Madhab}\ \bibnamefont {Neupane}}, \bibinfo {author} {\bibfnamefont {Guang}\ \bibnamefont {Bian}}, \bibinfo {author} {\bibfnamefont {Chenglong}\ \bibnamefont {Zhang}}, \bibinfo {author} {\bibfnamefont {Raman}\ \bibnamefont {Sankar}}, \bibinfo {author} {\bibfnamefont {Guoqing}\ \bibnamefont {Chang}}, \bibinfo {author} {\bibfnamefont {Zhujun}\ \bibnamefont {Yuan}}, \bibinfo {author} {\bibfnamefont {Chi-Cheng}\ \bibnamefont {Lee}}, \bibinfo {author} {\bibfnamefont {Shin-Ming}\ \bibnamefont {Huang}}, \bibinfo {author} {\bibfnamefont {Hao}\ \bibnamefont {Zheng}}, \bibinfo {author} {\bibfnamefont {Jie}\ \bibnamefont {Ma}}, \bibinfo {author} {\bibfnamefont {Daniel~S.}\ \bibnamefont {Sanchez}}, \bibinfo {author} {\bibfnamefont {BaoKai}\ \bibnamefont {Wang}},
  \bibinfo {author} {\bibfnamefont {Arun}\ \bibnamefont {Bansil}}, \bibinfo {author} {\bibfnamefont {Fangcheng}\ \bibnamefont {Chou}}, \bibinfo {author} {\bibfnamefont {Pavel~P.}\ \bibnamefont {Shibayev}}, \bibinfo {author} {\bibfnamefont {Hsin}\ \bibnamefont {Lin}}, \bibinfo {author} {\bibfnamefont {Shuang}\ \bibnamefont {Jia}}, \ and\ \bibinfo {author} {\bibfnamefont {M.~Zahid}\ \bibnamefont {Hasan}},\ }\bibfield  {title} {\enquote {\bibinfo {title} {Discovery of a weyl fermion semimetal and topological fermi arcs},}\ }\href {\doibase 10.1126/science.aaa9297} {\bibfield  {journal} {\bibinfo  {journal} {Science}\ }\textbf {\bibinfo {volume} {349}},\ \bibinfo {pages} {613–617} (\bibinfo {year} {2015}{\natexlab{b}})}\BibitemShut {NoStop}%
\bibitem [{\citenamefont {Zhang}\ \emph {et~al.}(2017)\citenamefont {Zhang}, \citenamefont {Yuan}, \citenamefont {Jiang}, \citenamefont {Tong}, \citenamefont {Zhang}, \citenamefont {Xie},\ and\ \citenamefont {Jia}}]{Zhang_et_al_2017(TaAs)}%
  \BibitemOpen
  \bibfield  {author} {\bibinfo {author} {\bibfnamefont {Cheng-Long}\ \bibnamefont {Zhang}}, \bibinfo {author} {\bibfnamefont {Zhujun}\ \bibnamefont {Yuan}}, \bibinfo {author} {\bibfnamefont {Qing-Dong}\ \bibnamefont {Jiang}}, \bibinfo {author} {\bibfnamefont {Bingbing}\ \bibnamefont {Tong}}, \bibinfo {author} {\bibfnamefont {Chi}\ \bibnamefont {Zhang}}, \bibinfo {author} {\bibfnamefont {X.~C.}\ \bibnamefont {Xie}}, \ and\ \bibinfo {author} {\bibfnamefont {Shuang}\ \bibnamefont {Jia}},\ }\bibfield  {title} {\enquote {\bibinfo {title} {Electron scattering in tantalum monoarsenide},}\ }\href {\doibase 10.1103/PhysRevB.95.085202} {\bibfield  {journal} {\bibinfo  {journal} {Phys. Rev. B}\ }\textbf {\bibinfo {volume} {95}},\ \bibinfo {pages} {085202} (\bibinfo {year} {2017})}\BibitemShut {NoStop}%
\bibitem [{\citenamefont {Xu}\ \emph {et~al.}(2015{\natexlab{c}})\citenamefont {Xu}, \citenamefont {Belopolski}, \citenamefont {Sanchez}, \citenamefont {Zhang}, \citenamefont {Chang}, \citenamefont {Guo}, \citenamefont {Bian}, \citenamefont {Yuan}, \citenamefont {Lu}, \citenamefont {Chang}, \citenamefont {Shibayev}, \citenamefont {Prokopovych}, \citenamefont {Alidoust}, \citenamefont {Zheng}, \citenamefont {Lee}, \citenamefont {Huang}, \citenamefont {Sankar}, \citenamefont {Chou}, \citenamefont {Hsu},\ and\ \citenamefont {Hasan}}]{Xu_et_al_2015(TaP)}%
  \BibitemOpen
  \bibfield  {author} {\bibinfo {author} {\bibfnamefont {Su-Yang}\ \bibnamefont {Xu}}, \bibinfo {author} {\bibfnamefont {Ilya}\ \bibnamefont {Belopolski}}, \bibinfo {author} {\bibfnamefont {Daniel}\ \bibnamefont {Sanchez}}, \bibinfo {author} {\bibfnamefont {Chenglong}\ \bibnamefont {Zhang}}, \bibinfo {author} {\bibfnamefont {Guoqing}\ \bibnamefont {Chang}}, \bibinfo {author} {\bibfnamefont {Cheng}\ \bibnamefont {Guo}}, \bibinfo {author} {\bibfnamefont {Guang}\ \bibnamefont {Bian}}, \bibinfo {author} {\bibfnamefont {Zhujun}\ \bibnamefont {Yuan}}, \bibinfo {author} {\bibfnamefont {Hong}\ \bibnamefont {Lu}}, \bibinfo {author} {\bibfnamefont {Tay-Rong}\ \bibnamefont {Chang}}, \bibinfo {author} {\bibfnamefont {Pavel}\ \bibnamefont {Shibayev}}, \bibinfo {author} {\bibfnamefont {Mykhailo}\ \bibnamefont {Prokopovych}}, \bibinfo {author} {\bibfnamefont {Nasser}\ \bibnamefont {Alidoust}}, \bibinfo {author} {\bibfnamefont {Hao}\ \bibnamefont {Zheng}}, \bibinfo {author} {\bibfnamefont {Chi-Cheng}\ \bibnamefont {Lee}},
  \bibinfo {author} {\bibfnamefont {Shin-Ming}\ \bibnamefont {Huang}}, \bibinfo {author} {\bibfnamefont {Raman}\ \bibnamefont {Sankar}}, \bibinfo {author} {\bibfnamefont {F.}~\bibnamefont {Chou}}, \bibinfo {author} {\bibfnamefont {Chuang-Han}\ \bibnamefont {Hsu}}, \ and\ \bibinfo {author} {\bibfnamefont {M.~Zahid}\ \bibnamefont {Hasan}},\ }\bibfield  {title} {\enquote {\bibinfo {title} {Experimental discovery of a topological weyl semimetal state in tap},}\ }\href {\doibase 10.1126/sciadv.1501092} {\bibfield  {journal} {\bibinfo  {journal} {Science Advances}\ }\textbf {\bibinfo {volume} {1}},\ \bibinfo {pages} {e1501092--e1501092} (\bibinfo {year} {2015}{\natexlab{c}})}\BibitemShut {NoStop}%
\bibitem [{\citenamefont {Xu}\ \emph {et~al.}(2016)\citenamefont {Xu}, \citenamefont {Weng}, \citenamefont {Lv}, \citenamefont {Matt}, \citenamefont {Park}, \citenamefont {Bisti}, \citenamefont {Strocov}, \citenamefont {Gawryluk}, \citenamefont {Pomjakushina}, \citenamefont {Conder}, \citenamefont {Plumb}, \citenamefont {Radovic}, \citenamefont {Autès}, \citenamefont {Yazyev}, \citenamefont {Fang}, \citenamefont {Dai}, \citenamefont {Qian}, \citenamefont {Mesot}, \citenamefont {Ding},\ and\ \citenamefont {Shi}}]{Xu_et_al_2016(TaP)}%
  \BibitemOpen
  \bibfield  {author} {\bibinfo {author} {\bibfnamefont {N.}~\bibnamefont {Xu}}, \bibinfo {author} {\bibfnamefont {Hongming}\ \bibnamefont {Weng}}, \bibinfo {author} {\bibfnamefont {B.}~\bibnamefont {Lv}}, \bibinfo {author} {\bibfnamefont {Christian}\ \bibnamefont {Matt}}, \bibinfo {author} {\bibfnamefont {J.}~\bibnamefont {Park}}, \bibinfo {author} {\bibfnamefont {Federico}\ \bibnamefont {Bisti}}, \bibinfo {author} {\bibfnamefont {V.}~\bibnamefont {Strocov}}, \bibinfo {author} {\bibfnamefont {Dariusz}\ \bibnamefont {Gawryluk}}, \bibinfo {author} {\bibfnamefont {Ekaterina}\ \bibnamefont {Pomjakushina}}, \bibinfo {author} {\bibfnamefont {K.}~\bibnamefont {Conder}}, \bibinfo {author} {\bibfnamefont {Nicholas}\ \bibnamefont {Plumb}}, \bibinfo {author} {\bibfnamefont {Milan}\ \bibnamefont {Radovic}}, \bibinfo {author} {\bibfnamefont {Gabriel}\ \bibnamefont {Autès}}, \bibinfo {author} {\bibfnamefont {Oleg}\ \bibnamefont {Yazyev}}, \bibinfo {author} {\bibfnamefont {Zanxi}\ \bibnamefont {Fang}}, \bibinfo {author}
  {\bibfnamefont {Xianzhu}\ \bibnamefont {Dai}}, \bibinfo {author} {\bibfnamefont {Tan}\ \bibnamefont {Qian}}, \bibinfo {author} {\bibfnamefont {J.}~\bibnamefont {Mesot}}, \bibinfo {author} {\bibfnamefont {Hanjie}\ \bibnamefont {Ding}}, \ and\ \bibinfo {author} {\bibfnamefont {Ming}\ \bibnamefont {Shi}},\ }\bibfield  {title} {\enquote {\bibinfo {title} {Observation of weyl nodes and fermi arcs in tantalum phosphide},}\ }\href {\doibase 10.1038/ncomms11006} {\bibfield  {journal} {\bibinfo  {journal} {Nature Communications}\ }\textbf {\bibinfo {volume} {7}},\ \bibinfo {pages} {11006} (\bibinfo {year} {2016})}\BibitemShut {NoStop}%
\bibitem [{\citenamefont {Xu}\ \emph {et~al.}(2015{\natexlab{d}})\citenamefont {Xu}, \citenamefont {Alidoust}, \citenamefont {Belopolski}, \citenamefont {Yuan}, \citenamefont {Bian}, \citenamefont {Chang}, \citenamefont {Zheng}, \citenamefont {Strocov}, \citenamefont {Sanchez}, \citenamefont {Chang}, \citenamefont {Zhang}, \citenamefont {Mou}, \citenamefont {Wu}, \citenamefont {Huang}, \citenamefont {Lee}, \citenamefont {Huang}, \citenamefont {Wang}, \citenamefont {Bansil}, \citenamefont {Jeng},\ and\ \citenamefont {Hasan}}]{Xu_et_al_2015(NbAs)}%
  \BibitemOpen
  \bibfield  {author} {\bibinfo {author} {\bibfnamefont {Su-Yang}\ \bibnamefont {Xu}}, \bibinfo {author} {\bibfnamefont {Nasser}\ \bibnamefont {Alidoust}}, \bibinfo {author} {\bibfnamefont {Ilya}\ \bibnamefont {Belopolski}}, \bibinfo {author} {\bibfnamefont {Zhujun}\ \bibnamefont {Yuan}}, \bibinfo {author} {\bibfnamefont {Guang}\ \bibnamefont {Bian}}, \bibinfo {author} {\bibfnamefont {Tay-Rong}\ \bibnamefont {Chang}}, \bibinfo {author} {\bibfnamefont {Hao}\ \bibnamefont {Zheng}}, \bibinfo {author} {\bibfnamefont {Vladimir}\ \bibnamefont {Strocov}}, \bibinfo {author} {\bibfnamefont {Daniel}\ \bibnamefont {Sanchez}}, \bibinfo {author} {\bibfnamefont {Guoqing}\ \bibnamefont {Chang}}, \bibinfo {author} {\bibfnamefont {Chenglong}\ \bibnamefont {Zhang}}, \bibinfo {author} {\bibfnamefont {Daixiang}\ \bibnamefont {Mou}}, \bibinfo {author} {\bibfnamefont {Yun}\ \bibnamefont {Wu}}, \bibinfo {author} {\bibfnamefont {Lunan}\ \bibnamefont {Huang}}, \bibinfo {author} {\bibfnamefont {Chi-Cheng}\ \bibnamefont {Lee}},
  \bibinfo {author} {\bibfnamefont {Shin-Ming}\ \bibnamefont {Huang}}, \bibinfo {author} {\bibfnamefont {Baokai}\ \bibnamefont {Wang}}, \bibinfo {author} {\bibfnamefont {Arun}\ \bibnamefont {Bansil}}, \bibinfo {author} {\bibfnamefont {Horng-Tay}\ \bibnamefont {Jeng}}, \ and\ \bibinfo {author} {\bibfnamefont {M.~Zahid}\ \bibnamefont {Hasan}},\ }\bibfield  {title} {\enquote {\bibinfo {title} {Discovery of a weyl fermion state with fermi arcs in niobium arsenide},}\ }\href {\doibase 10.1038/nphys3437} {\bibfield  {journal} {\bibinfo  {journal} {Nature Physics}\ }\textbf {\bibinfo {volume} {11}},\ \bibinfo {pages} {748--754} (\bibinfo {year} {2015}{\natexlab{d}})}\BibitemShut {NoStop}%
\bibitem [{\citenamefont {Li}\ \emph {et~al.}(2016)\citenamefont {Li}, \citenamefont {Kharzeev}, \citenamefont {Zhang}, \citenamefont {Huang}, \citenamefont {Pletikosić}, \citenamefont {Fedorov}, \citenamefont {Zhong}, \citenamefont {Schneeloch}, \citenamefont {Gu},\ and\ \citenamefont {Valla}}]{Li_et_al_2016}%
  \BibitemOpen
  \bibfield  {author} {\bibinfo {author} {\bibfnamefont {Qiang}\ \bibnamefont {Li}}, \bibinfo {author} {\bibfnamefont {Dmitri}\ \bibnamefont {Kharzeev}}, \bibinfo {author} {\bibfnamefont {Chuandi}\ \bibnamefont {Zhang}}, \bibinfo {author} {\bibfnamefont {Yuan}\ \bibnamefont {Huang}}, \bibinfo {author} {\bibfnamefont {Ivo}\ \bibnamefont {Pletikosić}}, \bibinfo {author} {\bibfnamefont {Alexei}\ \bibnamefont {Fedorov}}, \bibinfo {author} {\bibfnamefont {Ruidan}\ \bibnamefont {Zhong}}, \bibinfo {author} {\bibfnamefont {J.A.}\ \bibnamefont {Schneeloch}}, \bibinfo {author} {\bibfnamefont {G.D.}\ \bibnamefont {Gu}}, \ and\ \bibinfo {author} {\bibfnamefont {Tonica}\ \bibnamefont {Valla}},\ }\bibfield  {title} {\enquote {\bibinfo {title} {Chiral magnetic effect in zrte5},}\ }\href {\doibase 10.1038/nphys3648} {\bibfield  {journal} {\bibinfo  {journal} {Nature Physics}\ }\textbf {\bibinfo {volume} {12}} (\bibinfo {year} {2016}),\ 10.1038/nphys3648}\BibitemShut {NoStop}%
\bibitem [{\citenamefont {Moll}\ \emph {et~al.}(2016)\citenamefont {Moll}, \citenamefont {Nair}, \citenamefont {Helm}, \citenamefont {Potter}, \citenamefont {Kimchi}, \citenamefont {Vishwanath},\ and\ \citenamefont {Analytis}}]{Moll_et_al_2016}%
  \BibitemOpen
  \bibfield  {author} {\bibinfo {author} {\bibfnamefont {P.}~\bibnamefont {Moll}}, \bibinfo {author} {\bibfnamefont {Nityan}\ \bibnamefont {Nair}}, \bibinfo {author} {\bibfnamefont {Toni}\ \bibnamefont {Helm}}, \bibinfo {author} {\bibfnamefont {Andrew}\ \bibnamefont {Potter}}, \bibinfo {author} {\bibfnamefont {Itamar}\ \bibnamefont {Kimchi}}, \bibinfo {author} {\bibfnamefont {Ashvin}\ \bibnamefont {Vishwanath}}, \ and\ \bibinfo {author} {\bibfnamefont {James}\ \bibnamefont {Analytis}},\ }\bibfield  {title} {\enquote {\bibinfo {title} {Transport evidence for fermi-arc-mediated chirality transfer in the dirac semimetal cd3as2},}\ }\href {\doibase 10.1038/nature18276} {\bibfield  {journal} {\bibinfo  {journal} {Nature}\ }\textbf {\bibinfo {volume} {535}} (\bibinfo {year} {2016}),\ 10.1038/nature18276}\BibitemShut {NoStop}%
\bibitem [{\citenamefont {Turner}\ and\ \citenamefont {Vishwanath}(2013)}]{Turner_et_al_2013}%
  \BibitemOpen
  \bibfield  {author} {\bibinfo {author} {\bibfnamefont {Ari}\ \bibnamefont {Turner}}\ and\ \bibinfo {author} {\bibfnamefont {Ashvin}\ \bibnamefont {Vishwanath}},\ }\bibfield  {title} {\enquote {\bibinfo {title} {Beyond band insulators: Topology of semimetals and interacting phases},}\ }\href {\doibase 10.1016/B978-0-444-63314-9.00011-1} {\bibfield  {journal} {\bibinfo  {journal} {Contemporary Concepts of Condensed Matter Science}\ }\textbf {\bibinfo {volume} {6}},\ \bibinfo {pages} {293--324} (\bibinfo {year} {2013})}\BibitemShut {NoStop}%
\bibitem [{\citenamefont {Fang}\ \emph {et~al.}(2012)\citenamefont {Fang}, \citenamefont {Gilbert}, \citenamefont {Dai},\ and\ \citenamefont {Bernevig}}]{Fang_et_al_2012}%
  \BibitemOpen
  \bibfield  {author} {\bibinfo {author} {\bibfnamefont {Chen}\ \bibnamefont {Fang}}, \bibinfo {author} {\bibfnamefont {Matthew~J.}\ \bibnamefont {Gilbert}}, \bibinfo {author} {\bibfnamefont {Xi}~\bibnamefont {Dai}}, \ and\ \bibinfo {author} {\bibfnamefont {B.~Andrei}\ \bibnamefont {Bernevig}},\ }\bibfield  {title} {\enquote {\bibinfo {title} {Multi-weyl topological semimetals stabilized by point group symmetry},}\ }\href {\doibase 10.1103/PhysRevLett.108.266802} {\bibfield  {journal} {\bibinfo  {journal} {Phys. Rev. Lett.}\ }\textbf {\bibinfo {volume} {108}},\ \bibinfo {pages} {266802} (\bibinfo {year} {2012})}\BibitemShut {NoStop}%
\bibitem [{\citenamefont {Huang}\ \emph {et~al.}(2016)\citenamefont {Huang}, \citenamefont {Xu}, \citenamefont {Belopolski}, \citenamefont {Lee}, \citenamefont {Chang}, \citenamefont {Chang}, \citenamefont {Wang}, \citenamefont {Alidoust}, \citenamefont {Bian}, \citenamefont {Neupane}, \citenamefont {Sanchez}, \citenamefont {Zheng}, \citenamefont {Jeng}, \citenamefont {Bansil}, \citenamefont {Neupert}, \citenamefont {Lin},\ and\ \citenamefont {Hasan}}]{Huang_et_al_2016}%
  \BibitemOpen
  \bibfield  {author} {\bibinfo {author} {\bibfnamefont {Shin-Ming}\ \bibnamefont {Huang}}, \bibinfo {author} {\bibfnamefont {Su-Yang}\ \bibnamefont {Xu}}, \bibinfo {author} {\bibfnamefont {Ilya}\ \bibnamefont {Belopolski}}, \bibinfo {author} {\bibfnamefont {Chi-Cheng}\ \bibnamefont {Lee}}, \bibinfo {author} {\bibfnamefont {Guoqing}\ \bibnamefont {Chang}}, \bibinfo {author} {\bibfnamefont {Tay-Rong}\ \bibnamefont {Chang}}, \bibinfo {author} {\bibfnamefont {BaoKai}\ \bibnamefont {Wang}}, \bibinfo {author} {\bibfnamefont {Nasser}\ \bibnamefont {Alidoust}}, \bibinfo {author} {\bibfnamefont {Guang}\ \bibnamefont {Bian}}, \bibinfo {author} {\bibfnamefont {Madhab}\ \bibnamefont {Neupane}}, \bibinfo {author} {\bibfnamefont {Daniel}\ \bibnamefont {Sanchez}}, \bibinfo {author} {\bibfnamefont {Hao}\ \bibnamefont {Zheng}}, \bibinfo {author} {\bibfnamefont {Horng-Tay}\ \bibnamefont {Jeng}}, \bibinfo {author} {\bibfnamefont {Arun}\ \bibnamefont {Bansil}}, \bibinfo {author} {\bibfnamefont {Titus}\ \bibnamefont {Neupert}},
  \bibinfo {author} {\bibfnamefont {Hsin}\ \bibnamefont {Lin}}, \ and\ \bibinfo {author} {\bibfnamefont {M.~Zahid}\ \bibnamefont {Hasan}},\ }\bibfield  {title} {\enquote {\bibinfo {title} {New type of weyl semimetal with quadratic double weyl fermions},}\ }\href {\doibase 10.1073/pnas.1514581113} {\bibfield  {journal} {\bibinfo  {journal} {Proceedings of the National Academy of Sciences}\ }\textbf {\bibinfo {volume} {113}},\ \bibinfo {pages} {1180–1185} (\bibinfo {year} {2016})}\BibitemShut {NoStop}%
\bibitem [{\citenamefont {Ahn}\ \emph {et~al.}(2016)\citenamefont {Ahn}, \citenamefont {Hwang},\ and\ \citenamefont {Min}}]{Ahn_et_al_2016}%
  \BibitemOpen
  \bibfield  {author} {\bibinfo {author} {\bibfnamefont {Seongjin}\ \bibnamefont {Ahn}}, \bibinfo {author} {\bibfnamefont {E.~H.}\ \bibnamefont {Hwang}}, \ and\ \bibinfo {author} {\bibfnamefont {Hongki}\ \bibnamefont {Min}},\ }\bibfield  {title} {\enquote {\bibinfo {title} {Collective modes in multi-weyl semimetals},}\ }\href {\doibase 10.1038/srep34023} {\bibfield  {journal} {\bibinfo  {journal} {Scientific Reports}\ }\textbf {\bibinfo {volume} {6}} (\bibinfo {year} {2016}),\ 10.1038/srep34023}\BibitemShut {NoStop}%
\bibitem [{\citenamefont {Lai}(2015)}]{Lai_PRB_2015}%
  \BibitemOpen
  \bibfield  {author} {\bibinfo {author} {\bibfnamefont {Hsin-Hua}\ \bibnamefont {Lai}},\ }\bibfield  {title} {\enquote {\bibinfo {title} {Correlation effects in double-weyl semimetals},}\ }\href {\doibase 10.1103/PhysRevB.91.235131} {\bibfield  {journal} {\bibinfo  {journal} {Phys. Rev. B}\ }\textbf {\bibinfo {volume} {91}},\ \bibinfo {pages} {235131} (\bibinfo {year} {2015})}\BibitemShut {NoStop}%
\bibitem [{\citenamefont {Jian}\ and\ \citenamefont {Yao}(2015)}]{Jian_PRB_2015}%
  \BibitemOpen
  \bibfield  {author} {\bibinfo {author} {\bibfnamefont {Shao-Kai}\ \bibnamefont {Jian}}\ and\ \bibinfo {author} {\bibfnamefont {Hong}\ \bibnamefont {Yao}},\ }\bibfield  {title} {\enquote {\bibinfo {title} {Correlated double-weyl semimetals with coulomb interactions: Possible applications to ${\mathrm{hgcr}}_{2}{\mathrm{se}}_{4}$ and ${\mathrm{srsi}}_{2}$},}\ }\href {\doibase 10.1103/PhysRevB.92.045121} {\bibfield  {journal} {\bibinfo  {journal} {Phys. Rev. B}\ }\textbf {\bibinfo {volume} {92}},\ \bibinfo {pages} {045121} (\bibinfo {year} {2015})}\BibitemShut {NoStop}%
\bibitem [{\citenamefont {Huang}\ \emph {et~al.}(2015)\citenamefont {Huang}, \citenamefont {Liu}, \citenamefont {Zhang}, \citenamefont {Duan},\ and\ \citenamefont {Vanderbilt}}]{Huang_PRB_2015}%
  \BibitemOpen
  \bibfield  {author} {\bibinfo {author} {\bibfnamefont {Huaqing}\ \bibnamefont {Huang}}, \bibinfo {author} {\bibfnamefont {Zhirong}\ \bibnamefont {Liu}}, \bibinfo {author} {\bibfnamefont {Hongbin}\ \bibnamefont {Zhang}}, \bibinfo {author} {\bibfnamefont {Wenhui}\ \bibnamefont {Duan}}, \ and\ \bibinfo {author} {\bibfnamefont {David}\ \bibnamefont {Vanderbilt}},\ }\bibfield  {title} {\enquote {\bibinfo {title} {Emergence of a chern-insulating state from a semi-dirac dispersion},}\ }\href {\doibase 10.1103/PhysRevB.92.161115} {\bibfield  {journal} {\bibinfo  {journal} {Phys. Rev. B}\ }\textbf {\bibinfo {volume} {92}},\ \bibinfo {pages} {161115} (\bibinfo {year} {2015})}\BibitemShut {NoStop}%
\bibitem [{\citenamefont {Pyatkovskiy}\ and\ \citenamefont {Chakraborty}(2016)}]{Pyatkovskiy_PRB_2016}%
  \BibitemOpen
  \bibfield  {author} {\bibinfo {author} {\bibfnamefont {P.~K.}\ \bibnamefont {Pyatkovskiy}}\ and\ \bibinfo {author} {\bibfnamefont {Tapash}\ \bibnamefont {Chakraborty}},\ }\bibfield  {title} {\enquote {\bibinfo {title} {Dynamical polarization and plasmons in a two-dimensional system with merging dirac points},}\ }\href {\doibase 10.1103/PhysRevB.93.085145} {\bibfield  {journal} {\bibinfo  {journal} {Phys. Rev. B}\ }\textbf {\bibinfo {volume} {93}},\ \bibinfo {pages} {085145} (\bibinfo {year} {2016})}\BibitemShut {NoStop}%
\bibitem [{\citenamefont {Singh}\ \emph {et~al.}(2018)\citenamefont {Singh}, \citenamefont {Chang}, \citenamefont {Chang}, \citenamefont {Huang}, \citenamefont {Su}, \citenamefont {Lin}, \citenamefont {Lin},\ and\ \citenamefont {Bansil}}]{Singh_SCIREP_2018}%
  \BibitemOpen
  \bibfield  {author} {\bibinfo {author} {\bibfnamefont {Bahadur}\ \bibnamefont {Singh}}, \bibinfo {author} {\bibfnamefont {Guoqing}\ \bibnamefont {Chang}}, \bibinfo {author} {\bibfnamefont {Tay-Rong}\ \bibnamefont {Chang}}, \bibinfo {author} {\bibfnamefont {Shin-Ming}\ \bibnamefont {Huang}}, \bibinfo {author} {\bibfnamefont {Chenliang}\ \bibnamefont {Su}}, \bibinfo {author} {\bibfnamefont {Ming-Chieh}\ \bibnamefont {Lin}}, \bibinfo {author} {\bibfnamefont {Hsin}\ \bibnamefont {Lin}}, \ and\ \bibinfo {author} {\bibfnamefont {Arun}\ \bibnamefont {Bansil}},\ }\bibfield  {title} {\enquote {\bibinfo {title} {Tunable double-weyl fermion semimetal state in the srsi2 materials class},}\ }\href {\doibase 10.1038/s41598-018-28644-y} {\bibfield  {journal} {\bibinfo  {journal} {Scientific Reports}\ }\textbf {\bibinfo {volume} {8}},\ \bibinfo {pages} {10540} (\bibinfo {year} {2018})}\BibitemShut {NoStop}%
\bibitem [{\citenamefont {Guan}\ \emph {et~al.}(2015)\citenamefont {Guan}, \citenamefont {Lin}, \citenamefont {Yang}, \citenamefont {Shi}, \citenamefont {Ren}, \citenamefont {Li}, \citenamefont {Weng}, \citenamefont {Dai}, \citenamefont {Fang}, \citenamefont {Yan},\ and\ \citenamefont {Xiong}}]{Guan_PRL_2015}%
  \BibitemOpen
  \bibfield  {author} {\bibinfo {author} {\bibfnamefont {Tong}\ \bibnamefont {Guan}}, \bibinfo {author} {\bibfnamefont {Chaojing}\ \bibnamefont {Lin}}, \bibinfo {author} {\bibfnamefont {Chongli}\ \bibnamefont {Yang}}, \bibinfo {author} {\bibfnamefont {Youguo}\ \bibnamefont {Shi}}, \bibinfo {author} {\bibfnamefont {Cong}\ \bibnamefont {Ren}}, \bibinfo {author} {\bibfnamefont {Yongqing}\ \bibnamefont {Li}}, \bibinfo {author} {\bibfnamefont {Hongming}\ \bibnamefont {Weng}}, \bibinfo {author} {\bibfnamefont {Xi}~\bibnamefont {Dai}}, \bibinfo {author} {\bibfnamefont {Zhong}\ \bibnamefont {Fang}}, \bibinfo {author} {\bibfnamefont {Shishen}\ \bibnamefont {Yan}}, \ and\ \bibinfo {author} {\bibfnamefont {Peng}\ \bibnamefont {Xiong}},\ }\bibfield  {title} {\enquote {\bibinfo {title} {Evidence for half-metallicity in $n$-type ${\mathrm{hgcr}}_{2}{\mathrm{se}}_{4}$},}\ }\href {\doibase 10.1103/PhysRevLett.115.087002} {\bibfield  {journal} {\bibinfo  {journal} {Phys. Rev. Lett.}\ }\textbf {\bibinfo {volume} {115}},\
  \bibinfo {pages} {087002} (\bibinfo {year} {2015})}\BibitemShut {NoStop}%
\bibitem [{\citenamefont {Zhu}\ \emph {et~al.}(2018)\citenamefont {Zhu}, \citenamefont {Liu}, \citenamefont {Yu}, \citenamefont {Wang}, \citenamefont {Zhao}, \citenamefont {Feng}, \citenamefont {Sheng},\ and\ \citenamefont {Yang}}]{Zhu_PRB_2018}%
  \BibitemOpen
  \bibfield  {author} {\bibinfo {author} {\bibfnamefont {Ziming}\ \bibnamefont {Zhu}}, \bibinfo {author} {\bibfnamefont {Ying}\ \bibnamefont {Liu}}, \bibinfo {author} {\bibfnamefont {Zhi-Ming}\ \bibnamefont {Yu}}, \bibinfo {author} {\bibfnamefont {Shan-Shan}\ \bibnamefont {Wang}}, \bibinfo {author} {\bibfnamefont {Y.~X.}\ \bibnamefont {Zhao}}, \bibinfo {author} {\bibfnamefont {Yuanping}\ \bibnamefont {Feng}}, \bibinfo {author} {\bibfnamefont {Xian-Lei}\ \bibnamefont {Sheng}}, \ and\ \bibinfo {author} {\bibfnamefont {Shengyuan~A.}\ \bibnamefont {Yang}},\ }\bibfield  {title} {\enquote {\bibinfo {title} {Quadratic contact point semimetal: Theory and material realization},}\ }\href {\doibase 10.1103/PhysRevB.98.125104} {\bibfield  {journal} {\bibinfo  {journal} {Phys. Rev. B}\ }\textbf {\bibinfo {volume} {98}},\ \bibinfo {pages} {125104} (\bibinfo {year} {2018})}\BibitemShut {NoStop}%
\bibitem [{\citenamefont {Liu}\ and\ \citenamefont {Zunger}(2017)}]{Liu_PhysRevX.7.021019}%
  \BibitemOpen
  \bibfield  {author} {\bibinfo {author} {\bibfnamefont {Qihang}\ \bibnamefont {Liu}}\ and\ \bibinfo {author} {\bibfnamefont {Alex}\ \bibnamefont {Zunger}},\ }\bibfield  {title} {\enquote {\bibinfo {title} {Predicted realization of cubic dirac fermion in quasi-one-dimensional transition-metal monochalcogenides},}\ }\href {\doibase 10.1103/PhysRevX.7.021019} {\bibfield  {journal} {\bibinfo  {journal} {Phys. Rev. X}\ }\textbf {\bibinfo {volume} {7}},\ \bibinfo {pages} {021019} (\bibinfo {year} {2017})}\BibitemShut {NoStop}%
\bibitem [{\citenamefont {Murakami}\ and\ \citenamefont {Nagaosa}(2003)}]{Murakami_03}%
  \BibitemOpen
  \bibfield  {author} {\bibinfo {author} {\bibfnamefont {Shuichi}\ \bibnamefont {Murakami}}\ and\ \bibinfo {author} {\bibfnamefont {Naoto}\ \bibnamefont {Nagaosa}},\ }\bibfield  {title} {\enquote {\bibinfo {title} {Berry phase in magnetic superconductors},}\ }\href {\doibase 10.1103/PhysRevLett.90.057002} {\bibfield  {journal} {\bibinfo  {journal} {Phys. Rev. Lett.}\ }\textbf {\bibinfo {volume} {90}},\ \bibinfo {pages} {057002} (\bibinfo {year} {2003})}\BibitemShut {NoStop}%
\bibitem [{\citenamefont {Meng}\ and\ \citenamefont {Balents}(2012)}]{Meng_012}%
  \BibitemOpen
  \bibfield  {author} {\bibinfo {author} {\bibfnamefont {Tobias}\ \bibnamefont {Meng}}\ and\ \bibinfo {author} {\bibfnamefont {Leon}\ \bibnamefont {Balents}},\ }\bibfield  {title} {\enquote {\bibinfo {title} {Weyl superconductors},}\ }\href {\doibase 10.1103/PhysRevB.86.054504} {\bibfield  {journal} {\bibinfo  {journal} {Phys. Rev. B}\ }\textbf {\bibinfo {volume} {86}},\ \bibinfo {pages} {054504} (\bibinfo {year} {2012})}\BibitemShut {NoStop}%
\bibitem [{\citenamefont {Cho}\ \emph {et~al.}(2012)\citenamefont {Cho}, \citenamefont {Bardarson}, \citenamefont {Lu},\ and\ \citenamefont {Moore}}]{Cho-2012}%
  \BibitemOpen
  \bibfield  {author} {\bibinfo {author} {\bibfnamefont {Gil~Young}\ \bibnamefont {Cho}}, \bibinfo {author} {\bibfnamefont {Jens~H.}\ \bibnamefont {Bardarson}}, \bibinfo {author} {\bibfnamefont {Yuan-Ming}\ \bibnamefont {Lu}}, \ and\ \bibinfo {author} {\bibfnamefont {Joel~E.}\ \bibnamefont {Moore}},\ }\bibfield  {title} {\enquote {\bibinfo {title} {Superconductivity of doped weyl semimetals: Finite-momentum pairing and electronic analog of the ${}^{3}$he-$a$ phase},}\ }\href {\doibase 10.1103/PhysRevB.86.214514} {\bibfield  {journal} {\bibinfo  {journal} {Phys. Rev. B}\ }\textbf {\bibinfo {volume} {86}},\ \bibinfo {pages} {214514} (\bibinfo {year} {2012})}\BibitemShut {NoStop}%
\bibitem [{\citenamefont {Li}\ and\ \citenamefont {Haldane}(2018)}]{Li_018}%
  \BibitemOpen
  \bibfield  {author} {\bibinfo {author} {\bibfnamefont {Yi}~\bibnamefont {Li}}\ and\ \bibinfo {author} {\bibfnamefont {F.~D.~M.}\ \bibnamefont {Haldane}},\ }\bibfield  {title} {\enquote {\bibinfo {title} {Topological nodal cooper pairing in doped weyl metals},}\ }\href {\doibase 10.1103/PhysRevLett.120.067003} {\bibfield  {journal} {\bibinfo  {journal} {Phys. Rev. Lett.}\ }\textbf {\bibinfo {volume} {120}},\ \bibinfo {pages} {067003} (\bibinfo {year} {2018})}\BibitemShut {NoStop}%
\bibitem [{\citenamefont {Sun}\ \emph {et~al.}(2020)\citenamefont {Sun}, \citenamefont {Lee},\ and\ \citenamefont {Li}}]{Sun_020}%
  \BibitemOpen
  \bibfield  {author} {\bibinfo {author} {\bibfnamefont {C.}~\bibnamefont {Sun}}, \bibinfo {author} {\bibfnamefont {S.-P.}\ \bibnamefont {Lee}}, \ and\ \bibinfo {author} {\bibfnamefont {Y.}~\bibnamefont {Li}},\ }\bibfield  {title} {\enquote {\bibinfo {title} {Vortices in a monopole superconducting weyl semi-metal},}\ }\href {\doibase 10.48550/arXiv.1909.04179} {\bibfield  {journal} {\bibinfo  {journal} {arXiv.1909.04179v2}\ } (\bibinfo {year} {2020}),\ 10.48550/arXiv.1909.04179}\BibitemShut {NoStop}%
\bibitem [{\citenamefont {Mu\~noz}\ \emph {et~al.}(2020)\citenamefont {Mu\~noz}, \citenamefont {Soto-Garrido},\ and\ \citenamefont {Juri\ifmmode \check{c}\else \v{c}\fi{}i\ifmmode~\acute{c}\else \'{c}\fi{}}}]{Munoz-2020}%
  \BibitemOpen
  \bibfield  {author} {\bibinfo {author} {\bibfnamefont {Enrique}\ \bibnamefont {Mu\~noz}}, \bibinfo {author} {\bibfnamefont {Rodrigo}\ \bibnamefont {Soto-Garrido}}, \ and\ \bibinfo {author} {\bibfnamefont {Vladimir}\ \bibnamefont {Juri\ifmmode \check{c}\else \v{c}\fi{}i\ifmmode~\acute{c}\else \'{c}\fi{}}},\ }\bibfield  {title} {\enquote {\bibinfo {title} {Monopole versus spherical harmonic superconductors: Topological repulsion, coexistence, and stability},}\ }\href {\doibase 10.1103/PhysRevB.102.195121} {\bibfield  {journal} {\bibinfo  {journal} {Phys. Rev. B}\ }\textbf {\bibinfo {volume} {102}},\ \bibinfo {pages} {195121} (\bibinfo {year} {2020})}\BibitemShut {NoStop}%
\bibitem [{\citenamefont {Mu\~noz}\ \emph {et~al.}(2024)\citenamefont {Mu\~noz}, \citenamefont {Esparza}, \citenamefont {Braun},\ and\ \citenamefont {Soto-Garrido}}]{Muñoz_et_al_2024}%
  \BibitemOpen
  \bibfield  {author} {\bibinfo {author} {\bibfnamefont {Enrique}\ \bibnamefont {Mu\~noz}}, \bibinfo {author} {\bibfnamefont {Juan~Pablo}\ \bibnamefont {Esparza}}, \bibinfo {author} {\bibfnamefont {Jos\'e}\ \bibnamefont {Braun}}, \ and\ \bibinfo {author} {\bibfnamefont {Rodrigo}\ \bibnamefont {Soto-Garrido}},\ }\bibfield  {title} {\enquote {\bibinfo {title} {{Topological versus conventional superconductivity in a Weyl semimetal: A microscopic approach}},}\ }\href {\doibase https://doi.org/10.1016/j.supcon.2024.100132} {\bibfield  {journal} {\bibinfo  {journal} {Superconductivity}\ }\textbf {\bibinfo {volume} {12}},\ \bibinfo {pages} {100132} (\bibinfo {year} {2024})}\BibitemShut {NoStop}%
\bibitem [{\citenamefont {Roy}\ \emph {et~al.}(2017)\citenamefont {Roy}, \citenamefont {Goswami},\ and\ \citenamefont {Juri\ifmmode \check{c}\else \v{c}\fi{}i\ifmmode~\acute{c}\else \'{c}\fi{}}}]{Roy_PhysRevB_2017}%
  \BibitemOpen
  \bibfield  {author} {\bibinfo {author} {\bibfnamefont {Bitan}\ \bibnamefont {Roy}}, \bibinfo {author} {\bibfnamefont {Pallab}\ \bibnamefont {Goswami}}, \ and\ \bibinfo {author} {\bibfnamefont {Vladimir}\ \bibnamefont {Juri\ifmmode \check{c}\else \v{c}\fi{}i\ifmmode~\acute{c}\else \'{c}\fi{}}},\ }\bibfield  {title} {\enquote {\bibinfo {title} {Interacting weyl fermions: Phases, phase transitions, and global phase diagram},}\ }\href {\doibase 10.1103/PhysRevB.95.201102} {\bibfield  {journal} {\bibinfo  {journal} {Phys. Rev. B}\ }\textbf {\bibinfo {volume} {95}},\ \bibinfo {pages} {201102} (\bibinfo {year} {2017})}\BibitemShut {NoStop}%
\end{thebibliography}
%

\appendix
\section{Green Function}
\label{Appendix A: Green Function}

From the Hamiltonian $\hat{H}_{BdG} (\textbf{k})$, the finite temperature Green function in euclidean time satisfies
\begin{equation}
    \left( \frac{\partial}{\partial\tau} + \hat{H}_{BdG}(\textbf{k}) \right) \hat{\mathcal{G}}_{\textbf{k}}(\tau) = \delta(\tau),
\end{equation}
written in Matsubara frequency space
\begin{equation}
    \hat{\mathcal{G}}_{\textbf{k}}(\omega_{n}) = \int_{0}^{\beta} d\tau\, e^{i\omega_{n}\tau}\hat{\mathcal{G}}_{\textbf{k}}(\tau),
\end{equation}
where $\omega_{n} = (2n+1)\pi/\beta$, with $n\in\mathds{Z}$ is the fermionic Matsubara frequency. In our model we calculate the finite temperature Green function by means
\begin{equation}
    \hat{\mathcal{G}}_{\textbf{k}}(\omega_{n}) = \left[-i\omega_{n} + \hat{H}_{BdG}(\textbf{k})\right]^{-1},
\end{equation}
dividing the Green function in $2\times 2$ matrix blocks, i.e.
\begin{equation}
    \hat{\mathcal{G}}_{\textbf{k}}(\omega_{n}) = \left(
    \begin{matrix}
        \hat{\mathcal{G}}_{\textbf{k}}^{\text{intra},-} & \hat{\mathcal{G}}_{\textbf{k}}^{\text{inter}} \\
        (\hat{\mathcal{G}}_{\textbf{k}}^{\text{inter}})^{\dagger} & \hat{\mathcal{G}}_{\textbf{k}}^{\text{intra},+}
    \end{matrix}
    \right),
\end{equation}
and inverting the $4\times 4$ matrix $-i\omega_{n} + \hat{H}_{BdG}(\textbf{k})$, we obtain
\begin{widetext}
\begin{equation}
    \hat{\mathcal{G}}_{\textbf{k}}^{\text{intra}, \pm} =  \frac{(E_{\textbf{k}}^{2} \pm \delta\mu\bar{\xi}_{\textbf{k}}) \left[ i\omega_{n}\hat{\eta}_{0} + \xi^{\pm}_{\textbf{k}} \hat{\eta}_{3} + \Re\Delta^{\text{intra}}_{\textbf{k}}\hat{\eta}_{1} - \Im\Delta^{\text{intra}}_{\textbf{k}}\hat{\eta}_{2} \right] \pm\delta\mu |\Delta^{\text{inter}}_{\textbf{k}}|^{2} \hat{\eta}_{3} - 2B_{\textbf{k}}(\Re\Delta^{\text{inter}}_{\textbf{k}}\hat{\eta}_{1} - \Im\Delta^{\text{inter}}_{\textbf{k}}\hat{\eta}_{2}) }{E_{\textbf{k}}^{4} - 4B_{\textbf{k}}^{2} - \delta\mu^{2}(\bar{\xi}_{\textbf{k}}^{2} + |\Delta^{\text{inter}}_{\textbf{k}}|^{2})},
\end{equation}
\begin{equation}
\begin{split}
    \hat{\mathcal{G}}_{\textbf{k}}^{\text{inter}} = & \frac{ -2B_{\textbf{k}} \left[ \left( i\omega_{n} + \frac{\delta\mu}{2} \right)\hat{\eta}_{0} + \bar{\xi}_{\textbf{k}} \hat{\eta}_{3} + \Re\Delta^{\text{intra}}_{\textbf{k}}\hat{\eta}_{1} - \Im\Delta^{\text{intra}}_{\textbf{k}}\hat{\eta}_{2} \right] + \delta\mu \Delta^{\text{intra}}_{\textbf{k}} (\Delta^{\text{inter}}_{\textbf{k}})^{*} \hat{\eta}_{0} }{E_{\textbf{k}}^{4} - 4B_{\textbf{k}}^{2} - \delta\mu^{2}(\bar{\xi}_{\textbf{k}}^{2} + |\Delta^{\text{inter}}_{\textbf{k}}|^{2})} \\
    & + \frac{ \left( E_{\textbf{k}}^{2} - \frac{\delta\mu^{2}}{2} \right) (\Re\Delta^{\text{inter}}_{\textbf{k}}\hat{\eta}_{1} - \Im\Delta^{\text{inter}}_{\textbf{k}}\hat{\eta}_{2}) + \omega_{n} \delta\mu (\Im\Delta^{\text{inter}}_{\textbf{k}}\hat{\eta}_{1} + \Re\Delta^{\text{inter}}_{\textbf{k}}\hat{\eta}_{2}) }{E_{\textbf{k}}^{4} - 4B_{\textbf{k}}^{2} - \delta\mu^{2}(\bar{\xi}_{\textbf{k}}^{2} + |\Delta^{\text{inter}}_{\textbf{k}}|^{2})},
\end{split}
\end{equation}
where $()^{\dagger}$ is the conjugate transpose, considering $(i\omega_{n})^{*} = i\omega_{n}$.
\end{widetext}
\section{Gap Equations}
\label{Appendix B: Gap Eqns}

From the matrix elements of Green Function we obtain the following correlation functions
\begin{eqnarray}
    \langle \hat{\alpha}_{-}(\textbf{k}) \hat{\alpha}_{-}(-\textbf{k}) \rangle &=& [\hat{\mathcal{G}} ^{\text{intra},-}_{\textbf{k}}]_{12} \\
    &=& \frac{ \Delta^{\text{intra}}_{\textbf{k}} (E_{\textbf{k}}^{2} - \delta\mu \bar{\xi}_{\textbf{k}}) - 2 \Delta^{\text{inter}}_{\textbf{k}} B_{\textbf{k}} }{E_{\textbf{k}}^{4} - 4B_{\textbf{k}}^{2} - \delta\mu^{2}(\bar{\xi}_{\textbf{k}}^{2} + |\Delta^{\text{inter}}_{\textbf{k}}|^{2})}, \notag
\end{eqnarray}
\begin{eqnarray}
    &&\langle \hat{\alpha}_{-}(\textbf{k}) \hat{\alpha}_{+}(-\textbf{k}) \rangle = [\hat{\mathcal{G}} ^{\text{inter}}_{\textbf{k}}]_{12} \\
    &&= \frac{ \Delta^{\text{inter}}_{\textbf{k}} (E_{\textbf{k}}^{2} - \frac{\delta\mu^{2}}{2} ) - 2 \Delta^{\text{intra}}_{\textbf{k}} B_{\textbf{k}} - i\omega_{n}\delta\mu \Delta^{\text{inter}}_{\textbf{k}} }{E_{\textbf{k}}^{4} - 4B_{\textbf{k}}^{2} - \delta\mu^{2}(\bar{\xi}_{\textbf{k}}^{2} + |\Delta^{\text{inter}}_{\textbf{k}}|^{2})}. \notag
\end{eqnarray}
Substituting in the following gap equations
\begin{eqnarray}
    \Delta^{\text{intra}}_{\textbf{k}} &=& T\sum_{\textbf{k}',\omega_{n}} V_{\text{intra}}(\textbf{k},\textbf{k}') \langle \hat{\alpha}_{-}(\textbf{k}') \hat{\alpha}_{-}(-\textbf{k}') \rangle,  \\
    \Delta^{\text{inter}}_{\textbf{k}} &=& T\sum_{\textbf{k}',\omega_{n}} V_{\text{inter}}(\textbf{k},\textbf{k}') \langle \hat{\alpha}_{-}(\textbf{k}') \hat{\alpha}_{+}(-\textbf{k}') \rangle,
\end{eqnarray}
we obtain
\begin{equation}
    \Delta^{\text{intra}}_{\textbf{k}} = T\sum_{\textbf{k}',\omega_{n}} V_{\text{intra}}(\textbf{k},\textbf{k}') \frac{ \Delta^{\text{intra}}_{\textbf{k}'} (E_{\textbf{k}'}^{2} - \delta\mu \bar{\xi}_{\textbf{k}'}) - 2 \Delta^{\text{inter}}_{\textbf{k}'} B_{\textbf{k}'} }{E_{\textbf{k}'}^{4} - 4B_{\textbf{k}'}^{2} - \delta\mu^{2}(\bar{\xi}_{\textbf{k}'}^{2} + |\Delta^{\text{inter}}_{\textbf{k}'}|^{2})}
\end{equation}
\begin{equation}
    \Delta^{\text{inter}}_{\textbf{k}} = T\sum_{\textbf{k}',\omega_{n}} V_{\text{inter}}(\textbf{k},\textbf{k}') \frac{ \Delta^{\text{inter}}_{\textbf{k}'} (E_{\textbf{k}'}^{2} - \frac{\delta\mu^{2}}{2} ) - 2 \Delta^{\text{intra}}_{\textbf{k}'} B_{\textbf{k}'} }{E_{\textbf{k}'}^{4} - 4B_{\textbf{k}'}^{2} - \delta\mu^{2}(\bar{\xi}_{\textbf{k}'}^{2} + |\Delta^{\text{inter}}_{\textbf{k}'}|^{2})}
\end{equation}
where the term proportional to $i\omega_{n}$ was eliminated in the last equation, since it vanishes when adding in $\omega_{n}$. Then, we define
\begin{equation}
\begin{split}
    \Gamma_{\textbf{k}}^{2} &= \bar{\xi}_{\textbf{k}}^{2} + |\Delta^{\text{intra}}_{\textbf{k}}|^{2} + |\Delta^{\text{inter}}_{\textbf{k}}|^{2} + \frac{\delta\mu^{2}}{4} \\
    & + \sqrt{4B_{\textbf{k}}^{2} + \delta\mu^{2}(\bar{\xi}_{\textbf{k}}^{2} + |\Delta^{\text{inter}}_{\textbf{k}}|^{2})}, \\
\end{split}
\end{equation}
\begin{equation}
\begin{split}
    \gamma_{\textbf{k}}^{2} &= \bar{\xi}_{\textbf{k}}^{2} + |\Delta^{\text{intra}}_{\textbf{k}}|^{2} + |\Delta^{\text{inter}}_{\textbf{k}}|^{2} + \frac{\delta\mu^{2}}{4} \\
    & - \sqrt{4B_{\textbf{k}}^{2} + \delta\mu^{2}(\bar{\xi}_{\textbf{k}}^{2} + |\Delta^{\text{inter}}_{\textbf{k}}|^{2})},
\end{split}
\end{equation}
so the denominator of both gap equations can be written as
\begin{equation}
    E_{\textbf{k}}^{4} - 4B_{\textbf{k}}^{2} - \delta\mu^{2}(\bar{\xi}_{\textbf{k}}^{2} + |\Delta^{\text{inter}}_{\textbf{k}}|^{2}) = (\omega_{n}^{2} + \Gamma_{\textbf{k}}^{2})(\omega_{n}^{2} + \gamma_{\textbf{k}}^{2}),
\end{equation}
and we perform the decomposition into partial fractions to obtain
\begin{widetext}
\begin{equation}
\begin{split}
    \Delta^{\text{intra}}_{\textbf{k}} = \frac{T}{2} \sum_{\textbf{k}',\omega_{n}} V_{\text{intra}}(\textbf{k},\textbf{k}') & \left[ \left( \Delta^{\text{intra}}_{\textbf{k}'} + \frac{ 2 \Delta^{\text{inter}}_{\textbf{k}'} B_{\textbf{k}'} + \delta\mu \bar{\xi}_{\textbf{k}'} \Delta^{\text{intra}}_{\textbf{k}'} }{\sqrt{4B_{\textbf{k}'}^{2} + \delta\mu^{2}(\bar{\xi}_{\textbf{k}'}^{2} + |\Delta^{\text{inter}}_{\textbf{k}'}|^{2})}} \right) \frac{1}{\omega_{n}^{2} + \Gamma_{\textbf{k}'}^{2}} \right. \\ 
    & \left. + \left( \Delta^{\text{intra}}_{\textbf{k}'} - \frac{ 2 \Delta^{\text{inter}}_{\textbf{k}'} B_{\textbf{k}'} + \delta\mu \bar{\xi}_{\textbf{k}'} \Delta^{\text{intra}}_{\textbf{k}'} }{\sqrt{4B_{\textbf{k}'}^{2} + \delta\mu^{2}(\bar{\xi}_{\textbf{k}'}^{2} + |\Delta^{\text{inter}}_{\textbf{k}'}|^{2})}} \right) \frac{1}{\omega_{n}^{2} + \gamma_{\textbf{k}'}^{2}} \right],
\end{split}
\end{equation}
\begin{equation}
\begin{split}
    \Delta^{\text{inter}}_{\textbf{k}} = \frac{T}{2} \sum_{\textbf{k}',\omega_{n}} V_{\text{inter}}(\textbf{k},\textbf{k}') & \left[ \left( \Delta^{\text{inter}}_{\textbf{k}'} + \frac{ 2 \Delta^{\text{intra}}_{\textbf{k}'} B_{\textbf{k}'} + \frac{\delta\mu^{2}}{2}  \Delta^{\text{inter}}_{\textbf{k}'} }{\sqrt{4B_{\textbf{k}'}^{2} + \delta\mu^{2}(\bar{\xi}_{\textbf{k}'}^{2} + |\Delta^{\text{inter}}_{\textbf{k}'}|^{2})}} \right) \frac{1}{\omega_{n}^{2} + \Gamma_{\textbf{k}'}^{2}} \right. \\ 
    & \left. + \left( \Delta^{\text{inter}}_{\textbf{k}'} - \frac{ 2 \Delta^{\text{intra}}_{\textbf{k}'} B_{\textbf{k}'} + \frac{\delta\mu^{2}}{2}  \Delta^{\text{inter}}_{\textbf{k}'} }{\sqrt{4B_{\textbf{k}'}^{2} + \delta\mu^{2}(\bar{\xi}_{\textbf{k}'}^{2} + |\Delta^{\text{inter}}_{\textbf{k}'}|^{2})}} \right) \frac{1}{\omega_{n}^{2} + \gamma_{\textbf{k}'}^{2}} \right].
\end{split}
\end{equation}
\end{widetext}
Finally, we use the identity
\begin{equation}
    T\sum_{\omega_{n}} \frac{1}{\omega_{n}^{2} + c^{2}} = \frac{\tanh\left(\dfrac{c}{2T}\right)}{c},
\end{equation}
to obtain the gap equations of section \ref{Section: Gap Eqns}.
\section{Density of States Function}
\label{Appendix C: Continuum limit}

To take continuous limit of the form
\begin{equation}
    \sum_{\textbf{k}} f(\Omega_{\textbf{k}}, \xi_{\textbf{k}}) \approx \int\frac{d^{3}k}{(2\pi)^{3}} f(\Omega_{\textbf{k}}, \xi_{\textbf{k}}),
\end{equation}
we define the following density of states function
\begin{eqnarray}
    \rho(\xi) &\equiv & \int\frac{d^{3}k}{(2\pi)^{3}} \delta(\xi - \xi_{\textbf{k}}) \notag \\
    &= & \int\frac{d^{3}k}{(2\pi)^{3}} \delta\left(\xi + \mu - v_{f} \sqrt{k_{z}^{2} + \alpha^{2}k_{\perp}^{2\nu}} \right) \\
    &= & \frac{1}{(2\pi)^{2} v_{f}} \left( \frac{\xi+\mu}{\alpha v_{f}} \right)^{2/\nu} \frac{\beta(\frac{1}{\nu}, \frac{1}{2})}{\nu}. \notag
\end{eqnarray}
Then
\begin{equation}
    \int\frac{d^{3}k}{(2\pi)^{3}} f(\Omega_{\textbf{k}}, \xi_{\textbf{k}}) = \int_{-\omega_{D}}^{\omega_{D}} d\xi \int\frac{d^{3}k}{(2\pi)^{3}} \delta(\xi - \xi_{\textbf{k}}) f(\Omega_{\textbf{k}}, \xi).
\end{equation}
Using cylindrical coordinates $d^{3}k = k_{\perp} dk_{\perp} d\phi dk_{z}$, we use the properties of the $\delta$ function to solve the integral in $dk_{z}$, and then we make the change of variable $x = \left( \frac{\alpha v_{f}}{\xi+\mu} \right)^{1/\nu} k_{\perp}$ to obtain
\begin{widetext}
\begin{equation}
\begin{split}
    \int\frac{d^{3}k}{(2\pi)^{3}} f(\Omega_{\textbf{k}}, \bar{\xi}_{\textbf{k}}) &= 2 \int_{-\omega_{D}}^{\omega_{D}}  \frac{d\xi}{(2\pi)^{2} v_{f}} \left( \frac{\xi+\mu}{\alpha v_{f}} \right)^{2/\nu} \int_{0}^{1}\frac{dx x}{\sqrt{1-x^{2\nu}}} \int_{0}^{2\pi} \frac{d\phi}{2\pi} f(\theta_{x}, \phi, \xi) \\
    &= \frac{2\nu}{\beta(\frac{1}{\nu}, \frac{1}{2})} \int_{-\omega_{D}}^{\omega_{D}} d\xi \rho(\xi) \int_{0}^{1}\frac{dx x}{\sqrt{1-x^{2\nu}}} \int_{0}^{2\pi} \frac{d\phi}{2\pi} f(\theta_{x}, \phi, \xi),
\end{split}
\end{equation}
\end{widetext}
where $\theta_{x}$ is defined by
\begin{equation}
    \tan\theta_{x} = \frac{cx}{\sqrt{1-x^{2\nu}}},
\end{equation}
with
\begin{equation}
    c = \alpha^{-1/\nu} \left( \frac{\xi+\mu}{v_{f}} \right)^{\frac{1}{\nu} - 1} \approx \alpha^{-1/\nu} \left( \frac{\mu}{v_{f}} \right)^{\frac{1}{\nu} - 1}.
\end{equation}

We make the approximation of evaluating $\rho(\xi)$ at $\xi=0$, since the density of states is centered at that value. Finally the integral remains
\begin{widetext}
\begin{equation}
    \int\frac{d^{3}k}{(2\pi)^{3}} f(\Omega_{\textbf{k}}, \bar{\xi}_{\textbf{k}}) = \frac{\rho(0)}{B_{\nu}} \int_{0}^{1}\frac{dx x}{\sqrt{1-x^{2\nu}}} \int_{0}^{2\pi} \frac{d\phi}{2\pi} \int_{-\omega_{D}}^{\omega_{D}} d\xi f(\theta_{x}, \phi, \xi),
\end{equation}
with $B_{\nu} = \beta(\frac{1}{\nu}, \frac{1}{2})/(2\nu)$
\end{widetext}

\end{document}